\documentclass[]{jfm}

\usepackage{graphicx}
\usepackage{newtxtext}
\usepackage{newtxmath}
\usepackage{natbib}
\usepackage[hidelinks]{hyperref}
\hypersetup{
    colorlinks = true,
    urlcolor   = blue,
    citecolor  = blue
}
\usepackage[justification=justified,
   format=plain,width=\textwidth]{caption}

\newcommand{\RomanNumeralCaps}[1]

\graphicspath{{figures/}}
\usepackage{amsmath}
\usepackage[inkscapeformat=png]{svg}
\usepackage{calrsfs}

\DeclareMathOperator{\nus}{\mathit{Nu}}

\DeclareMathOperator{\ra}{\mathit{Ra}}

\usepackage{color}

\newcommand{\pd}{{\mathsf{PD}}}

\usepackage{rotating}
\usepackage{pdflscape}
\usepackage{afterpage}

\usepackage[inline]{trackchanges}
\addeditor{MDP}
\addeditor{LK}

\title[Flow morphology and patterns in porous media convection]{Flow morphology and patterns in porous media convection: A persistent homology analysis}

\author{
Marco De Paoli\aff{1,2}\corresp{\email{marco.de.paoli@tuwien.ac.at}},
Sergio Pirozzoli\aff{3},
Catherin Neena Lalu\aff{4} and 
Lou Kondic\aff{5}
}
\affiliation{
\aff{1}Institute of Fluid Mechanics and Heat Transfer, TU Wien, 1060 Vienna, Austria
\aff{2}Physics of Fluids Group and Max Planck Center for Complex Fluid Dynamics and J. M. Burgers Centre for Fluid Dynamics, University of Twente, P.O. Box 217 7500AE Enschede, The Netherlands
\aff{3}Dipartimento di Ingegneria Meccanica e Aerospaziale, Sapienza Università di Roma, Rome, Italy
\aff{4}Department of Physics, New Jersey Institute of Technology, Newark, New Jersey 07102, USA
\aff{5}Department of Mathematical Sciences, New Jersey Institute of Technology, Newark, New Jersey 07102, USA
}
\shortauthor{De Paoli M. et al.}

\begin{document}
\maketitle

\begin{abstract}
Convective mixing in porous media is crucial in both geophysical and industrial fields, spanning applications ranging from carbon dioxide sequestration to geothermal energy extraction. Key processes are affected by convective heat transport or diffusion of chemical species in porous formations. Intense convection flow and mixing create complex, dynamic patterns that are difficult to predict and measure. The present work focuses on the use of topological data analysis, in particular, the measures emerging from the growing field of persistent homology (PH), to quantify these patterns.  These measures are objective and quantify structures across all temperature or concentration values simultaneously. These techniques, when applied to classical porous media setups, such as one-sided and Rayleigh-Bénard flow configurations, provide new insights into the system's structure, flow patterns, and macroscopic mixing properties. Using large datasets we make publicly available, comprising original simulations as well as those presented in previous works, we correlate the behaviour of the heat transport rate (quantified by the Nusselt number) with the evolution of the flow structures (quantified by the PH measures). Finally, we provide a detailed analysis of the flow evolution over a wide range of governing parameters, namely the Rayleigh-Darcy number and the domain size.
\end{abstract}


\section{Introduction}\label{sec:rt_intro}
Transport of heat and chemical species in porous media is relevant to natural and industrial flows: from latent heat thermal energy storage systems \citep{trelles2003numerical,xu2017evaluation} to the formation of sea ice \citep{feltham2006,wells2019mushy}, several key processes are controlled by the redistribution of heat and solutes in confined domains. When the motion is driven by density gradients within the fluid layer, and the density field depends on the local distribution of the scalar (e.g., temperature or solute concentration), the flow is controlled by natural convection.
In porous-media convection, the governing equations for thermally- and solute-driven flows are equivalent once expressed in dimensionless form, with temperature and solute concentration playing analogous roles as buoyancy-driving scalars. As a result, insights obtained from one formulation can be directly transferred to the other. In the present work, we adopt a thermal formulation of the problem, which provides a convenient and widely used framework while retaining full relevance for solutal convection applications.
In convective flows, local density differences are contrasted by the dissipative mechanisms of friction (due to narrow pore spaces) and diffusion (which reduce the gradients of the scalar field), which, in turn, affect the flow field. The relative importance of driving (convection) and dissipative (diffusion, friction) mechanisms is quantified by the Rayleigh-Darcy number, $\ra$ (hereinafter defined as Rayleigh number). A similar dynamics occurs in the presence of key geophysical systems, where buoyancy is induced by solute concentration gradients rather than temperature variations, e.g., geothermal flows in underground sites \citep{hu2023effects}, thawing of permafrost \citep{wang2025permafrost}, dispersion of contaminants in groundwater flows \citep{simmons2001variable,depaoli2025solute}, and storage of carbon dioxide (CO$_2$) in saline aquifers. The latter, in particular, has been extensively studied in recent decades, due to its enormous relevance in mitigating the effects of climate change \citep{metz2005carbon}. Geological sequestration of carbon dioxide involves injecting large amounts of CO$_2$ into underground geological porous formations for permanent storage \citep{Emami-Meybodi2015,jin2024pore}. 
A key question in assessing the suitability of potential sequestration sites is determining the CO$_2$ mixing rate in brine. 

Although the physical motivation for this study is rooted in solutal convection relevant to CO$_2$ sequestration, the analysis presented here is carried out using a thermal formulation. Owing to the formal equivalence between thermal and solutal convection in porous media, all results can be directly reinterpreted in terms of mass transfer, with the dimensionless governing and response parameters retaining the same physical meaning.
The archetypal system used to analyse this problem is the one-sided (or semi-infinite) configuration \citep{hidalgo2012scaling}: the flow is idealized as a rectangular box, initially filled with low-density fluid (representing brine). 
At the top boundary (representing the CO$_2$-brine interface) the fluid density is maximum \citep{jin2024pore}. At the bottom boundary, the domain is assumed to be closed due to the presence of an impermeable rock layer. 
The flow dynamics that takes place is marked by a transient behaviour, controlled by the convective flow structures that populate the system. 
Numerical simulations have been extensively deployed to analyse these flows \citep{hewitt2013convective,slim2014solutal,wen_2018}. Despite these efforts, several research questions remain unanswered. For sufficiently large domains and high values of Rayleigh numbers, the flow evolution is still transient, but it is also independent of the domain size and $\ra$ \citep{depaoli2025grl}.  What is the minimal flow unit required to achieve such a self-similar behaviour? And what is the effect of the domain size on the mixing properties of the flow? In this work, we answer these questions by systematically analysing the effects of the domain size and of the Rayleigh number $\ra$ of the emerging flow patterns.

The one-sided configuration described above presents a major drawback: it is intrinsically transient, which remarkably complicates any theoretical analysis of the flow. In contrast, the ``two-sided'' (Rayleigh-B\'enard) configuration, in which the fluid density is fixed at top and also at the bottom, attains a (statistically) steady state, which makes it more suitable for theoretical analysis. In addition, it has been shown \citep{hewitt2013convective} that the two-sided flow is directly related to the one-sided configuration during the late-stage dynamics, indicated as the ``shutdown of convection'' and corresponding to a well-mixed bulk flow configuration. The two-sided system is well-defined in terms of boundary conditions, and during the statistically steady state, it is possible to derive exact equations that describe the flow transport properties as a function of the governing parameters \citep{grossmann2001thermal,grossmann2000scaling,lohse2024ultimate}. For these reasons, a number of works \citep{hewitt2012ultimate,wen2015structure,hewitt2014high,depaoli2016influence,pirozzoli2021towards,depaoli2022strong,hu2023effects,depaoli2024heat} have investigated the two-sided configuration. However, several questions remain unanswered in this case as well. The near-wall flow structure organizes into multiple hierarchical levels, from the small-scale (near-wall) plumes nesting into large-scale structures \citep[labelled as `supercells',][]{depaoli2022strong}, which have been only partially characterized by analyzing scalar fields after applying ad-hoc threshold filtering. Can the supercells be identified more robustly and univocally? Can their formation be predicted based on the values of the domain size and $\ra$ considered? Can a $\ra$-independent description of the near-wall small-scale structures be derived?  What is the flow structure at small and large $\ra$ in the presence of large domains?

As a result of the complex interplay between diffusion and buoyancy, the flow structures organize into tangled, dynamic patterns, making it extremely challenging to predict and quantify their behaviour. Initial efforts in this direction include using Fourier-based analysis \citep{hewitt2014high,pirozzoli2021towards,depaoli2025grl} as well as utilizing cell sizes as a measure of emerging patterns \citep{fu2013pattern,depaoli2022strong}. However, there is much more that can be done in this direction. One approach that appears appropriate is based on topological measures, which provide an efficient and detailed analysis of structures characterized by a value of a scalar field (in this case, solute concentration). While several approaches of varying complexity can be implemented for this purpose, measures based on the exploration of the connectivity of the considered structures appear to be the most appropriate. Such approaches, based on computational topology discipline, persistent homology, have been used to analyse a variety of systems emerging from materials science, such as granular matter~\citep{epl12,physicaD14,pre14,D4SM00560K}, Rayleigh-B\'enard convection~\citep{kramar_physD_2016}, suspensions~\citep{gameiro_prf_2020}, and porous media~\citep{Suzuki2021}, among others.  One advantage of using persistence-homology-based measures is that they are objective in the sense that they do not require specifying a threshold value (such as average temperature or a similar quantity), but instead, they quantify the structure across all thresholds at once. We will see that persistence-homology-based approaches provide significant new information about the structures developing in the considered system, their connection to heat transport,  and in particular, that they allow for answering some questions formulated so far.

The paper is organized as follows: in \S\ref{sec:darcy} we present the governing equations and flow configurations considered, and discuss the equivalence between thermal and solutal formulations in porous-media convection.
In~\S\ref{sec:topo}, we provide an overview of the persistent homology methods used to quantify the emerging patterns. The results of the one-sided and two-sided systems analysed are presented in~\S\ref{sec:OS} and~\S\ref{sec:RB}, respectively. Finally, an overview of the work and conclusions are presented in \S\ref{sec:concl}.

\section{Governing equations}\label{sec:darcy}
We consider convective porous media flows at the Darcy scale, i.e., the flow properties averaged over a Reference Elementary Volume (REV) \citep{whitaker1998method,nield2017convection}. In particular, we focus on density-driven flows in which the source of buoyancy is due to a temperature-induced density field, and with the solid locally in thermal equilibrium with the liquid phase.  We consider thermally driven flows, as they are relevant to both the one-sided and two-sided configurations~\citep{hu2023effects}. However, the same dimensionless equations apply to solutal convection. 
The governing equations in dimensional and dimensionless form are presented in \S\ref{sec:problem} and in \S\ref{sec:adimeq}, respectively, and the data analysed are discussed in \S\ref{sec:num}.

\subsection{Problem formulation}\label{sec:problem}
\begin{figure}
    \centering
    \includegraphics[width=0.85\linewidth]{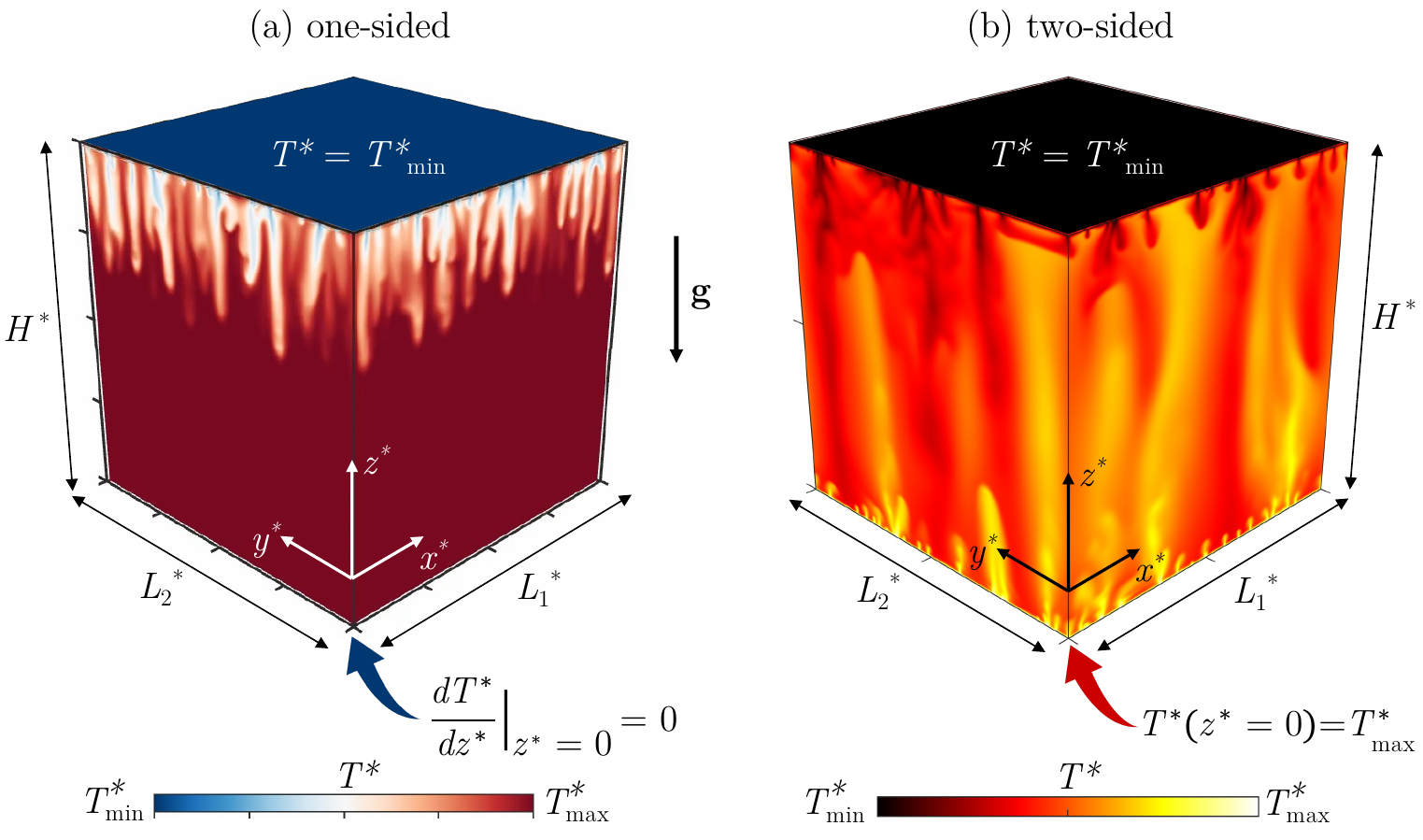}
    \caption{
    Temperature distribution over the surface of the domain, with indication of the dimensional extension in each direction ($L_1^*,L_2^*,H^*$) and of the boundary conditions. (a)~One-sided flow configuration, discussed in \S\ref{sec:OS}.
    (b)~Two-sided flow configuration, discussed in \S\ref{sec:RB}.
    }
    \label{fig:fig1}
\end{figure}

We consider a 3D porous domain with uniform porosity $\phi$ and permeability $K$.  The medium is fully saturated with fluid, whose density depends on the temperature field, $T^*$. The local gradients of fluid density within the system drive the flow. Figure~\ref{fig:fig1} shows a sketch of the domain; here, two different flow configurations, discussed in detail in the following, are described. 

We indicate by $x^*$ and $y^*$ the horizontal directions, by $z^*$ the vertical direction (perpendicular to the walls) along which the gravitational acceleration $\mathbf {g}$ is aligned. The temperature field $T^*$, averaged over the REV, varies between $T^*_\text{min}$ and $T^*_\text{max}$, and it is controlled by the advection-diffusion equation \citep{nield2017convection}:
\begin{equation}
    [(1-\phi)(\rho c)_s + \phi(\rho c)_f] \frac{\partial T^*}{\partial t^*} + (\rho c)_f \mathbf{u}^*\cdot \nabla^* T^* = \nabla^*\cdot([(1-\phi)\lambda_s + \phi \lambda_f] \nabla^* T^*),
    \label{eq:nb2017a}
\end{equation}
where $s$ and $f$ refer to the solid and fluid phases, respectively, $c$ is the specific heat (considered at constant pressure for the fluid) and $\lambda$ the thermal conductivity (the superscript $^{*}$ indicates dimensional variables). 
In this case, we consider a system in local thermal equilibrium ($T_f = T_s = T$) and without heat production.
In addition, when the heat capacity of the solid is negligible ($(\rho c)_s\to0$), Eq.~\eqref{eq:nb2017a} reduces to 
\citep{hewitt2020vigorous}
\begin{equation}
\phi\frac{\partial T^*}{\partial t^*}+\mathbf{u}^{*}\nabla^{*} \cdot T^* =\nabla^{*} \cdot (\kappa \nabla^{*} T^*)\text{  ,} 
\label{eq:eqadim3}
\end{equation}
where $t^*$ is time, $\mathbf{u}^{*}=(u^{*},v^{*},w^{*})$ is the volume-averaged velocity field, $\phi$ is the porosity of the medium, and $\kappa = [(1-\phi)\lambda_s + \phi \lambda_f]/(\rho c)_f$ is the thermal diffusivity of the medium, considered constant here. 
The fluid density, $\rho^{*}$, is assumed to depend linearly on temperature:
\begin{equation}
\rho^{*}(T^*)=\rho^{*}(T^*_\text{min})-\Delta\rho^{*}\frac{T^*-T^*_\text{min}}{T^*_\text{max}-T^*_\text{min}}
\text{  ,} 
\label{eq:eqadim4}
\end{equation}
with $\Delta\rho^{*}=\rho^{*}(T^*_\text{min})-\rho^{*}(T^*_\text{max})$ being the maximum density contrast within the domain. Assuming the validity of the Boussinesq approximation \citep{landman2007heat}, the flow field is described by the continuity and the Darcy equations,
\begin{equation}
\nabla^{*}\cdot\mathbf{u}^{*}=0\quad,\quad
\mathbf{u}^{*}=-\frac{K}{\mu}\left(\nabla^{*} P^{*}+\rho^{*}g \mathbf{k}\right) \text{  ,} 
\label{eq:eqadim2}
\end{equation}
respectively, with $\mu$ the (constant) fluid viscosity, $P^{*}$ the pressure and $\mathbf{k}$ the vertical unit vector. The walls are considered impermeable to the fluid, i.e., the following boundary conditions for the flow field apply:
\begin{equation}
    w^*(z^*=0)=0\quad,\quad 
    w^*(z^*=H^{*})=0 . 
\label{eq:bcnopenet}
\end{equation}
Note that a slip at the walls is possible, due to the Darcy formulation adopted. Periodicity is considered across the vertical boundaries (directions $x,y$) for all flow variables.

Regarding the temperature field, at the upper and lower walls, both Dirichlet ($T^*$ fixed) or Neumann ($\partial T^*/\partial z^*$ fixed) boundary conditions are considered:
\begin{itemize}
    \item[(A)] Semi-infinite domain: the temperature is fixed at the top and no flux condition is assumed at the bottom (see figure~\ref{fig:fig1}a), namely:
\begin{equation}
T^*(z^*=H^*)=T^*_\text{min} \quad,\quad 
\frac{\partial T^*}{\partial z^*}\biggr|_{z^*=0} =0 .
    \label{eq:bcos}
\end{equation}
In addition, the domain is initially filled with still fluid and at temperature $T^*_\text{max}$.
Thus, the velocity and temperature fields are initialized as $\mathbf{u^*}=0$ and
    \begin{equation}
    T^*(x^*,y^*,z^*<H^*,t_0^*)=T^*_\text{min} + (T^*_\text{max}-T^*_\text{min})\,\text{erf}\left[\sqrt{\frac{\phi}{\kappa}}\frac{(H^*-z^*)}{2\sqrt{t^*_0}}\right],
    \label{eq:osic1a}
    \end{equation}
respectively. 
The initial condition~\eqref{eq:osic1a} is obtained evaluating at time $t_0^*$ the self-similar solution of~\eqref{eq:eqadim3} in the absence of convection ($\mathbf{u}^*=0$) in a semi-infinite domain $(H^*\to\infty)$ \citep{depaoli2017dissolution}.
The instant considered to initialize the simulation is $t_0^*=250 \phi\kappa (g \Delta \rho^{*} K / \mu)^{-2}$, corresponding to a dimensionless time of $\widehat{t}_0=250$ or $t_0=250/\ra$, expressed in diffusive or convective units, respectively (see~\S\ref{sec:adimeq} for further details).
Finally, a random perturbation (white noise) is added.
The amplitude of this noise controls the time at which the plumes form, i.e., the onset of convection occurs \citep{depaoli2025grl}. This flow configuration, also labelled as ``one-sided'', is representative of a domain cooled from above and insulated from below, and will be analysed in \S\ref{sec:OS}. \item[(B)] Two-sided configuration: the temperature is imposed at both boundaries (see figure~\ref{fig:fig1}b), such that the corresponding density profile is unstable: 
\begin{equation}
T^*(z^*=H^*)=T^*_\text{min}\quad,\quad T^*(z^*=0)=T^*_\text{max}.
    \label{eq:bcrb}
\end{equation}
The velocity field is again initialized as $\mathbf{u^*}=0$, while a linear profile is used for the temperature distribution,
\begin{equation}
T^*(x^*,y^*,z^*,t^*=0) = T^*_\text{max} - \frac{z^*}{H^*}(T^*_\text{max}-T^*_\text{min})\, ,
    \label{eq:bcrb2}
\end{equation}
corresponding to the initial condition labelled as IC1 in \citet{hewitt2014high}. This configuration, labelled as ``Rayleigh-B\'enard'' or ``two-sided'', is analysed in \S\ref{sec:RB}.
\end{itemize}

\subsection{Dimensionless equations}\label{sec:adimeq}
Depending on whether the domain height $H^*$ or an intrinsic diffusive length scale $\ell^*$ (defined later) is used to make the flow quantities dimensionless, two different formulations can be derived (convective or diffusive units, both used in this work), described in \S\ref{sec:dimeq34} and \S\ref{sec:dimeq35}, respectively.

\subsubsection{Convective units}\label{sec:dimeq34}
Natural flow scales relevant to the convective system considered are the buoyancy velocity, $\mathcal{U}^{*}=g \Delta \rho^{*} K / \mu$, and the domain height, $H^{*}$. Using the following set of dimensionless variables,
\begin{equation}
T=\frac{T^*-T^*_\text{min}}{T^*_\text{max}-T^*_\text{min}},\quad x=\frac{x^{*}}{H^{*}},\quad \mathbf{u}=\frac{\mathbf{u}^{*}}{\mathcal{U}^{*}},
\label{eq:eqadim5aaa}
\end{equation}
\begin{equation}
t=\frac{t^*}{\phi H^{*}/\mathcal{U}^{*}},\quad p=\frac{p^{*}}{\Delta \rho^{*}gH^{*}} ,
\label{eq:eqadim5}
\end{equation}
and introducing the reduced pressure $p^{*}=P^*-\rho^*(T^*_\text{min})gz^*$, we obtain the dimensionless form of the governing equations~\eqref{eq:eqadim3}-\eqref{eq:eqadim2}:
\begin{equation}
\label{eq:equ1bis1}
\frac{\partial T}{\partial t} + \mathbf{u} \cdot \nabla T = \frac{1}{\ra} \nabla^2 T ,
\end{equation}
\begin{equation}
\nabla\cdot\mathbf{u}=0,
\label{eq:equ1bis2}
\end{equation}
\begin{equation}
\mathbf{u}=-\left( \nabla p - T \mathbf{k}\right) ,
\label{eq:equ1bis3}
\end{equation}
where  
\begin{equation}
\ra=\frac{g \Delta \rho^{*} K H^{*} }{ \kappa \mu}=\frac{\mathcal{U}^{*} H^{*}}{\kappa}.
\label{eq:rada}
\end{equation}
The flow is controlled by three dimensionless parameters: the Rayleigh number $\ra$, and the domain width in horizontal directions, $L_1=L^*_1/H^{*}$ and $L^*_2/H^{*}$.

A distinction should be made depending on the behaviour of the solid phase concerning the transported scalar. When the solid phase is impermeable to the scalar field considered, e.g., in case of mass transport problems, the definition~\eqref{eq:rada} of the Rayleigh number changes as $\ra=\mathcal{U^*}H^*/(\kappa\phi)$.  However, the dimensionless formulation of the problem is the same for both permeable and impermeable behaviour, provided that when the permeable case is considered, the two phases are in local thermal equilibrium. Further details on this matter are provided by \citet{depaoli2023review}.

\subsubsection{Diffusive units}\label{sec:dimeq35}
An alternative nondimensionalization approach involves using $\ell^*=\kappa/\mathcal{U}^*$ as the reference length scale. This strategy was introduced by \citet{fu2013pattern} and is suitable for describing local flow dynamics that is not influenced by the largest scales of the flow. Defining the dimensionless variables as
\begin{equation}
T=\frac{T^*-T^*_\text{min}}{T^*_\text{max}-T^*_\text{min}},\quad \widehat{x}=\frac{x^{*}}{\ell^{*}},\quad \mathbf{u}=\frac{\mathbf{u}^{*}}{\mathcal{U}^{*}},
\label{eq:eqadim5aaab}
\end{equation}
\begin{equation}
\widehat{t}=\frac{t^*}{\phi \ell^{*}/\mathcal{U}^{*}},\quad \widehat{p}=\frac{p^{*}}{\Delta \rho^{*}g \ell^{*}} ,
\label{eq:eqadim5b}
\end{equation}
we obtain the following dimensionless form of the governing equations~\eqref{eq:eqadim3},~\eqref{eq:eqadim2}:
\begin{equation}
\label{eq:equ1bis1b}
\frac{\partial T}{\partial \widehat{t}} + \mathbf{u} \cdot \widehat{\nabla} T = \widehat{\nabla}^2 T ,
\end{equation}
\begin{equation}
\widehat{\nabla}\cdot \mathbf{u}=0 , 
\label{eq:equ1bis2b}
\end{equation}
\begin{equation}
\mathbf{u}=-\left( \widehat{\nabla} \widehat{p} - T \mathbf{k}\right) ,
\label{eq:equ1bis3b}
\end{equation}
where $\widehat{\cdot}$ indicates variables made dimensionless with respect to diffusive units. Note that in this case $\ra$ does not appear explicitly in the equations; instead, it represents the dimensionless height of the system, $\ra=\mathcal{U}^* H^{*}/ \kappa =H^*/\ell^*$.

\subsection{Data analysed}\label{sec:num}
The data analysed in this work consist of a collection of results presented in previous works \citep{pirozzoli2021towards,depaoli2022strong,depaoli2025grl} and original simulations presented here for the first time (see Appendix~\ref{sec:appA} for a detailed description of the database).  The relevant part of the data analysed is made freely available via \citet{databaseC}.

The original simulations presented here (simulations `A' in table~\ref{tab:os}) are obtained as follows.  The governing equations~\eqref{eq:equ1bis1}~-~\eqref{eq:equ1bis3} are solved numerically with the aid of an (open-source) second-order finite difference solver AFiD-Darcy \citep{depaoli2025code}, which was previously employed to simulate convective porous media flows \citep{depaoli2025solute,depaoli2025afid}. The code comprises an efficient parallel solver for simulating convective, incompressible, and wall-bounded flows in porous media. This solver is based on the initial version of AFiD developed for turbulent flows \citep{van2015pencil}. The algorithm utilizes a pressure-correction scheme in conjunction with an efficient Fast Fourier Transform-based solver. Additional details on the algorithm are described by~\citet{depaoli2025afid}.

\section{Overview of topological methods used for quantification of the emerging patterns: Persistent Homology (PH)}\label{sec:topo}

For the present purposes, one can think of Persistent Homology PH as a tool describing a complicated spatial pattern in the form of so-called persistence diagrams ($\pd$s). These diagrams are obtained by filtration, i.e.\ thresholding the scalar field of interest (such as temperature), in such a manner that only the areas characterized by the value above the specified threshold appear. Then, the simplest $\pd$, which we refer to as $\beta_0$ $\pd$, essentially traces how the regions of the high value of the considered scalar field appear as a filtration level is decreased, or disappear as two regions merge ($\beta_0$ stands here for the 0th Betti number that essentially counts the number of regions that do not involve holes). In two spatial dimensions (2D) (the case that will be of interest to us in the present work, since we will consider temperature distributions on planes extracted from 3D domains), there are two Betti numbers, $\beta_0$ and $\beta_1$ corresponding to components and loops, and therefore also two $\pd$s (in the interest of simplicity of notation, we use, e.g., $\beta_0$ to refer to both Betti number (the number of components at a given threshold value), and to the concept of a connected component). In 3D, there is an additional  $\beta_2$, counting enclosed 3D structures, and therefore also additional $\pd$.  In the present work we focus on 2D structures only; the consideration of 3D aspects of the results is left for future work.

We note, and will discuss later in this section, that $\pd$s contain information that is significantly more detailed than the Betti numbers, since they reveal the connectivity of the components over all thresholds, while Betti numbers are threshold-dependent. The reader is referred to~\citet{pre14,physicaD14} for a more extensive discussion of $\pd$s in the context of materials science problems.  We note that Betti numbers were already used for morphology characterization~\citep{blunt2017multiphase}; an alternative approach based on Euclidean Distance Transform quantifying pore geometry was considered in \citet{Suzuki2021}. 

\subsection{Persistent Homology: an example}
\label{sec:toy}

\begin{figure*}
\centering
\includegraphics[width=0.9\columnwidth]{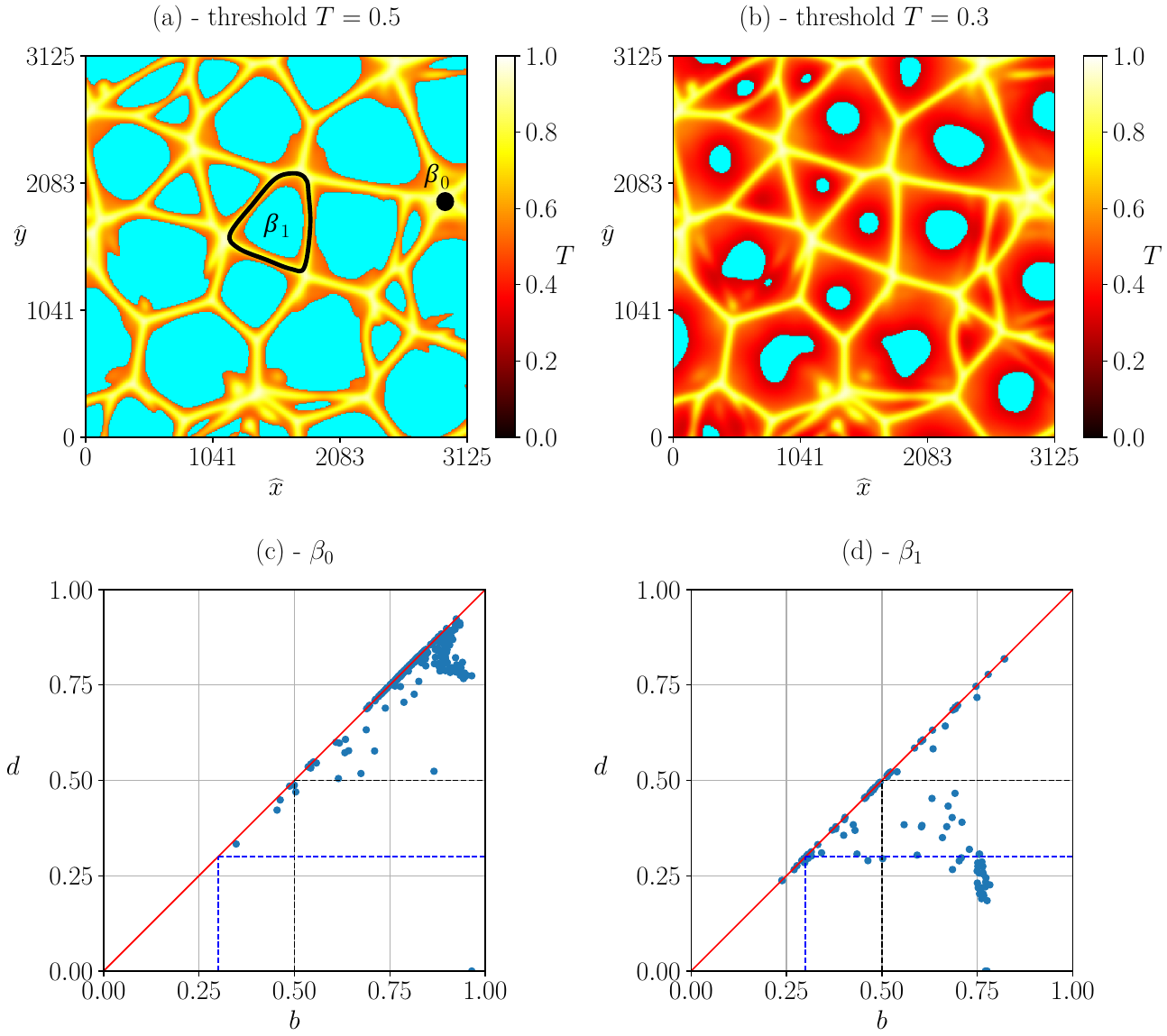}
\caption{2D example illustrating $\pd$s for a function of two spatial coordinates, $x$ and $y$  (here we use the temperature results that will be discussed later in the text). Parts~(a,~b) show the results for chosen thresholds of $T = 0.5$~(a) and $T = 0.3$~(b) - the light-blue areas are the ones where $T$ is less than the specified threshold value. Plots~(c) and~(d) show $\pd$s corresponding to $\beta_0$ (components) and to $\beta_1$ (holes), respectively, and schematic examples of which are reported in~(a); here the $T = 0.5$ (black dashed line) and $T = 0.3$ (blue dashed line) show the threshold levels. The fields shown in (a,~b) are two snapshots from simulation C5 (see table~\ref{tab:rb} for additional details). We refer to movie~\textcolor{red}{S1} for a visual interpretation of the threshold-dependent temperature field and of the corresponding $\pd$s.}
\label{fig:2D_toy}
\end{figure*}

Here we discuss a two-dimensional example. The basic geometric structures of interest are connected components and loops (holes), denoted by $\beta_0$ and $\beta_1$, respectively. To illustrate the topological quantities describing such structures, we use the simple 2D example shown in figure~\ref{fig:2D_toy}. The reader is referred to \citet{pre13,physicaD14} for a more in-depth discussion, along with simpler one-dimensional examples; additional toy examples in the context of granular media analysis can be found in~\citet{pre14,D4SM00560K} and in the context of porous media flow in~\citet{Illingworth26}.  Specifics related to computations of $\pd$s are discussed later in \S\ref{sec:PH_spec}.

 The example shown in figure~\ref{fig:2D_toy} is a two-sided convection simulation from \S\ref{sec:RB}, discussed later in the text (settings: $\ra=500$, case~C5 in table~\ref{tab:rb}, field taken at the boundary layer), with the colour map showing the temperature.   Figure~\ref{fig:2D_toy} shows two threshold values, namely $0.5$ in figure~\ref{fig:2D_toy}(a) and $0.3$ in figure~\ref{fig:2D_toy}(b); the parts of the domain characterized by the temperature values lower than the specified threshold are not shown and appear light blue. By comparing parts~\ref{fig:2D_toy}(a) and~\ref{fig:2D_toy}(b), we observe a change in connectivity and structure as the threshold value changes.  

The change in connectivity, although visually apparent, needs to be quantified.
Such quantification can be established by considering persistence diagrams, $\pd$s.  Let us consider first the persistence diagram corresponding to components, $\beta_0$ $\pd$, shown in figure~\ref{fig:2D_toy}c (see also movie~\textcolor{red}{S1} in the online supplementary material). As one descends from high values of temperature to lower ones, separated areas of high temperature appear (are born); each of these areas is called a component. Decreasing the threshold causes these components to merge; whenever two components merge, the `younger' one (born at a lower temperature value) disappears (dies). The coordinates of each point (generator) in $\beta_0$ $\pd$ specify birth ($b$) and death ($d$) temperature for each component; the difference between birth and death values is called lifespan, $\cal L$, and shows the range of temperatures over which a component `lived' before merging with an older one. Therefore, the generators near the diagonal are associated with minor spatial variations in temperature. 
All key trends were verified to be insensitive to variations of the persistence threshold and to the number of temporal samples (Appendix B2), confirming that the observed behavior is not an artifact of post-processing choices.
In contrast, the ones far from the diagonal are related to components that remained separated from others over an extensive range of temperatures (one could use a landscape analogy and say that there is a deep `valley' surrounding the local temperature maximum). One generator that is treated slightly differently is the one that corresponds to the component that appeared first (at the largest value of temperature); this component never dies and therefore has a death coordinate equal to zero; inspecting carefully $\beta_0$ $\pd$ shows that this particular generator has the birth coordinate close to unity in the example shown in figure~\ref{fig:2D_toy}. Large `clump' of generators in $\beta_0$ $\pd$, with the coordinates close to $(0.8, 0.8)$ shows that there are multiple regions of the temperatures larger than 0.8 which merged at similar values of temperature threshold; checking figure~\ref{fig:2D_toy}a, we observe that these generators are due to large temperatures of the narrow regions connecting the temperature `nodes' (inspecting movie~\textcolor{red}{S1} shows that this is indeed the case; the advantage of considering the $\pd$ instead is that this information is provided in a much simpler form).

Still considering $\beta_0$ $\pd$, but shifting our focus to lower temperature values shows much less activity in the particular example shown in figure~\ref{fig:2D_toy}, suggesting there are few topological changes at these temperatures. This diagram also shows that there are no topological changes (considering components only) below the temperature value of $\approx 0.3$; the temperature field is fully connected. The reader familiar with the concept of Betti numbers (counting simply the number of components) will note that Betti number $\beta_0$ can be trivially obtained from the $\beta_0$ $\pd$; opposite is however not true, since a $\pd$ contains significantly more detailed information about connectivity of the considered scalar field than the corresponding Betti number; most importantly, a $\pd$ includes the information over all thresholds (temperature values), while Betti numbers are threshold dependent.  Furthermore, $\pd$s are stable with respect to noise (small perturbations), while the same cannot be claimed for Betti numbers~\citep{physicaD14}.  We also note in passing that some works present the information contained in $\pd$s using different visual representations, such as bar-codes~\citep{carlsson}, or sometimes consider sub-level thresholding instead of super-level as considered here; the connection between different graphical representations and thresholding approaches is straightforward.  

In the two-dimensional example considered in figure~\ref{fig:2D_toy}, we also have $\beta_1$ $\pd$, corresponding to holes (loops) in the temperature field, shown in figure~\ref{fig:2D_toy}(d). A loop appears (is born) when the temperature threshold is decreased sufficiently that a closed structure, enclosing the area of lower temperatures, forms.  A loop dies when the threshold is decreased to the degree that the whole interior of the loop is filled up.  As expected, both birth and death coordinates of loops are typically lower than those of components; this is observed by inspecting the two $\pd$s shown in figures~\ref{fig:2D_toy}(c)-(d). The movie~\textcolor{red}{S1} in the online supplementary material helps fully grasp the process of loop formation and disappearance.  

We conclude the discussion of the presented example by noting that $\pd$s provide significant data reduction, as they condense the information about the complicated two-variable function $T(x,y)$ into a point cloud ($\pd$).  This data reduction entails some information loss, primarily due to the loss of geometrical information ($\pd$ includes connectivity information but not the size (spatial extent) of each component). If such information is of importance, then one wants to combine the information available by considering $\pd$s by a complementary measure.  Furthermore, although $\pd$s provide reasonably complete information about the connectivity of the considered field, they are still point clouds, and one can come up with measures summarizing a $\pd$. While many elaborate approaches are being considered to achieve this goal, in this work we resort to a simple approach and account for three measures only (separately for $\beta_0$ and $\beta_1$): the number of generators, $N_g$, the average lifespan $\cal L$ (the average distance of the generators to the diagonal), and total persistence, $TP = {\cal L} N_g$. One could think of $TP$ as a simple measure describing (admittedly in a vague manner) the `intensity' of a considered $\pd$.  We also note that to focus on the main features of the results, we exclude from consideration generators with ${\cal L} < 0.01$, as the fluctuations of the temperature field on such a scale are not of interest.  Such exclusion does not influence the statistics of the (relevant) long-lived generators. In Secs.~\ref{sec:phres1} and \ref{sec:Ra} we comment in more detail on the influence of the specific value (0.01) on the results.

\subsection{Persistent Homology: specifics}
\label{sec:PH_spec}

Over the past decade, several research groups have developed and released open-source software tools for computing $\pd$s and related topological measures. All computations and results presented in this work were carried out using the GUDHI library~\citep{gudhi}.  In our calculations, we use the results obtained from the simulations directly, without any additional post-processing. Consistent with the simulation setup, we assume periodic boundary conditions when computing topological measures (though using non-periodic boundaries yields only minor changes in the results).   As illustrated by our toy example, figure~\ref{fig:2D_toy}, we use super-level filtration (meaning that the features that appear above the chosen threshold value are considered).  For the considered data, the use of computing resources is very reasonable; e.g., for one data slice of B7 simulation (see table\ref{tab:os} in Appendix~\ref{sec:appA}), the computing time to build a cubical complex and compute persistence is $\approx 1.41$ sec (on M2 Mac Studio with 128 GB of RAM). An example Python script for interacting with the GUDHI library is included in the supplementary material.

\section{Results: One-sided convection}\label{sec:OS}

The one-sided flow configuration is representative of a domain that is cooled from above, insulated at the bottom, and initially filled with hot fluid, $T(x,y,z,t=0)=1$, corresponding to the boundary conditions~\eqref{eq:bcos} and the initial condition \eqref{eq:osic1a}, respectively. A sketch of the flow configuration with explicit indication of the boundary conditions is shown in figure~\ref{fig:fig1}(a), with quantities expressed in dimensional form. The flow (and then the evolution of the flow morphology) is controlled by the following dimensionless parameters: the domain size in horizontal directions ($\widehat{L}_1,\widehat{L}_2$) and the Rayleigh number ($\ra$).  In this work, we investigate the influence of these parameters independently: First we describe in \S\ref{sec:flowdyn} the flow evolution in case of large $\ra$ and large domain size ($\widehat{L}_1=\widehat{L}_2=10^4$), and then we analyse in detail in \S\ref{sec:phres1} how the flow morphology is influenced by the governing parameters, considering the effect of size and $\ra$ independently. The details of all the cases analysed are listed in table~\ref{tab:os} (see~Appendix~\ref{sec:appA}), where it is indicated which simulations have been specifically performed for this study \citep[simulations~A, data available via][]{databaseC} and which are obtained from previous works \citep[simulations~B, presented by][]{depaoli2025grl}.

\subsection{Flow dynamics}\label{sec:flowdyn}

\begin{figure}
    \centering
    \includegraphics[width=0.95\columnwidth]{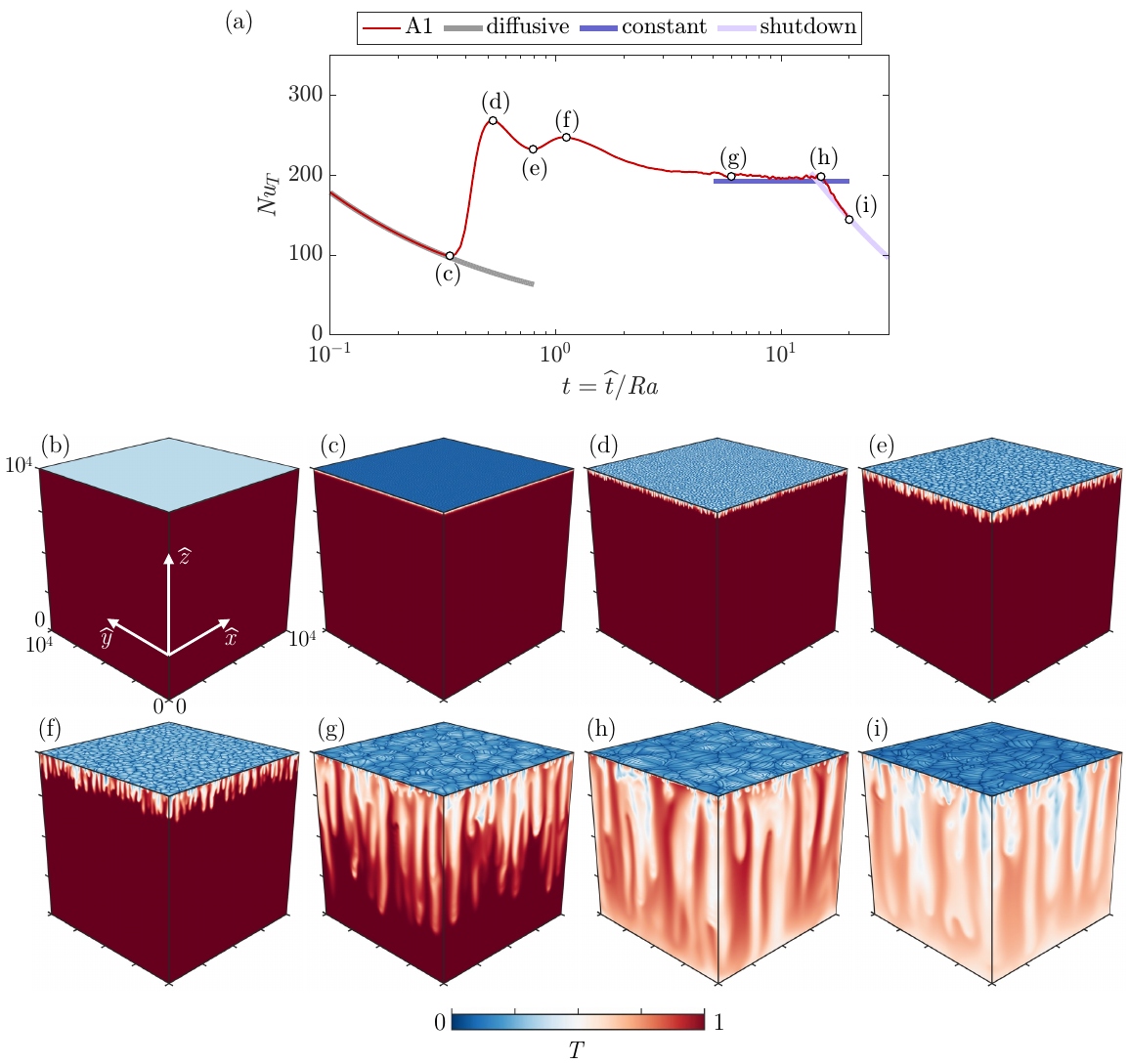}
    \caption{\label{fig:qualit}
    Evolution of a one-sided system (simulation A1, see table~\ref{tab:os} for the details).(a)~Flux at the top wall $\nus_T$, defined in Eq.~\eqref{eq:flux}. The flux is reported as a function of the convective time (note the logarithmic scale for the time variable). The analytical predictions of the flux during the initial (diffusive, Eq.~\eqref{eq:nudiff}), intermediate (constant, Eq.~\eqref{eq:nussconst}) and the late (shutdown, Eq.~\eqref{eq:nusssd}) stages are also reported. (b-i)~Temperature distribution at different times, from the initial condition~(b) through the shutdown of convection~(i). The time instants corresponding to panels~(c-i) are also indicated in~(a). See \S\ref{sec:flowdyn} for additional details.
    }
\end{figure}

The evolution of the flow in one-sided systems has been previously studied in 2D by \citet{slim2014solutal} and in 3D by \citet{depaoli2025grl}, among others \citep{slim2013dissolution,hewitt2013convective,depaoli2017dissolution,wen_2018,fu2013pattern,dhar2022convective}. The global response parameter representative of the state of the flow is the Nusselt number evaluated at the top boundary,
\begin{equation}
\nus_T=-\frac{1}{L_1L_2}\int_0^{L_1}\int_0^{L_2} \frac{\partial T}{\partial z}\biggr|_{z=1}\text{ d}x\text{d}y,
\label{eq:flux}
\end{equation}
which is a dimensionless measure of the importance of convective relative to diffusive heat transport \citep{nield2017convection}. The transient nature of these systems has been described and modelled in detail, \cite[see ][and references therein]{hewitt2013convective,slim2014solutal,depaoli2025grl}. In the following, we provide a brief description of the flow dynamics during each phase of the mixing process, with emphasis on the convective flux $\nus_T$ (reported in figure~\ref{fig:qualit}a) and on the near-wall flow structures (figures~\ref{fig:qualit}b-i). We consider the simulation labelled as A1 in table~\ref{tab:os} (see~Appendix~\ref{sec:appA}) to discuss this dynamics.

The domain is initially filled with hot ($T=1$) fluid at rest ($\mathbf{u}=0$), see figure~\ref{fig:qualit}(b) for the temperature distribution. Under these conditions, mixing is controlled by diffusion at the boundary layer located at the upper wall; an analytical self-similar solution for the evolution of the temperature distribution can be obtained \citep{slim2014solutal,depaoli2025grl}, and in dimensionless convective units reads $T=\text{erf}[(1-z)\ra/\sqrt{4t\ra}]$. The corresponding evolution of $\nus_T$ is inferred using~\eqref{eq:flux}, and it yields \citep{depaoli2017dissolution}
\begin{equation}
    \nus_T=\left(\frac{\pi t}{\ra}\right)^{-1/2}.
    \label{eq:nudiff}
\end{equation}
The diffusive behaviour describes well the one observed numerically and reported in figure~\ref{fig:qualit}(a) for early times ($t\le0.35$). 

When a sufficiently thick layer of dense (cold) fluid forms beneath the top boundary, the initial temperature fluctuations amplify, leading to the formation of thermal plumes (figure~\ref{fig:qualit}c) \citep{riaz2006onset,ennis2005onset}, which eventually grow and extend vertically bringing downward dense fluid, and enhancing transport due to convection (figure~\ref{fig:qualit}d) \citep{slim2013dissolution}. Afterwards, the dynamics of the plumes stops being individual, and the interaction among neighbouring flow structures is such that small plumes merge into larger descending plumes \citep{riaz2006onset,backhaus2011convective}. This phase occurs over several generations of plumes, and corresponds to a reduction of the flux (figures~\ref{fig:qualit}e-f). The space left by the plumes that merged is eventually filled with dense fluid \citep{slim2014solutal}, and the process of plumes formation and merging continues in a statistically-steady fashion (figure~\ref{fig:qualit}g). Eventually, the plumes grow and reach the lower boundary (at $t\approx7$), and the average fluid density in the domain increases progressively (i.e., the bulk flow temperature diminishes). During this regime, the Nusselt number is nearly constant and equal to \citep{depaoli2025grl}:
\begin{equation}
    \nus_T = 0.01926 \times \ra.
    \label{eq:nussconst}
\end{equation}
At a later stage ($t\approx 14-15$ in 3D, figure~\ref{fig:qualit}h), this process leads to an increase of the fluid density in the upper layer, corresponding to a reduction of the density contrast between the fluid in the upper layer and the upper boundary.  This marks the start of the shutdown of convection~\citep{hewitt2013convective}: a sudden reduction of the driving force corresponds to a decrease of $\nus_T$, the evolution of which has been accurately predicted by physical models \citep{hewitt2013convective,slim2014solutal,depaoli2025grl}. We choose here to employ the formulation proposed by \citet{depaoli2025grl} (with the same parameters) to model the shutdown phase, given by
\begin{equation}
    \nus_T = \frac{14.4\times \ra}{[0.7303\times (t-14)+27.0]^2} .
    \label{eq:nusssd}
\end{equation}
Also in this late-stage phase, the simulation and the model predictions are in excellent agreement (see figure~\ref{fig:qualit}a).

\begin{figure}
    \centering
    \includegraphics[width=0.99\linewidth]{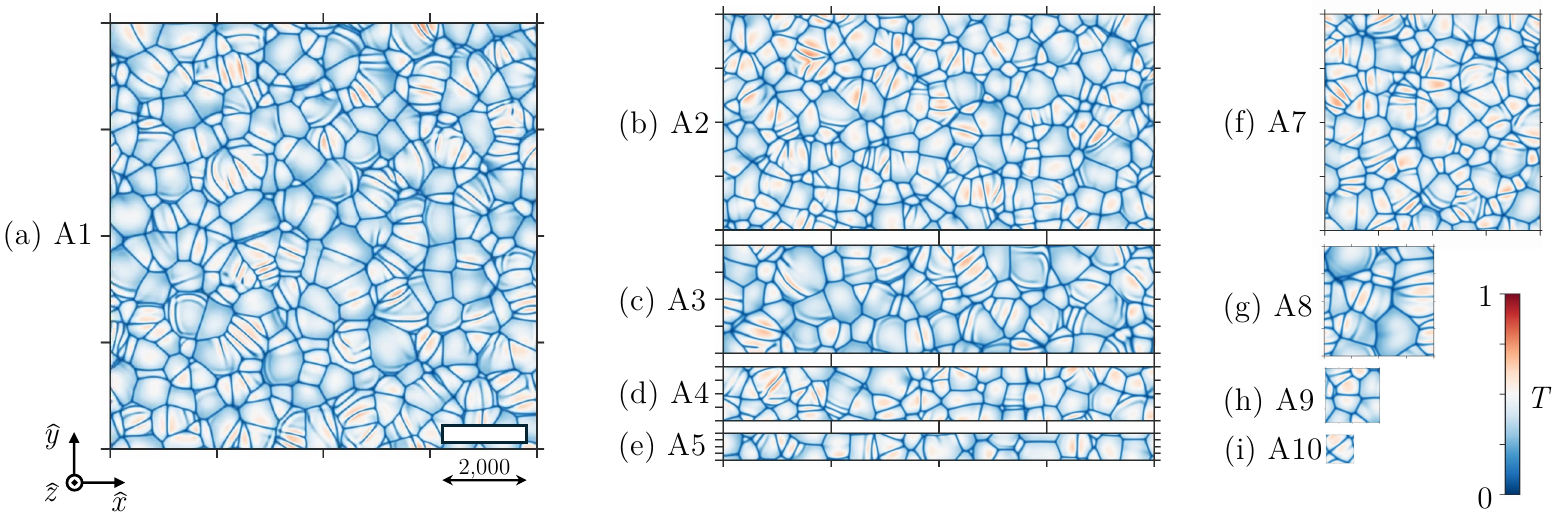}
    \includegraphics[width=0.85\linewidth]{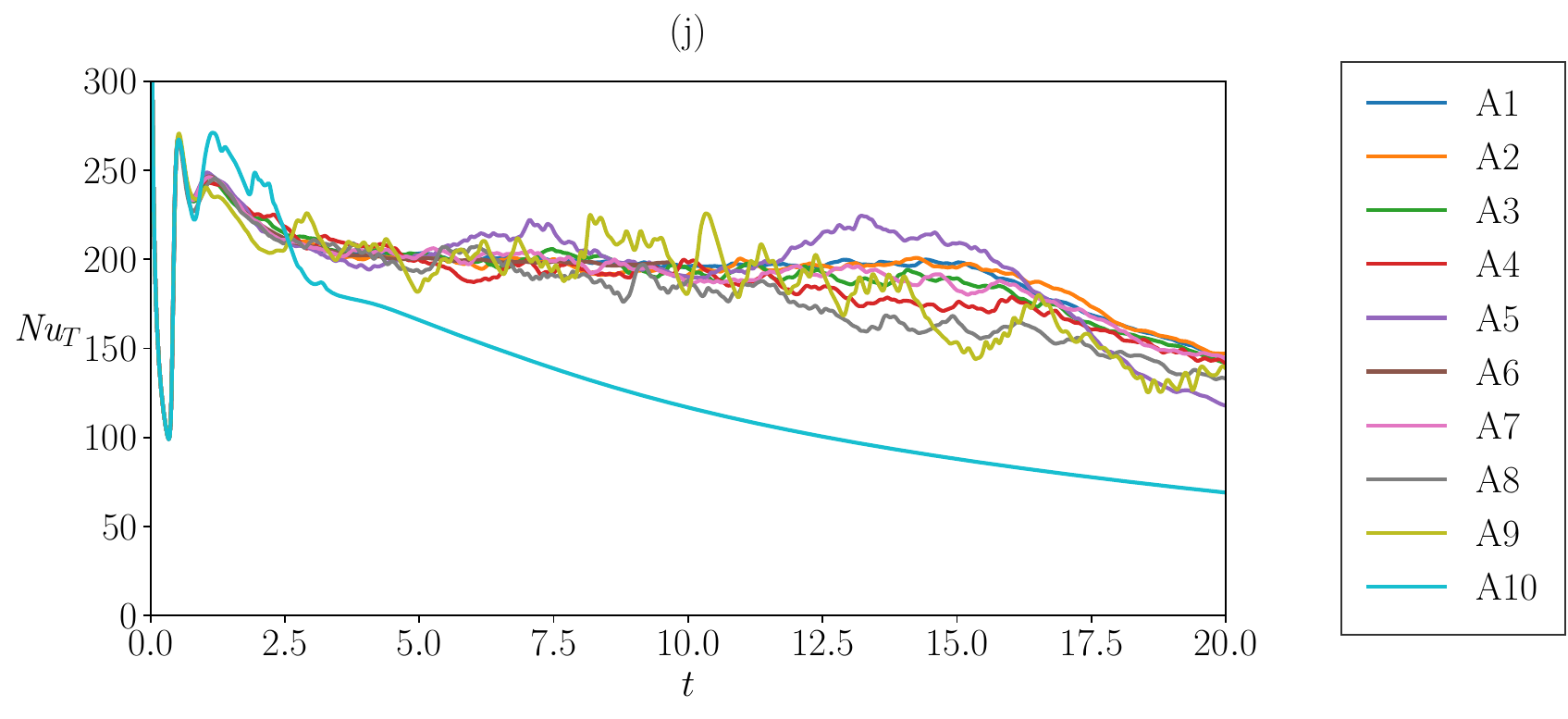}
    \caption{
    (a)-(i)~Examples of temperature fields taken near the upper wall ($z=0.998$) at time $t\approx0.8$, for the A simulations listed in table~\ref{tab:os} with the exception of A6, the domain size of which is too large to be shown to scale.    The full domain size is shown, and a scale bar (in diffusive units) is reported in panel~(a) for reference. We refer to movie~\textcolor{red}{S2} in the online supplementary material for the time-dependent evolution of these patterns. Panel~(j) shows the time evolution of the flux at the top wall, $\nus_T$, defined by Eq.~\eqref{eq:flux}, for the domains considered and listed in table~\ref{tab:os} (time is displayed in convective units). Note the different behaviour of the A10 configuration caused by the small domain size.
    }
    \label{fig:fields}
\end{figure}

In what follows, we quantitatively analyse the morphology evolution of the flow structures near the upper boundary. First we consider the effect of the domain size for a fixed (and large) value of $\ra$, and then we investigate the role of $\ra$. In table~\ref{tab:os} we report the details of the simulations analysed. Simulations indicated with `A', are presented here for the first time, and are used to investigate the effect of the domain size.  These simulations are performed at constant Rayleigh number, $\ra=10^4$, and variable domain extension in horizontal directions, $L_1$ and $L_2$. An overview of the system's evolution, as well as examples of the flow fields, are shown in figures~\ref {fig:qualit} and~\ref{fig:fields}. For simulations indicated by `B', initially presented in \citet{depaoli2025grl}, the domains are of square cross-section. These data, obtained at different values of $\ra$ between $10^2$ and $8\times10^4$, are used to investigate the effect of the driving parameter on the flow morphology. The domain extension is chosen to be sufficiently large, allowing for the neglect of any confinement- or periodicity-induced influence on the development of the flow structures.  We refer to \citet{depaoli2025grl} for a detailed discussion on the minimal domain size to be employed at $\ra\ge10^4$.

\subsection{Connecting the flux to the flow structure next to the top wall}\label{sec:phres1}

In this section, we focus on quantifying the morphology of the flow structure that develops next to the top and the flux throughout the domain, and on showing the connection between the measures describing this structure and the heat flux.  We start by presenting an example of results obtained from persistence diagrams, $\pd$s, for one flow configuration (A1), and then proceed to illustrate the correlation between the computed topological measures and the heat flux across all considered configurations. 

\begin{figure}
    \centering    
    \includegraphics[width=1.0\linewidth]{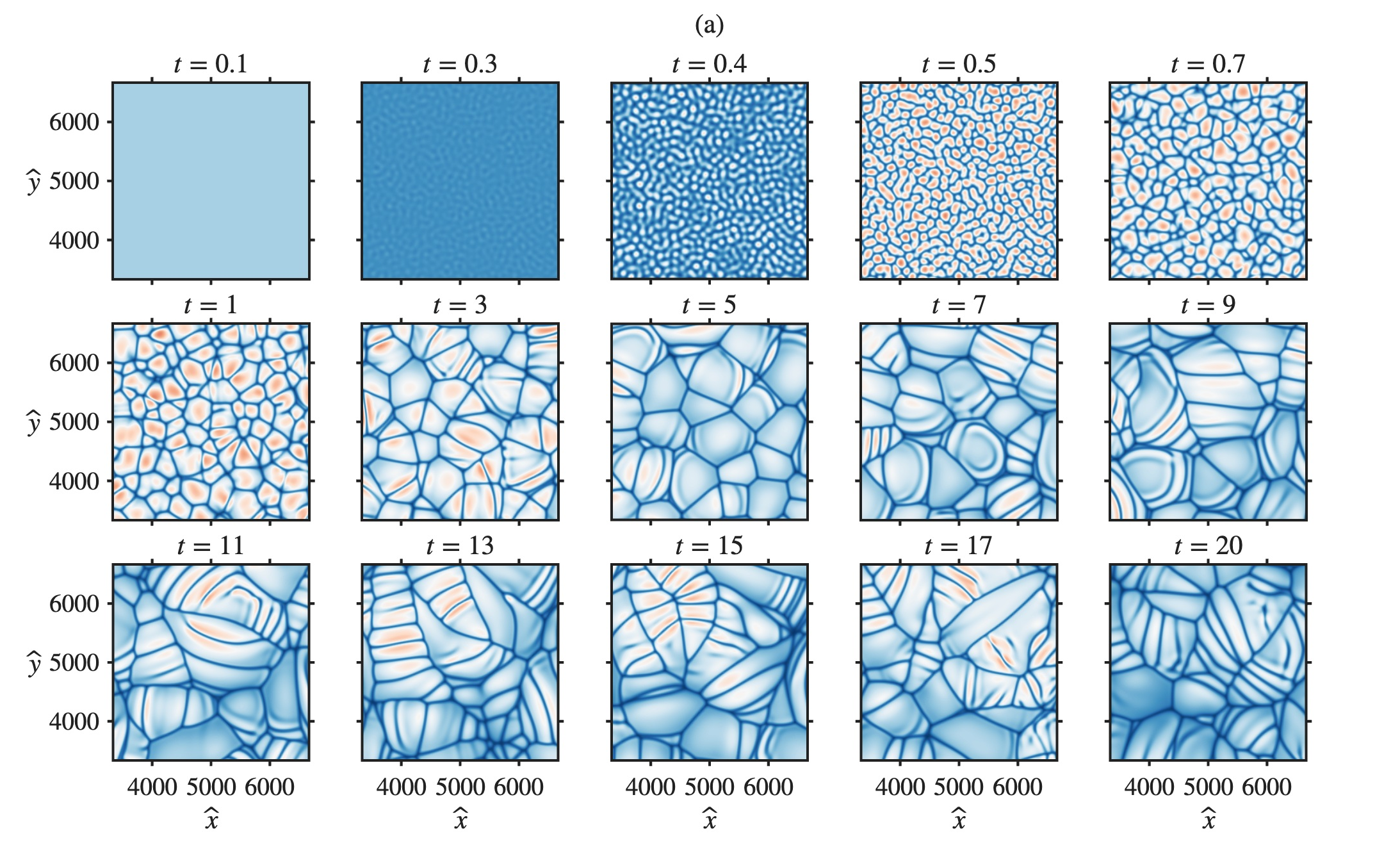}
    \includegraphics[width=0.9\linewidth]{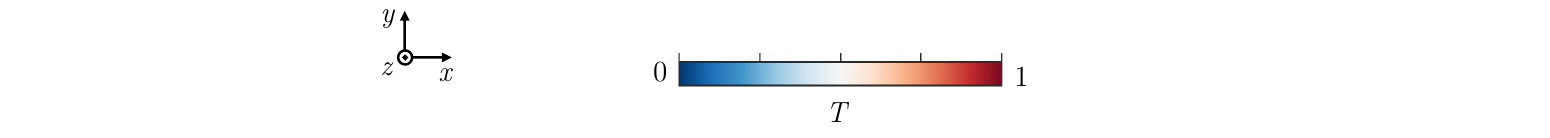}    
    \vskip 0.1in
    \includegraphics[width=0.65\linewidth]{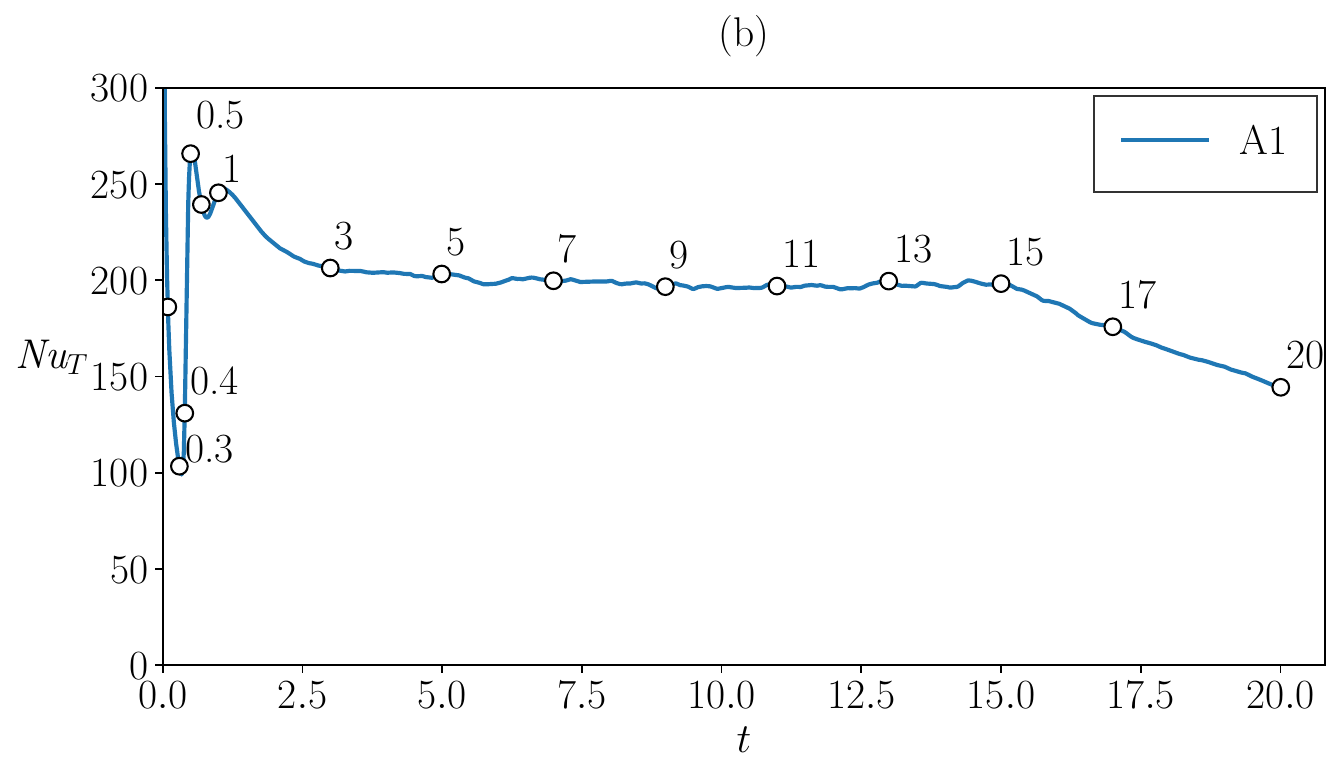}    
    \caption{
    Pattern formation evolution for the simulation A1. (a)~Temperature distribution close to the top wall (a zoomed portion of the domain is shown, corresponding to $ \widehat{L}_1  \le x,~y\le 2 \widehat{L}_1/3 $; the plots' titles show the time instances at which the temperature field was recorded. (b)~Heat flux measured by $\nus_T$. The time instances from (a) are shown as well. }
    \label{fig:structure}
\end{figure}

Figure~\ref{fig:structure} shows the evolution of pattern formation for the simulation A1, focusing on early times.  The related heat flux (Nusselt number) is also shown. We observe complex evolving structure, which is initially maze-like. As time passes, this structure gradually transitions into a well-defined cellular network that eventually stabilizes. When comparing these pattern changes to the flux variations shown in figure~\ref{fig:structure}(b), a clear relationship emerges: the evolution of the cellular pattern correlates closely with the changes in flux.

\begin{figure}  
    \centering
    \includegraphics[width=0.95\linewidth]{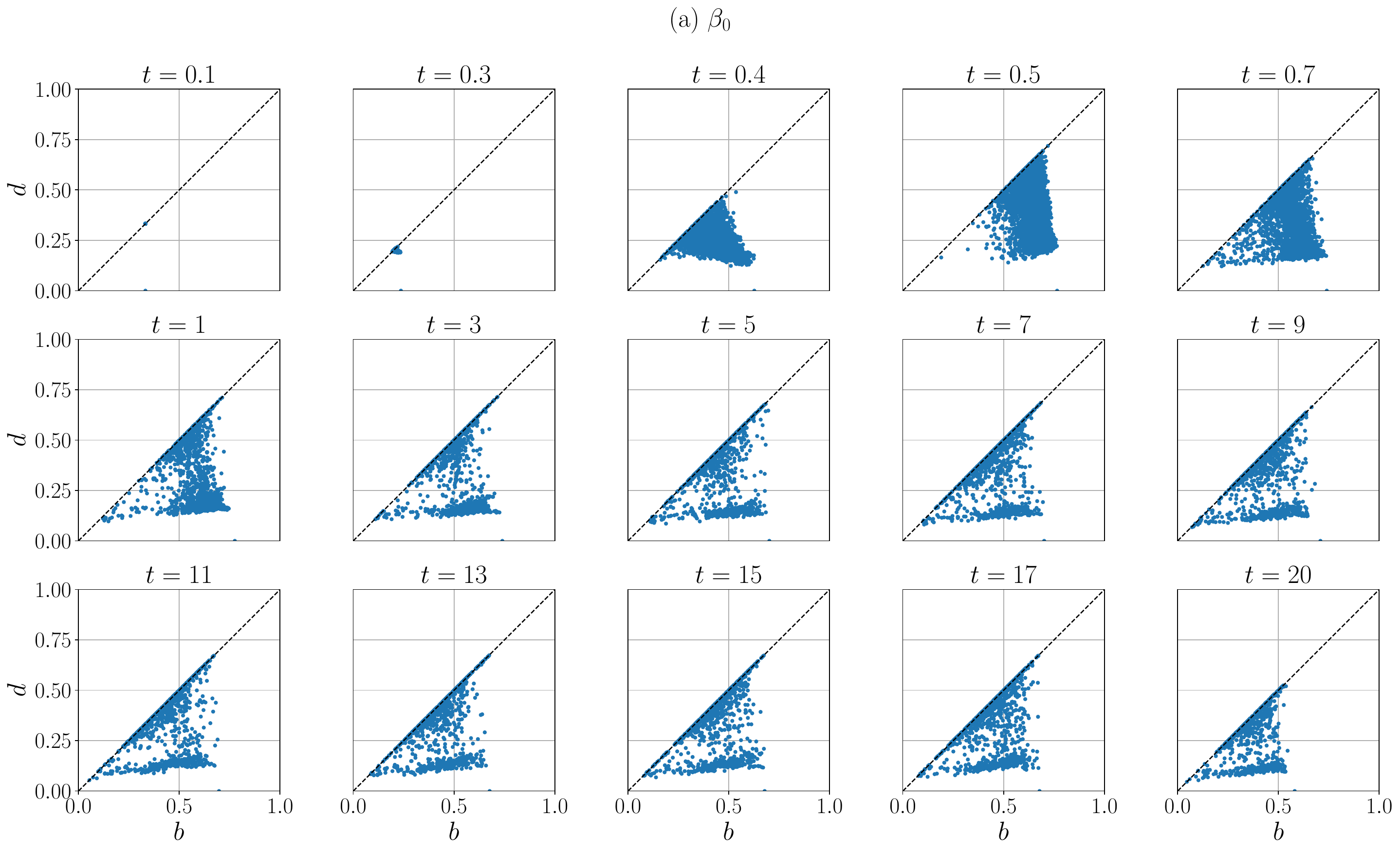}   
     \centering
    \includegraphics[width=0.95\linewidth]{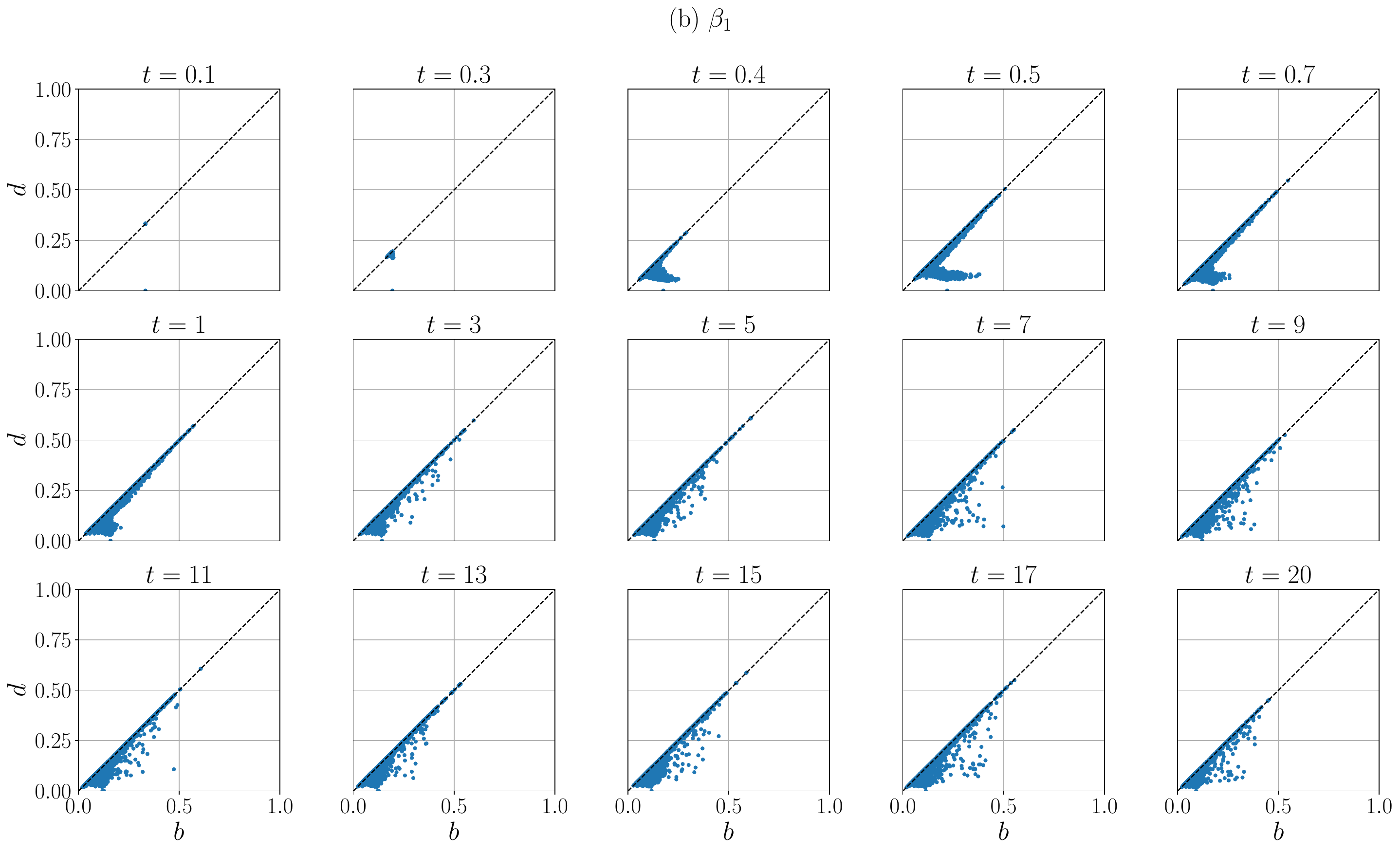}     
    \caption{
    $\pd$s corresponding to the temperature field shown in figure~\ref{fig:structure}(a). (a) connected components, $\beta_0$, and (b) loops, $\beta_1$. Here, `$b$' stands for birth and `$d$' for death of the considered features.} 
    \label{fig:pds}
\end{figure}

Our next goal is to quantify the structures shown in figure~\ref{fig:structure}(a) and correlate them with the observed heat flux in figure~\ref{fig:structure}(b). $\pd$s shown in figure~\ref{fig:pds} provide a detailed description of the structure at various time instances. These diagrams show the evolution of connected components, $\beta_0$, and of loops (cycles), $\beta_1$.  Carefully comparing the temperature plots, figure~\ref{fig:structure}(a), and corresponding diagrams, figure~\ref{fig:pds}, illustrates the ability of $\pd$s to quantify complex structures in a manner that captures the main topological features while also providing a clear visual and (as we will see) quantitative description of the results.  

Focusing first on connected components, figure~\ref{fig:pds}(a), we observe how the topological properties of the temperature patterns develop as time progresses. For very early times, $t\sim 0.1$, the temperature is nearly uniform and we note a single generator in the corresponding $\pd$. More elaborate patterns start developing around $t=0.3~-~0.4$ around the temperature $T\sim0.3$, and then rapidly, by $t\sim 0.7$, the structures develop for the temperatures as high at $T \sim 0.7$. Around this time, the generators begin to gradually separate into upper and lower groups. The upper group (close to the diagonal) reflects `noise' (due to small temperature variations, to be discussed further below), while the lower group corresponds to (hot) cell interiors, indicating that cells remain separated over wide range of temperature thresholds (that is, the areas of high temperature appear already at $T\sim 0.7$, but they connect (merge) only at the temperatures $T \sim 0.2$). 

Further insight is reached by considering loops, see figure~\ref{fig:pds}(b).  Here we note the formation of loops with a significant birth temperature ($T\sim 0.3$) only at relatively early times ($t\lesssim 0.7$), as illustrated by the elongated tails in the corresponding $\pd$s. As the pattern stabilizes, these tails disappear, indicating that loops form only at very low temperatures.  Overall, the diagrams capture the topological transition from the maze phase to an ordered cellular network.

\begin{figure}
    \centering
    \includegraphics[width=0.4\linewidth]{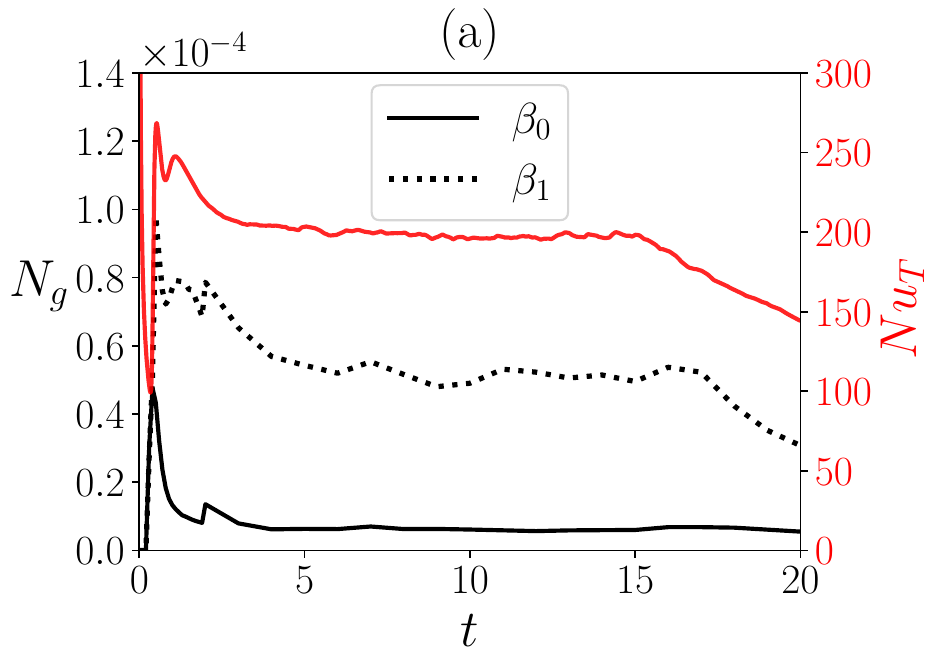}
    \hspace{0.05\linewidth}
    \includegraphics[width=0.4\linewidth]{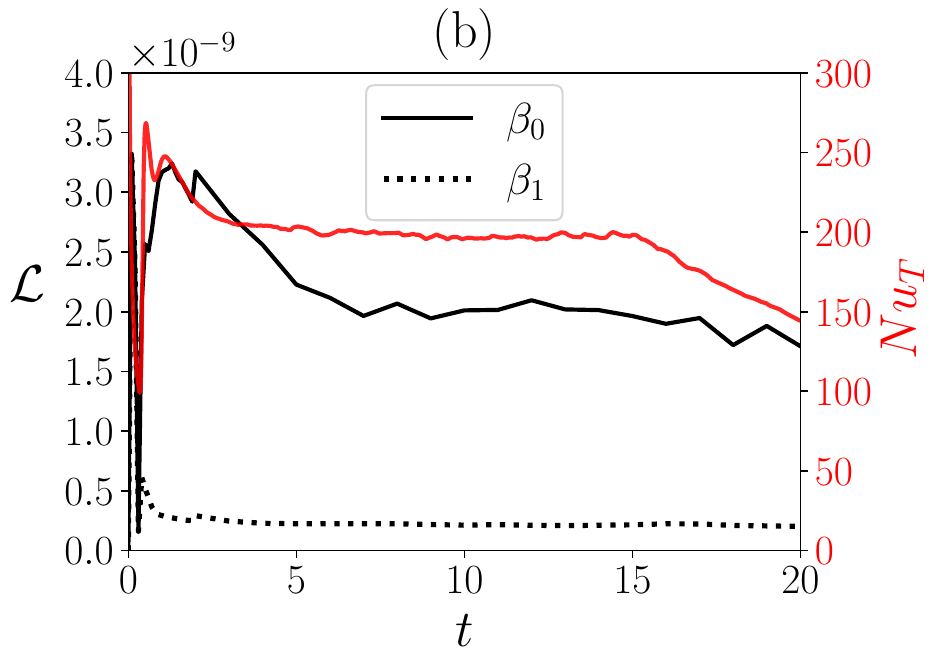}
    \caption{\label{fig:TP}
      Flux and topological measures for A1 simulation, normalized by the area ($\widehat{L}_1\times \widehat{L}_2$).
      Flux is reported in red (right axis). 
      Number of generators $N_g$~(a) and average lifespan $\mathcal{L}$~(b) are indicated for components ($\beta_0$, solid lines) and loops ($\beta_1$, dotted lines).}
\end{figure}

Figure~\ref{fig:TP} demonstrates a strong similarity in the time-dependent behaviour of heat flux (figure~\ref{fig:TP}a), total persistence (figure~\ref{fig:TP}b), number of generators (figure~\ref{fig:TP}c), and average lifespans (figure~\ref{fig:TP}d). All considered quantities exhibit consistent trends, suggesting that the evolution of heat flux and topological measures is closely correlated. One peculiarity of the particular set of patterns considered here is illustrated by the number of generators, $N_g$, and lifespans, $\cal{L}$, for components ($\beta_0$) and loops ($\beta_1$): $N_g$ is much larger for the loops, and $\cal L$ for the components.
Taken together, these results suggest the existence of small variations of temperature leading to loop formation (very close to the diagonal, as suggested by small $\cal L$): these flow structures correspond to the footprint of the small plumes near the upper wall, see e.g. figures~\ref{fig:qualit}(d-f). 
While Fourier analysis provides a dominant wavelength, it cannot distinguish between disconnected plumes and interconnected cellular networks with similar spectral content. PH, instead, resolves these differences through topological measures such as generator count and lifespan.
On the other hand, $\cal L$ is significantly larger for components, showing the existence of isolated areas of elevated temperature which persist over a significant temperature range. These areas of elevated temperature are the feature of the results, which is of real physical interest; regarding small variations of temperature, further inspection shows that these variations occur on the computational grid scale, consistently with the tail of the Fourier spectrum for the large wavenumbers for the results discussed in previous works~\citep{depaoli2025grl}. They are therefore not relevant for our purposes and we remove them from future discussion, as mentioned in \S\ref{sec:toy}. The noise band width used for this removal is a post-processing choice tied to the temperature-field scale and numerical resolution, and not a universal value.  We note that the numerical resolution employed is appropriate and does not affect the main PH trends: extensive validation has been performed in previous works \citep{depaoli2022strong,depaoli2025grl}, showing that numerical noise is well-separated from physically-meaningful flow features.

We now offer a more quantitative interpretation of the correlation between heat flux and topological metrics. 
Initially, when heat transfer is purely conductive, no generators are detected, neither components nor loops, and $\nus_T$ diminishes as in Eq.~\eqref{eq:nudiff}. 
This process leads to a thickening of the thermal boundary layer beneath the interface. 
When the temperature fluctuations contained within grow sufficiently (leading to the formation of topological structures, the boundary layer becomes unstable and plumes form.
These plumes remove heat from the top wall much more efficiently than conduction, and as a result, a sudden increase in the flux is observed. This process leads to the sudden emergence of topological generators with significant lifespans, linked to the formed plumes, followed by a subsequent reduction ($t>1$). 
The increase in average lifespan reflects the progressive merging of plumes into larger coherent structures, which reduces interfacial area and modifies the effective transport pathways, thereby explaining the observed variation in Nusselt number.
This dynamics suggests a rearrangement of the flow structures, i.e., a process of plumes merging, which also leads to a reduction of $\nus_T$: they cannot keep growing independently, as they interact with neighbouring plumes and with the hot rising fluid in the interplume spacing \citep{slim2014solutal}. This dynamic constrains plume displacement and reduces the amount of heat that can be removed compared with an undisturbed plume.
Using similar reasoning, one can conclude that when $\nus_T$ is constant, there is no major difference in the flow morphology, i.e., the number of plumes and their organization (as characterized by $N_g$ and $\cal L$) remain statistically unchanged. 
Finally, for longer times ($t>15$) during the shutdown phase, plumes reduce in number, and so does the $\nus_T$ due to the progressively reducing density contrast resulting from the saturation of the domain.

\begin{figure}
    \centering
    \includegraphics[height=0.28\linewidth]{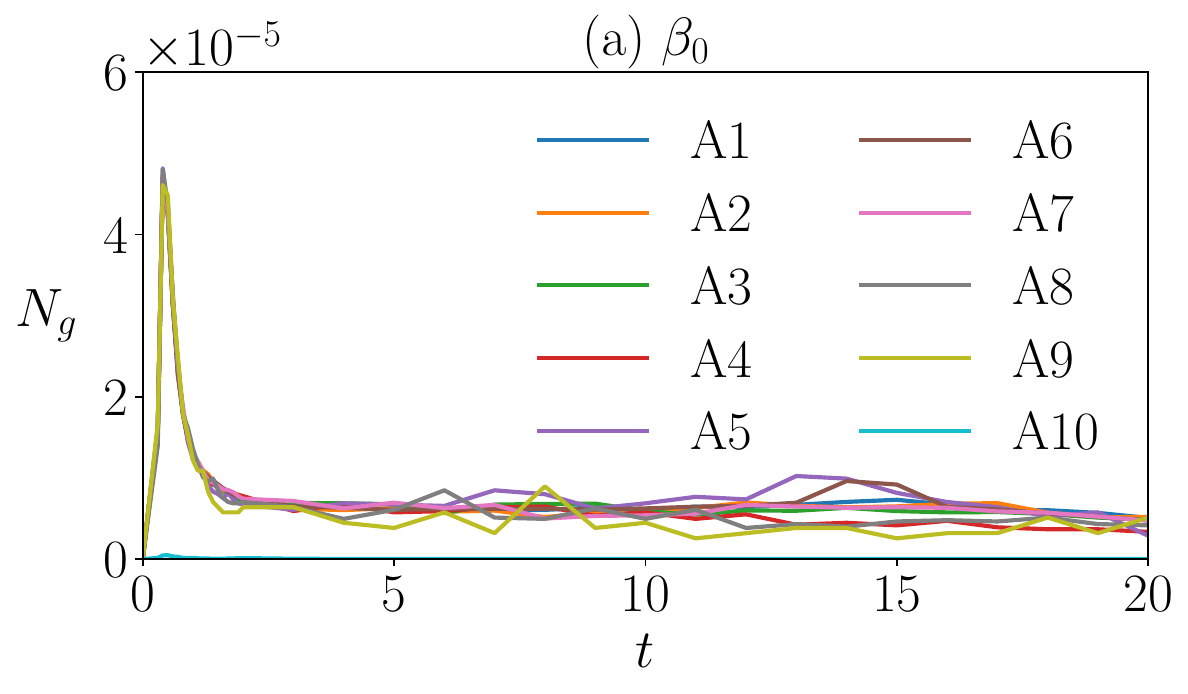}\hspace{0.2cm}
    \includegraphics[height=0.28\linewidth]{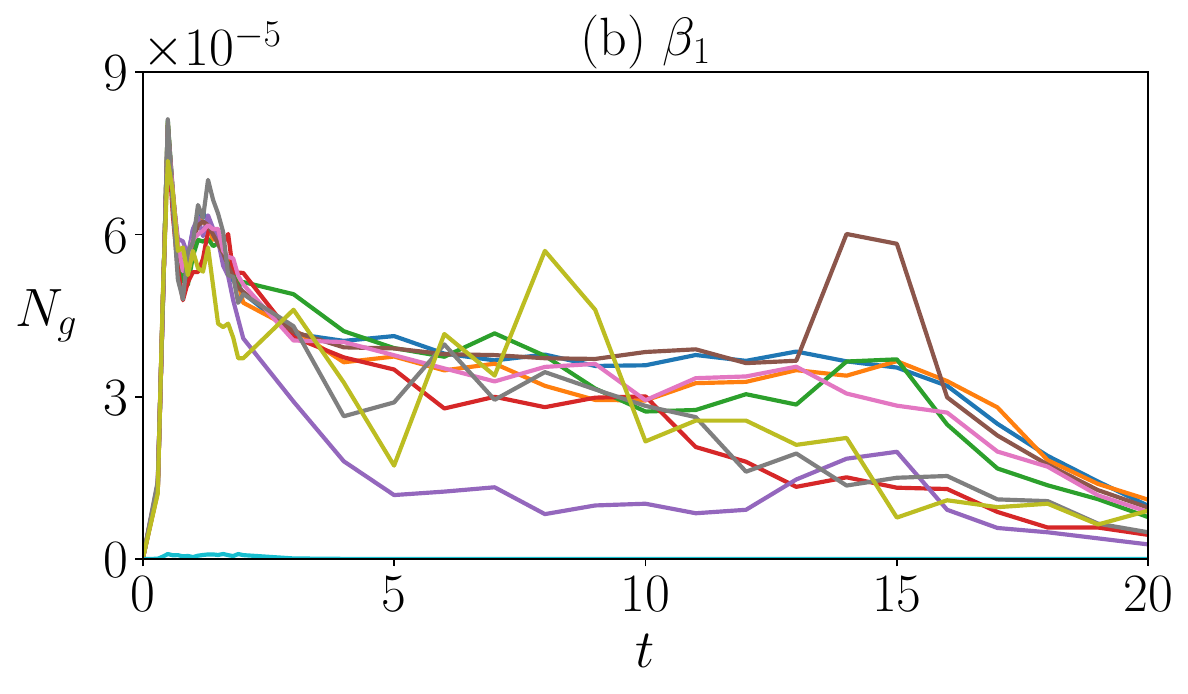}
    \includegraphics[height=0.28\linewidth]{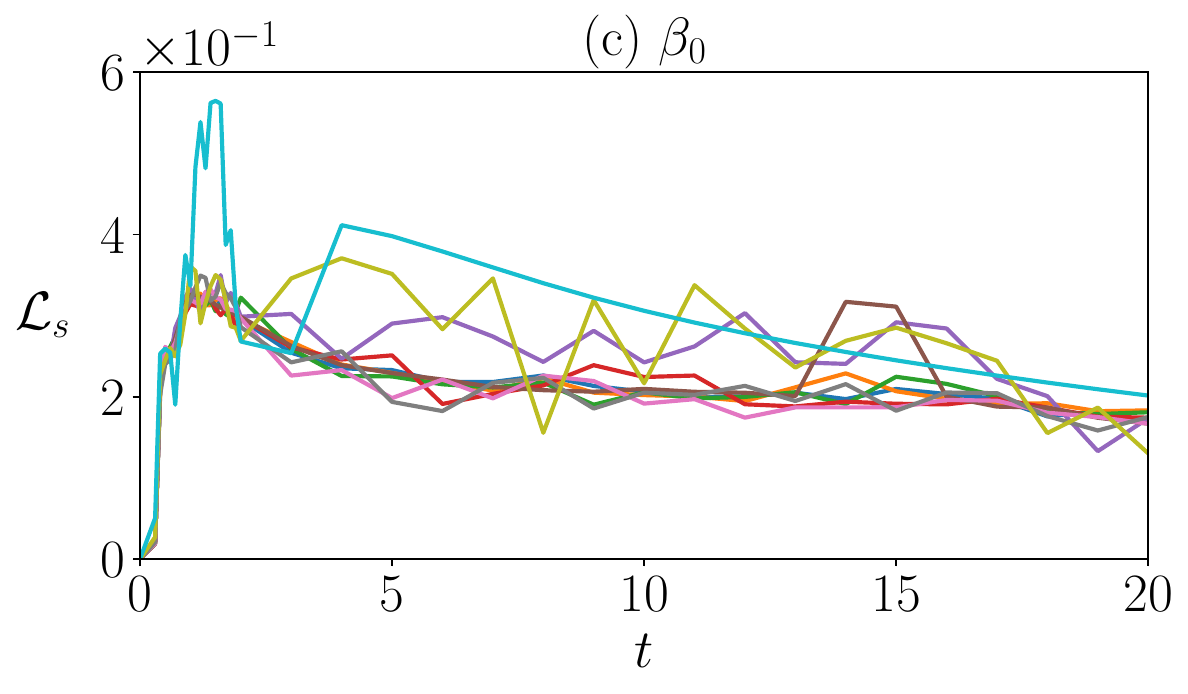}\hspace{0.2cm}
    \includegraphics[height=0.28\linewidth]{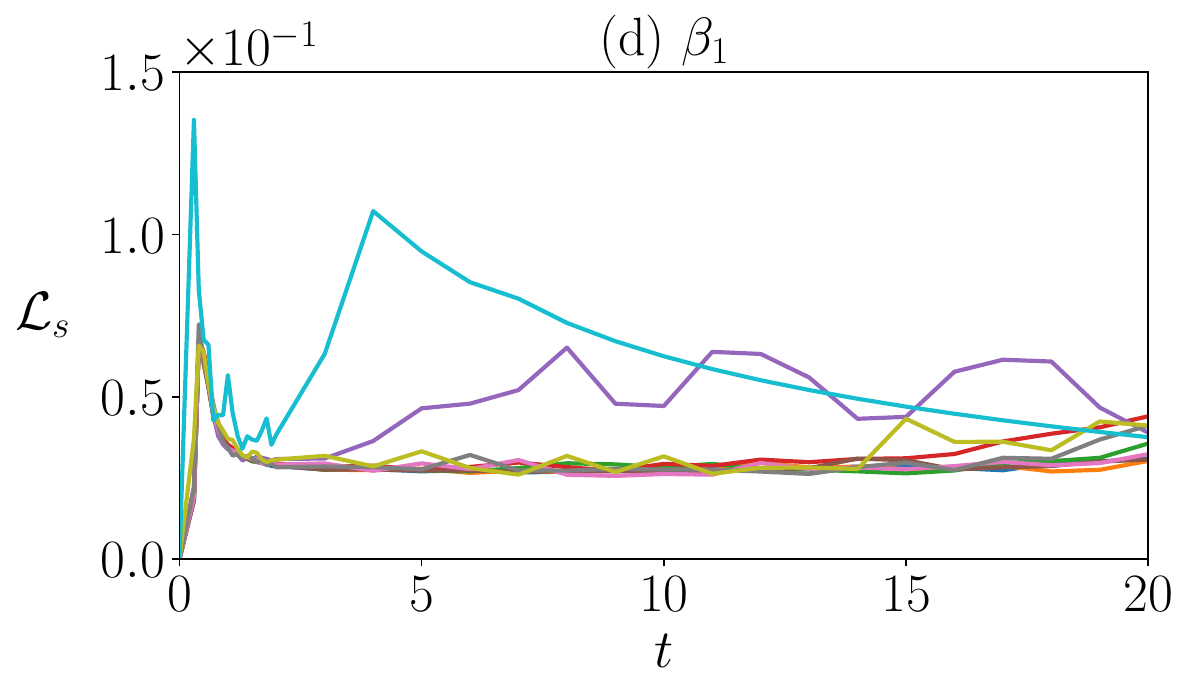}
    \caption{Number of Generators $N_g$ (a,b) (per area) and average lifespan $\mathcal{L_s}$ (c,d) (with band of width 0.01 removed), for simulations A1-A10. 
    Components ($\beta_0$) and loops ($\beta_1$) are reported in panels~(a,c) and~(b,d), respectively.  
    Note that for simulation A10, $N_g$ reaches only very small values (barely distinguishable from 0 on the scale plotted.)}
    \label{fig:result3}
\end{figure}

Figure~\ref{fig:result3} shows the results for all `A' simulations from table~\ref{tab:os}.  While there is some noise in the results, we find that they are essentially independent of domain size for most domains considered. The main exception is simulation A10, which involves a very small domain. 

\subsection{Influence of the Rayleigh number on the pattern morphology\label{sec:Ra}}

Here, we focus on the influence of the value of $\ra$, and therefore on the ``B'' simulations.  These data, obtained at different values of $\ra$ ranging from $10^2$ to $8\times10^4$, are used to investigate the effect of the driving parameter on the flow morphology. The domain size is chosen to be sufficiently large to neglect any influence of confinement and periodicity on the development of the flow structures. We refer to \citet{depaoli2025grl} for a detailed discussion on the minimal domain size to be employed at $\ra\ge10^4$.

\begin{figure}
    \centering
    \includegraphics[width=0.95\linewidth]{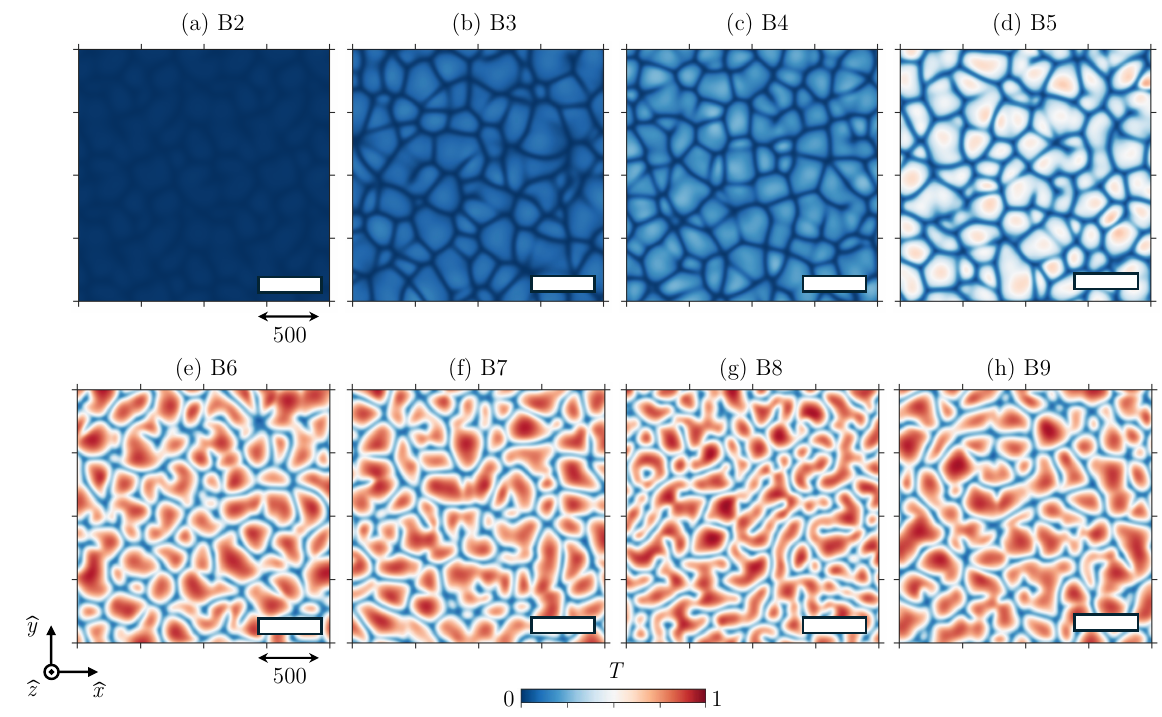}
    \caption{
    Temperature distribution over a horizontal $(\hat{x},\hat{y})$ plane near the upper wall at time $\hat{t}=5\times10^3$ for simulations B2-B9 (see table~\ref{tab:os}).
    Only a squared portion of the domain having side $2\times10^3$ is reported (note that the temperature field shown for B2 is repeated in a periodic manner, because the original domain, of side $10^3$ diffusive units, is smaller than the domain shown here, $2\times10^3$).
    A scale bar of length $500$ is shown as a reference.
    The morphology of the flow appears to be dependent of the Rayleigh number considered for B2-B5, while no macroscopic change is observed for B6-B9.
    }
    \label{fig:fieldsOS}
\end{figure}

Figure~\ref{fig:fieldsOS} shows the emerging patterns at time $\hat t = 5\times 10^3$ ($t = 5\times 10^3/Ra$).  We observe that, while the patterns for the simulations obtained with smaller values of $Ra$ ($\lesssim 2000$, simulations B2~-~B5), there are significant differences between the patterns, for larger values of Ra, simulations B6~-~B9, the emerging patterns become more similar, at least visually.

\begin{figure}
    \centering
    \includegraphics[height=0.28\linewidth]{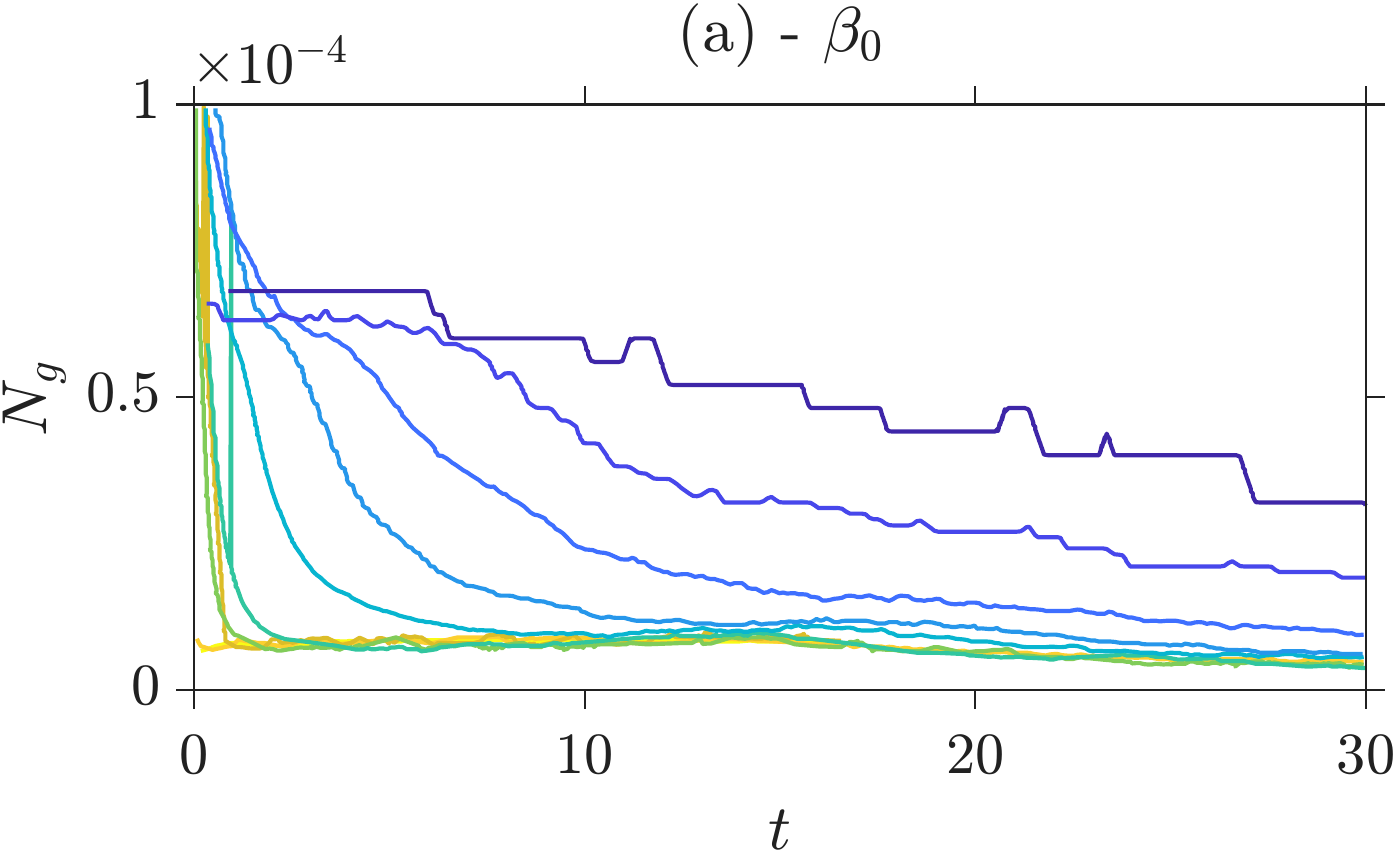}
    \includegraphics[height=0.28\linewidth]{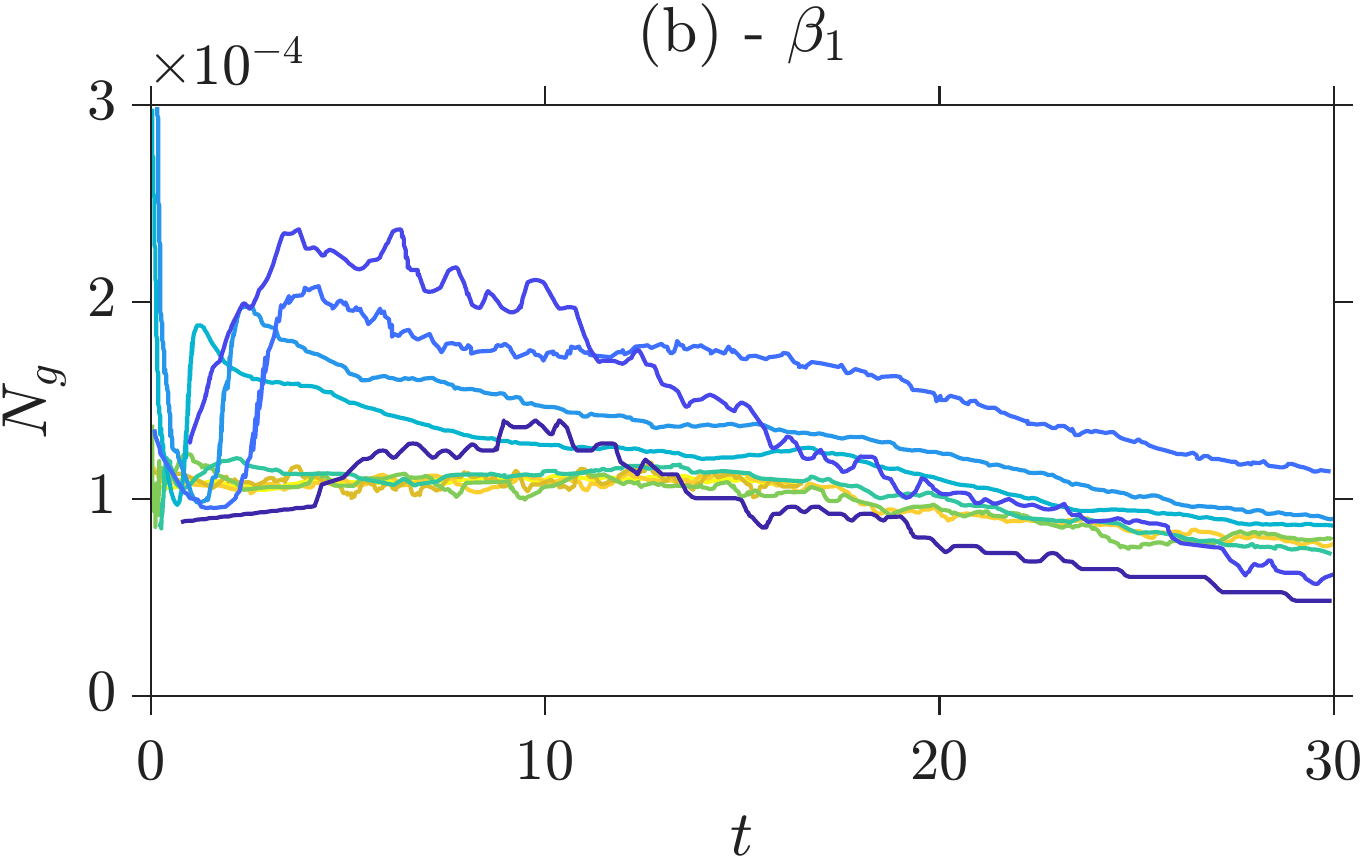}\\
    \vspace{0.5cm}
    \includegraphics[height=0.28\linewidth]{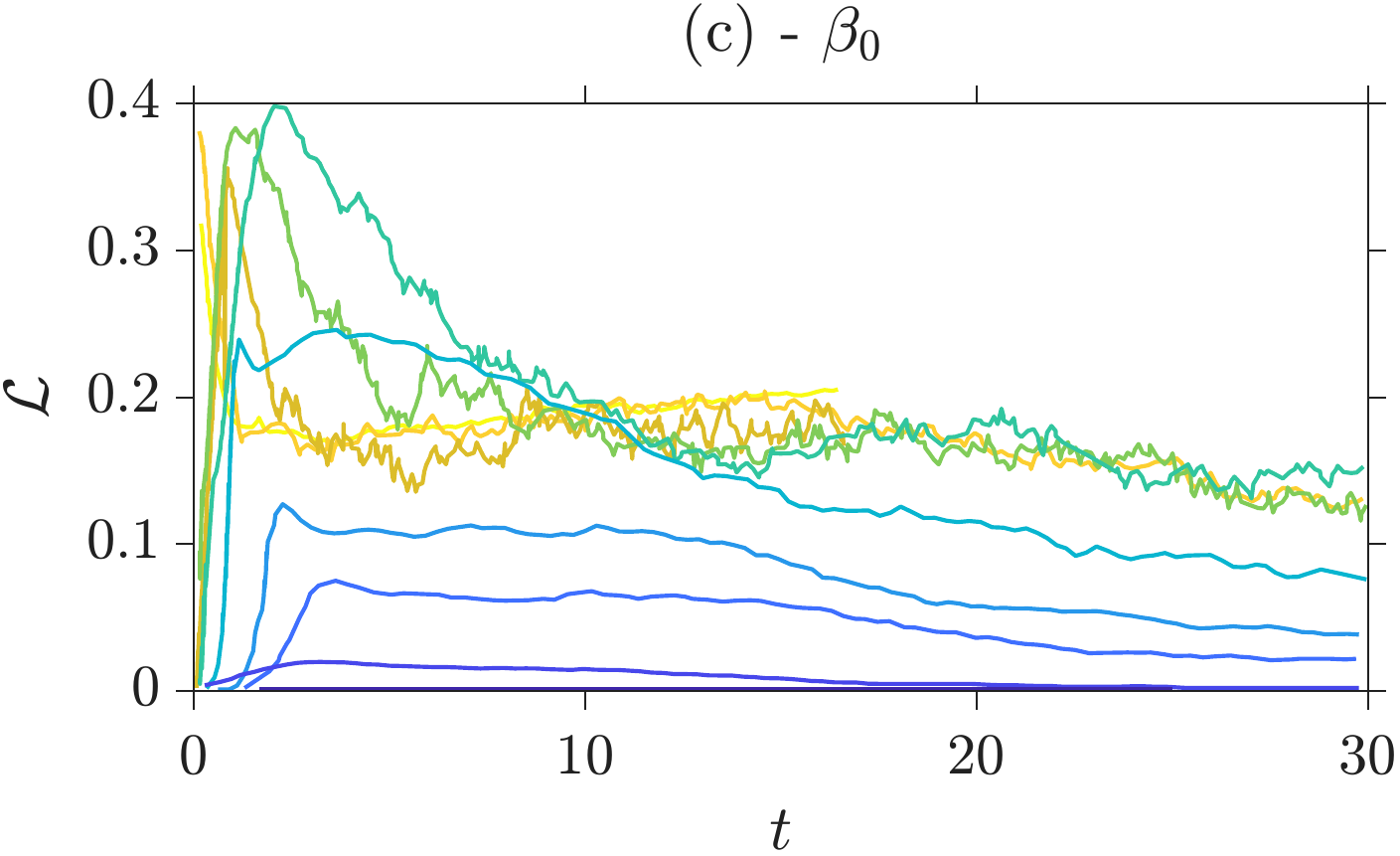}
    \includegraphics[height=0.28\linewidth]{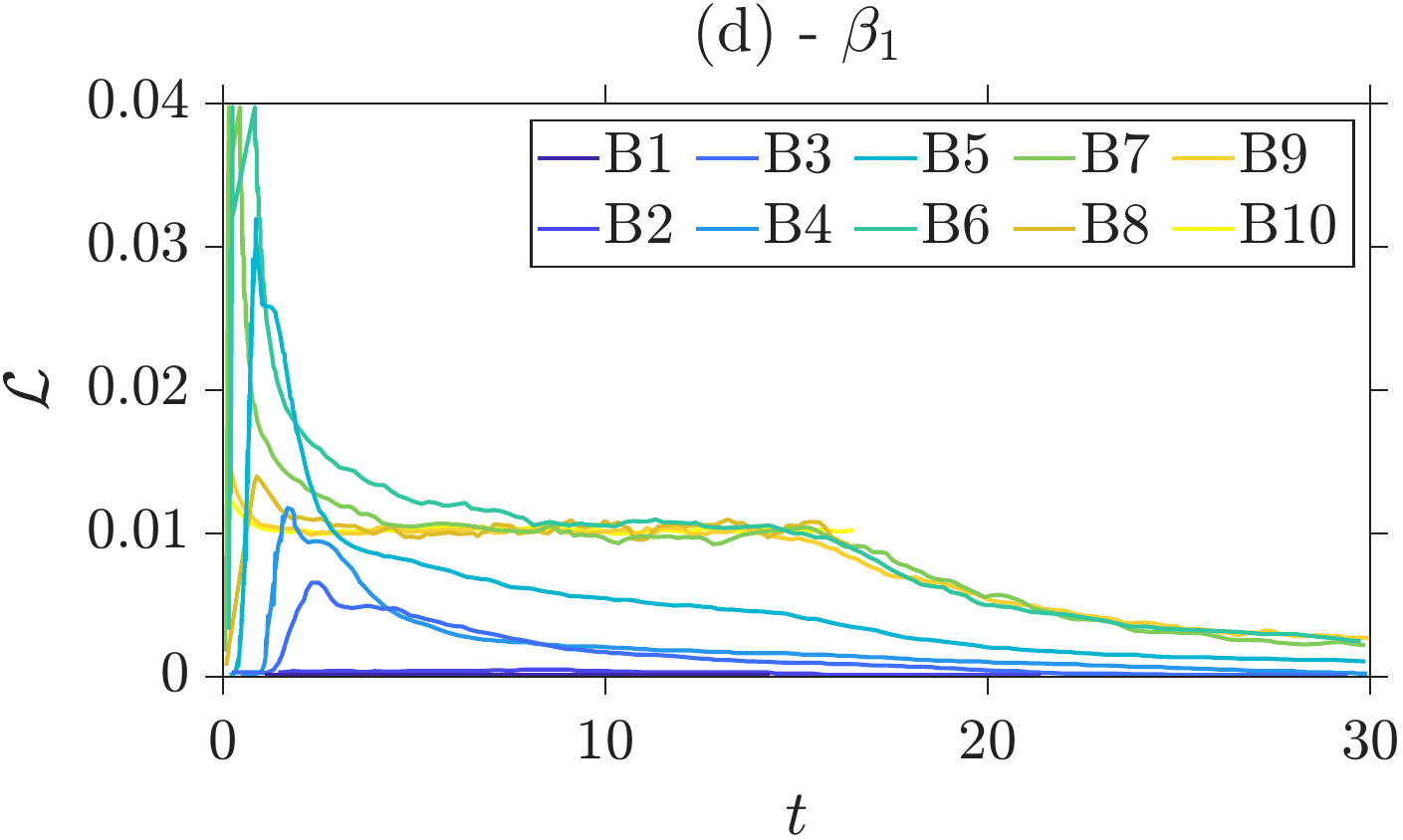}    
    \caption{Number of Generators $N_g$ (a, b) (per area) and average lifespan $\mathcal{L_s}$ (c, d) (with noise band 0.01 removed), for the simulations B1~-~B9.
    Note that simulation B10 is shorter than the others due to the limited amount of data available. }
    \label{fig:B-sim-total}
\end{figure}

Figure~\ref{fig:B-sim-total} puts the statements above on a firmer footing.  This figure shows the topological measures obtained from the corresponding persistence diagrams for components and loops.  The main observation is that for the considered measures (number of generators, $N_g$, and average lifespan, ${\cal L}$), the results for B6~-~B10 simulations are essentially identical, while for smaller values of $Ra$ we see that the results depend on the value of $\ra$. This dependence is not obvious from the snapshot of the patterns, such as shown on figure~\ref{fig:fieldsOS}, and it is much easier to quantify by considering PDs shown in figure~\ref{fig:B-sim-total}.  For example, considering components, shown in figure~\ref{fig:B-sim-total}(a) and~(c), one can observe that the number of generators (therefore, features in the temperature field) per area is larger for smaller values of $\ra$, however, these features are much less prominent (on average) since their lifespan, ${\cal L}$, becomes progressively smaller as $\ra$ decreases.  Regarding the loops, we observe a clear correlation between their lifespans, shown in figure~\ref{fig:B-sim-total}(d) and the onset of shutdown stage, see figure~\ref{fig:qualit}(a) and figure~\ref{fig:structure}.

Persistent diagrams also allow addressing how emerging length scales depend on the value of $\ra$.  This question was discussed in previous works by \cite{fu2013pattern} and \cite{depaoli2025grl}, and here we find these values using $\pd$s; as we will see, additional information can be extracted since $\pd$s encode detailed information about the topology of the temperature field.  In particular, it is straightforward to extract the number of `features', measured by  Betti numbers, from $\pd$s - essentially, for a given threshold value, one counts how many generators were born, but have not died yet - this is the number of features or objects of interest. Such objects could be components, measured by $\beta_0$, or loops, measured by $\beta_1$. As we will now show, our findings illustrate the fact that the emerging length scales are strongly influenced by both the temperature threshold and by the type of features considered. Before proceeding, we note that the results that follow are essentially independent of the width of the band of generators next to the diagonal that we remove (as long as the width is small, corresponding to typical lifespans); those generators (close to the diagonal) do not influence the number of components or loops, except for very small thresholds as we discuss further below. 

\begin{figure}
    \centering
    \includegraphics[height=0.36\linewidth]{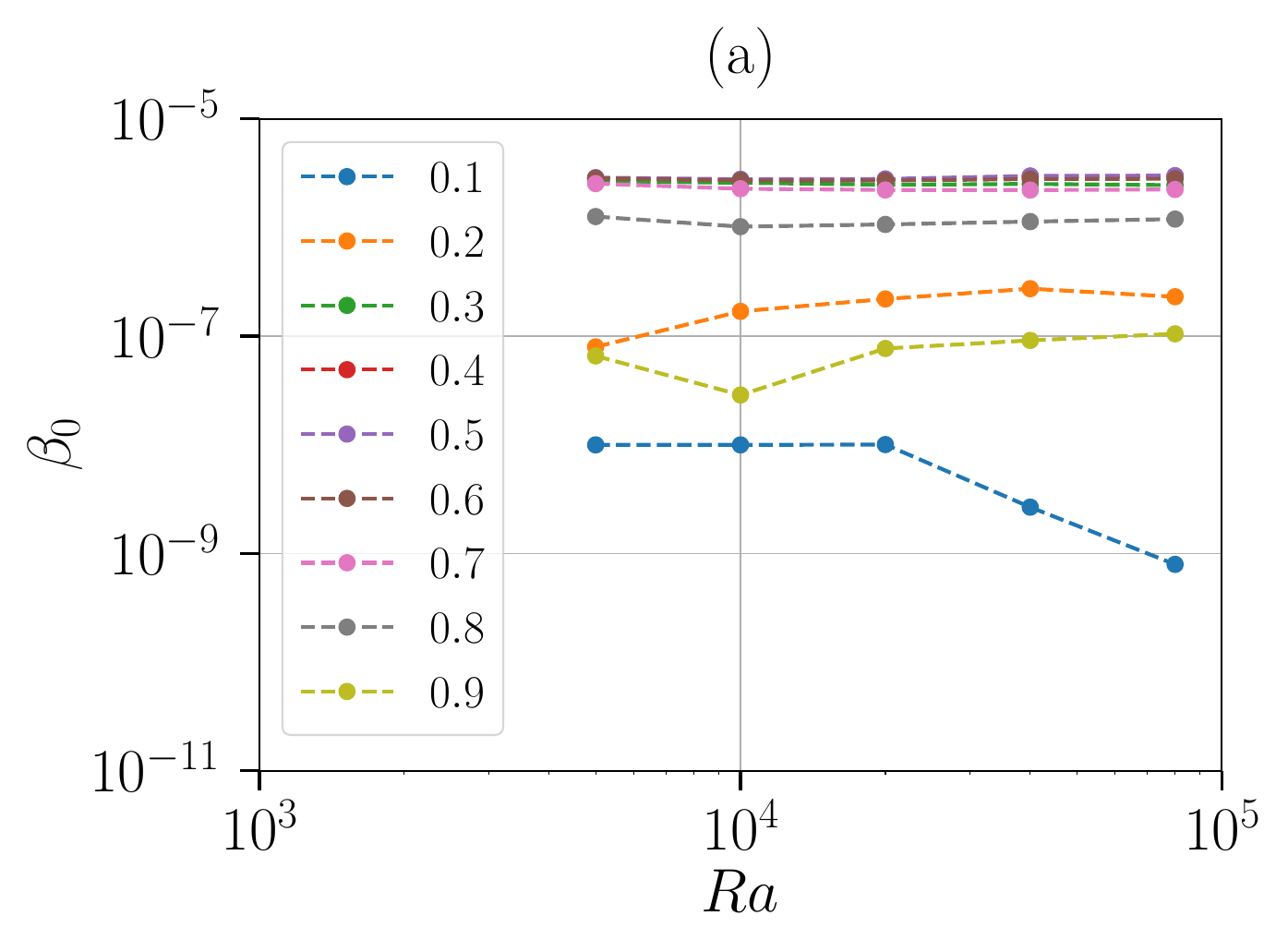}
    \includegraphics[height=0.36\linewidth]{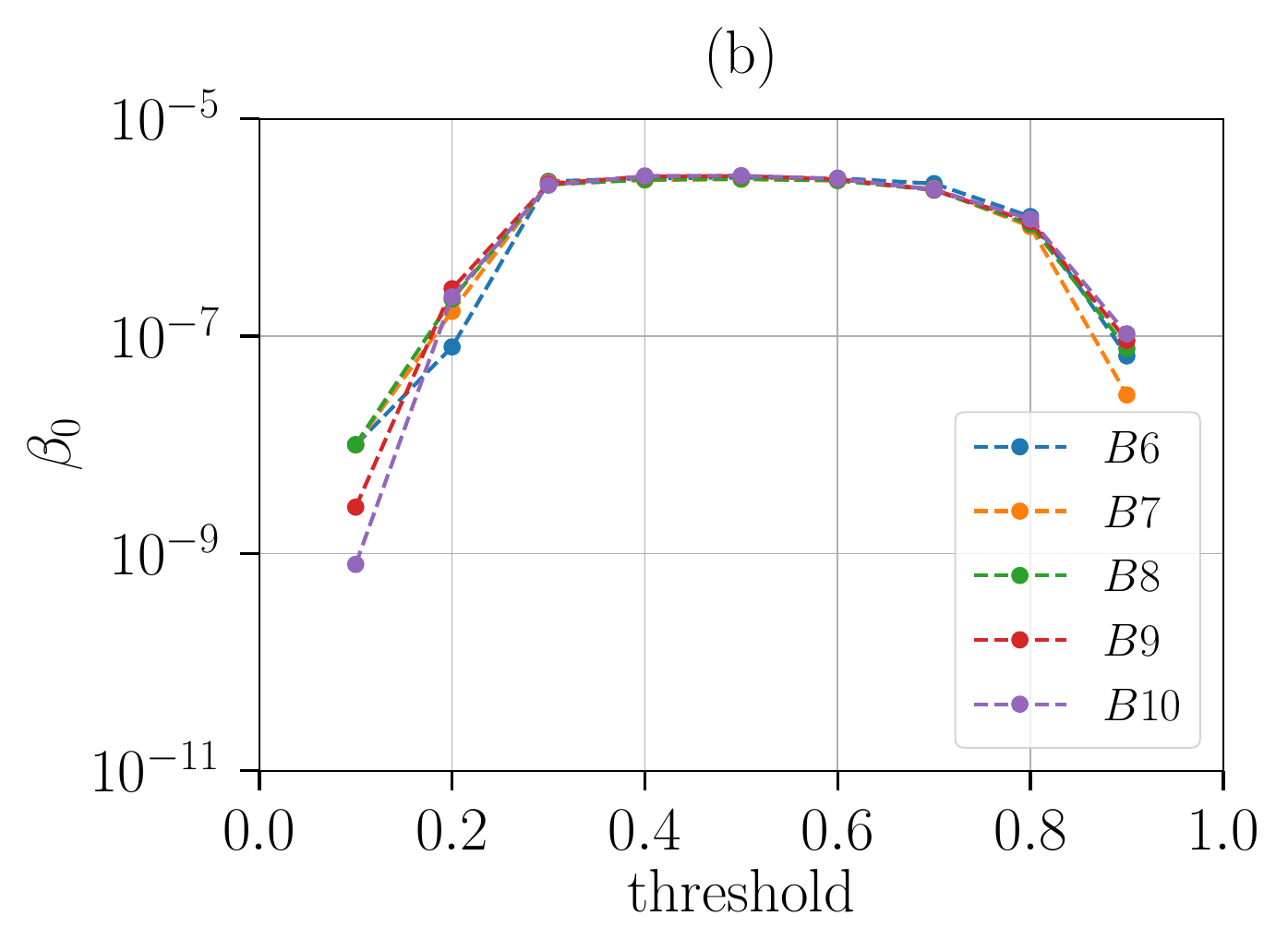}\\
    \vspace{0.5cm}
    \includegraphics[height=0.36\linewidth]{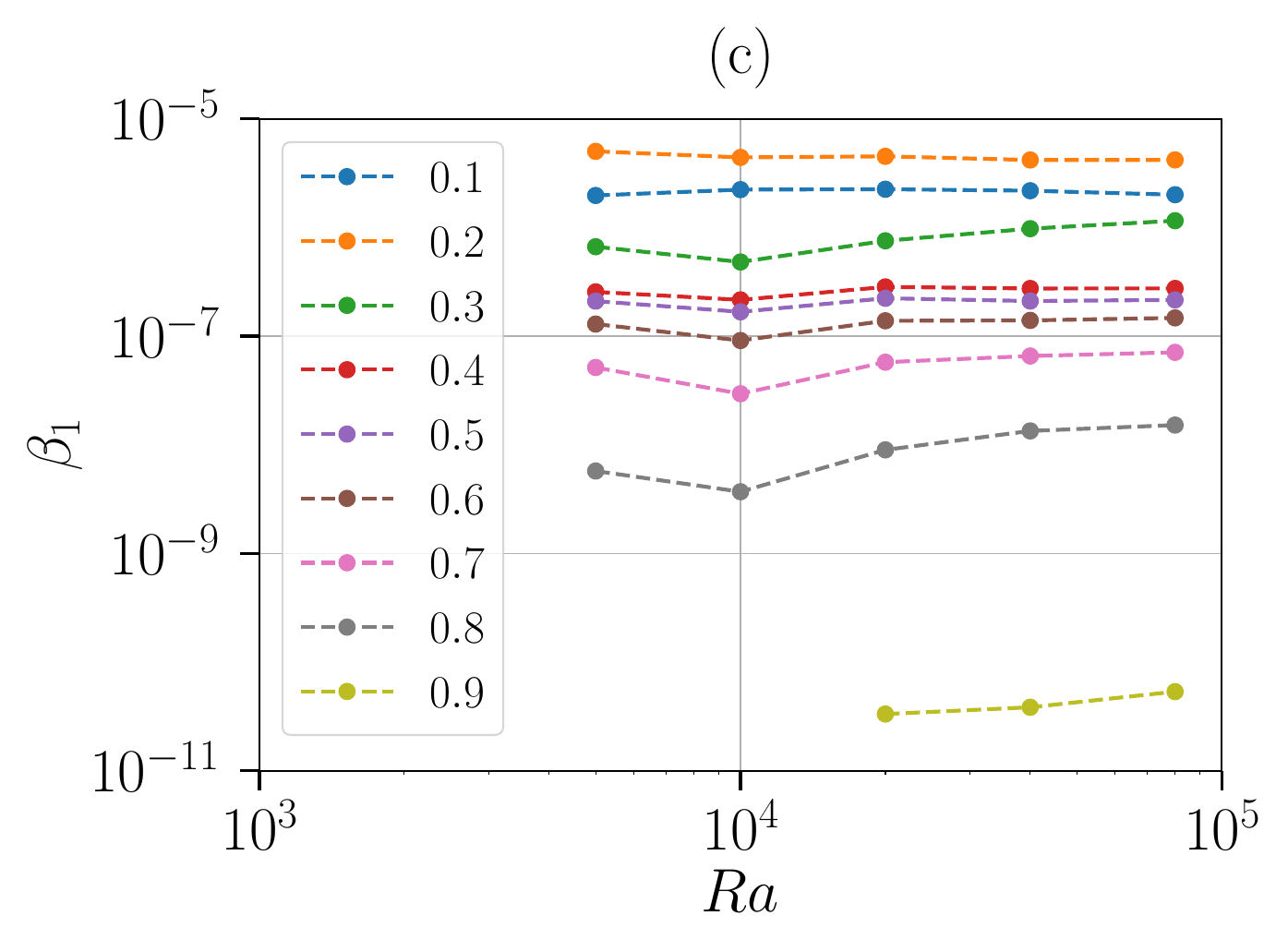}
    \includegraphics[height=0.36\linewidth]{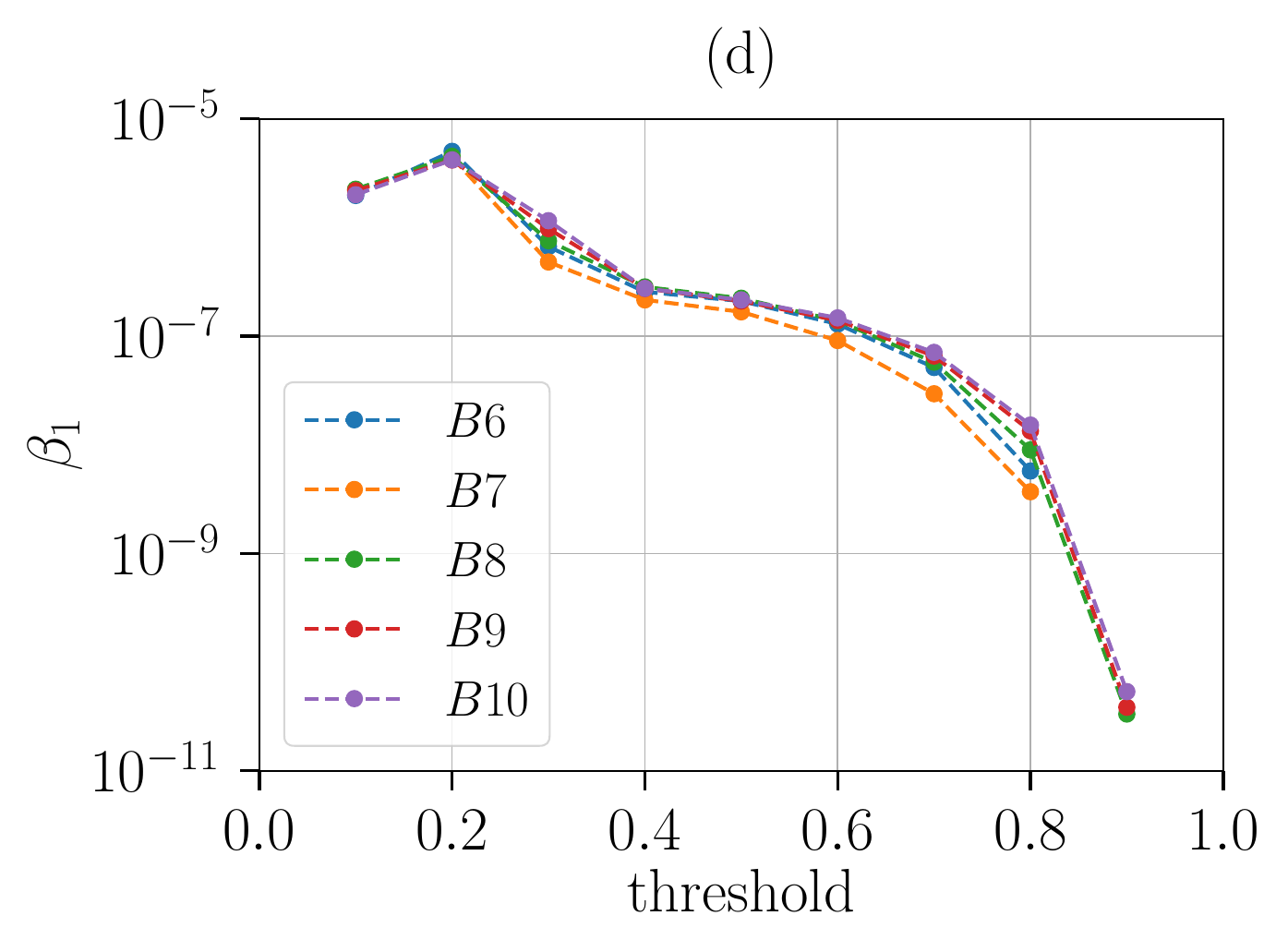}    
    \caption{Betti numbers (per area) for B6~-~B10 simulations: $\beta_0$ (a~-~b), and $\beta_1$ (c~-~d).  The parts (a,~c) show how Betti numbers depend on $\ra$ for different thresholds, and parts (b,~d) show how Betti numbers depend on the threshold value for all values of $\ra$ considered.}
    \label{fig:B6-B10Betti}
\end{figure}

Figure~\ref{fig:B6-B10Betti} shows $\beta_0$ and $\beta_1$ per area for different values of $\ra$ (corresponding to B6 - B10 simulations), and for a set of thresholds; to help understanding of the results, we also plot how Betti numbers depend on the threshold for each value of $\ra$. 

Focusing first on $\beta_0$, figure~\ref{fig:B6-B10Betti}(a) we note that the results are non-monotonous when considering different threshold; the reason for the non-monotonicity is clear from figure~\ref{fig:B6-B10Betti}(b) which shows that the values of $\beta_0$ are small for either very small or for very large thresholds: For small threshold values, $\beta_0$ is small since the areas of low temperature are all connected, so their number is small.  For large threshold values, the number of areas with high temperatures is small, leading again to small values of $\beta_0$.  For intermediate thresholds, the values of $\beta_0$ reach their maximum. We note that the decrease in $\beta_0$ values for the two largest considered values of $\ra$ at the threshold of 0.1 suggests increased connectivity of areas with temperature values of at least 0.1.  We also note that figure~\ref{fig:B6-B10Betti}(b) illustrates that the dependence of $\beta_0$ on the threshold value is robust across all considered values of $\ra$.

Considering $\beta_1$, figure~\ref{fig:B6-B10Betti}(c-d), we note a different trend of the results as the considered temperature threshold is modified.  Here, see part (d): there is an (almost) monotonic trend in $\beta_1$, since the number of loops (per area) decreases as the threshold increases. Recall that the loops are born when the weakest link appears (as the threshold is decreased), and therefore, for lower thresholds, there are more loops. For very large threshold values, there may not even be any loops, as can be seen in figure~\ref{fig:B6-B10Betti}(d) for the two smallest values of $\ra$ considered. The observed trend (decrease of $\beta_1$) persists for all but the smallest threshold; the reason for this change of trend is that for very small thresholds, the loops become filled up (recall the toy example shown in figure~\ref{fig:2D_toy}) and therefore their number decreases. 

A brief comment is in place regarding the band of generators next to the diagonal, which is excluded from the consideration.  As mentioned previously, if the bandwidth is small compared to a typical lifespan, these generators can be safely ignored.  This requirement is satisfied for all but the smallest threshold (0.1) considered; additional results (not shown for brevity) show that indeed the Betti numbers are larger for very small thresholds if such a band is included, however, the trend of the results remains the same.  We also note that minor deviations in the trends for $\beta_0$ and $\beta_1$ are expected due to the finite domain size considered. We also note in passing that similar trends in Betti number behaviour were observed when considering interaction networks in particulate-based systems~\citep{epl12}.  

\begin{figure}
    \centering
    \includegraphics[height=0.37\linewidth]{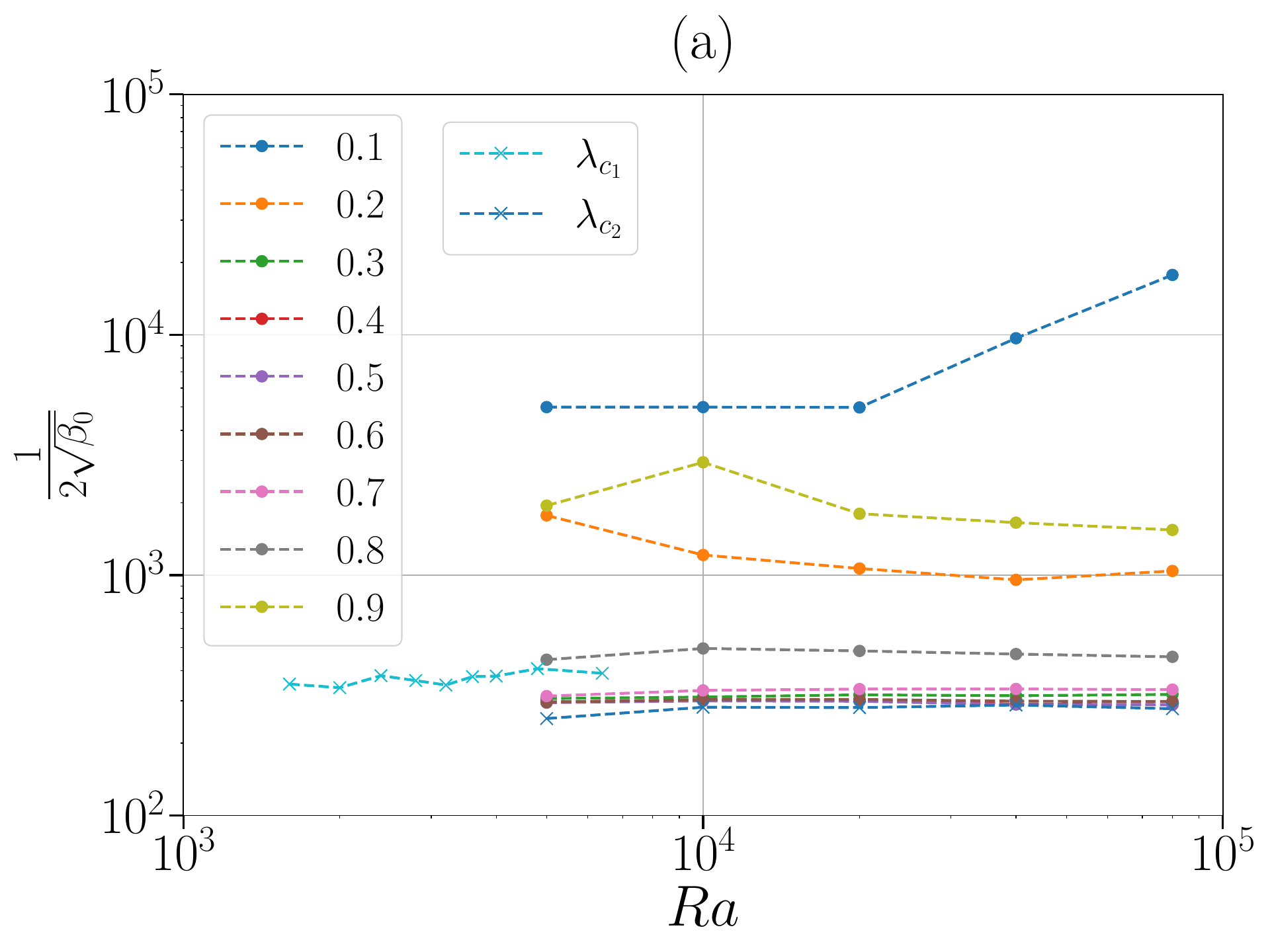}
    \includegraphics[height=0.37\linewidth]{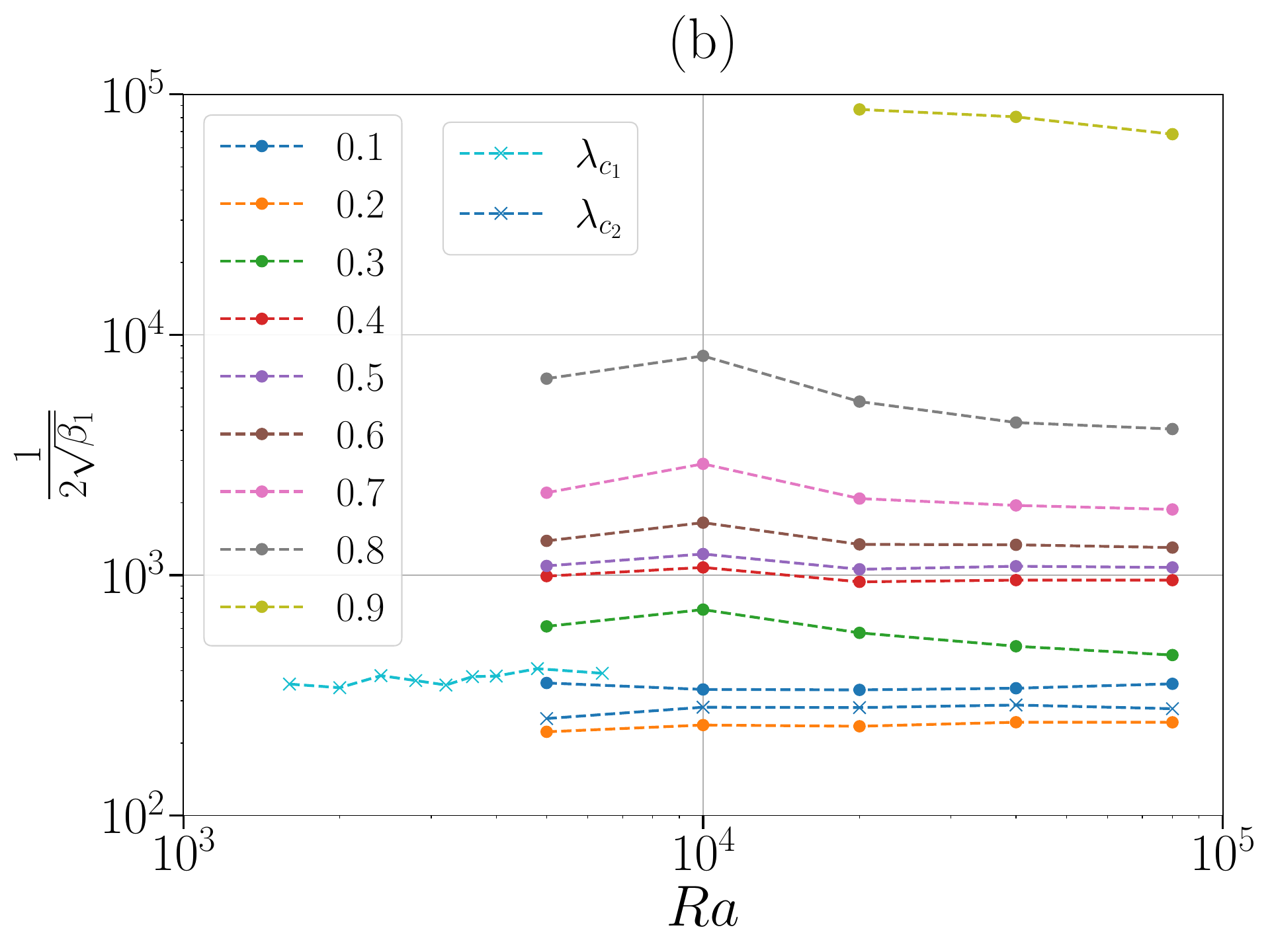}   
    \caption{The length scales for B6~-~B10 simulations obtained using Betti numbers, $\beta_0$ (a), and $\beta_1$ (b) as the considered temperature threshold is varied. We also show (by crosses) the values obtained by~\cite{fu2013pattern} ($\lambda_{c1}$) and by~\cite{depaoli2025grl} ($\lambda_{c2}$).}
    \label{fig:B6-B10lengthscale}
\end{figure}

To obtain a typical length scale corresponding to Betti numbers shown in figure~\ref{fig:B6-B10Betti}, we consider the following question: assuming that $N$ features (components or loops) are randomly distributed in a rectangular domain, what is the typical minimal distance between such features?  In two spatial dimensions, it is easy to see that for domain of linear dimensions $\widehat{L}_1$ and $\widehat{L}_2$, such distance is given by $\sqrt{\widehat{L}_1 \widehat{L}_2/N}/2$, (this is an asymptotic expression obtained in the limit of large $N$ and ignoring boundary effects).  Figure~\ref{fig:B6-B10lengthscale} shows the corresponding results obtained when substituting  $\beta_0$ (a) or $\beta_1$ (b) for $N$ (per area) in the above expression.  For $\beta_0$, we find non-monotonous behaviour for different thresholds, as expected since $\beta_0$'s are non-monotonous as well.  For intermediate values of thresholds (0.3 - 0.7), the computed length scale is close to the one found in~\cite{depaoli2025grl} and slightly lower than the one reported by~\cite{fu2013pattern}.  When considering $\beta_1$, figure \ref{fig:B6-B10lengthscale}(b), we find agreement with \citet{fu2013pattern} and \citet{depaoli2025grl} for low thresholds (0.1 - 0.2), for which almost all loops present in the considered data set are formed. In Appendix~\ref{sec:appC} we show that the results are essentially independent of the width of the excluded noise band, see figure~\ref{fig:noise1}. 

To summarize, the topological analysis based on persistent homology enhances our understanding of pattern formation in the system under consideration, illustrating, in particular, that results for emerging length scales depend on the approach used to compute them. 

\section{Two-sided convection}\label{sec:RB}

\subsection{Convection regimes}\label{sec:convreg}
We consider here a system heated from below and cooled from above, where a high (low) temperature is fixed at the bottom (top) wall, corresponding to the boundary conditions specified in~\eqref{eq:bcrb}. The domain is periodic in the horizontal directions, and the flow is initialized with a linear temperature distribution as in~\eqref{eq:bcrb2}. After an initial transient phase, the flow attains a statistically steady state. The boundary conditions and an example of temperature distribution on the lateral boundaries of the domain are illustrated by figure~\ref{fig:fig1}(b).

In free fluids, i.e. in the absence of a porous medium, the two-sided flow is also called Rayleigh-B\'enard flow, and it is particularly suitable to be analysed theoretically, due to its well-defined boundary conditions and its statistically-steady nature: it represents a perfect candidate to develop new concepts on instabilities and dynamical systems \citep{lohse2024ultimate}. For the same reasons, the porous counterpart (also called Rayleigh-B\'enard-Darcy or Rayleigh flow, where the free fluid is replaced by a fluid-saturated porous medium) has been widely investigated. In particular, extensive studies focused on the onset of the flow instabilities \cite[$0<\ra<O(10^2)$, see][]{horton1945convection,lapwood1948convection} and on the dynamics at moderate- to high-$\ra$ \cite[$O(10^2)<\ra<O(10^4)$, see][]{graham1994plume,otero2004high}. In these systems, the global response parameter is the Nusselt number $\nus$ defined as in~\eqref{eq:flux} which, for $\ra>10^3$,  scales as $\nus\sim\ra$ plus sublinear corrections \citep{depaoli2024heat}. As a result, an increase of $\ra$ by a factor 10 would require substantially increasing the number of degrees of freedom required to resolve the flow, namely by a factor of $O(10^3)$ in 3D, corresponding to an even larger increase of the computational cost. In recent decades, however, numerical advancements have allowed to explore in detail the flow dynamics in 3D and at large $\ra$ \citep{hewitt2014high,pirozzoli2021towards,depaoli2022strong,hu2023effects}, and a rich flow morphology has been observed. A review for $\ra\le740$ is provided by \citet{hewitt2014high}, which we summarize and extend in the following.

\begin{figure}
    \centering
    \includegraphics[width=0.95\linewidth]{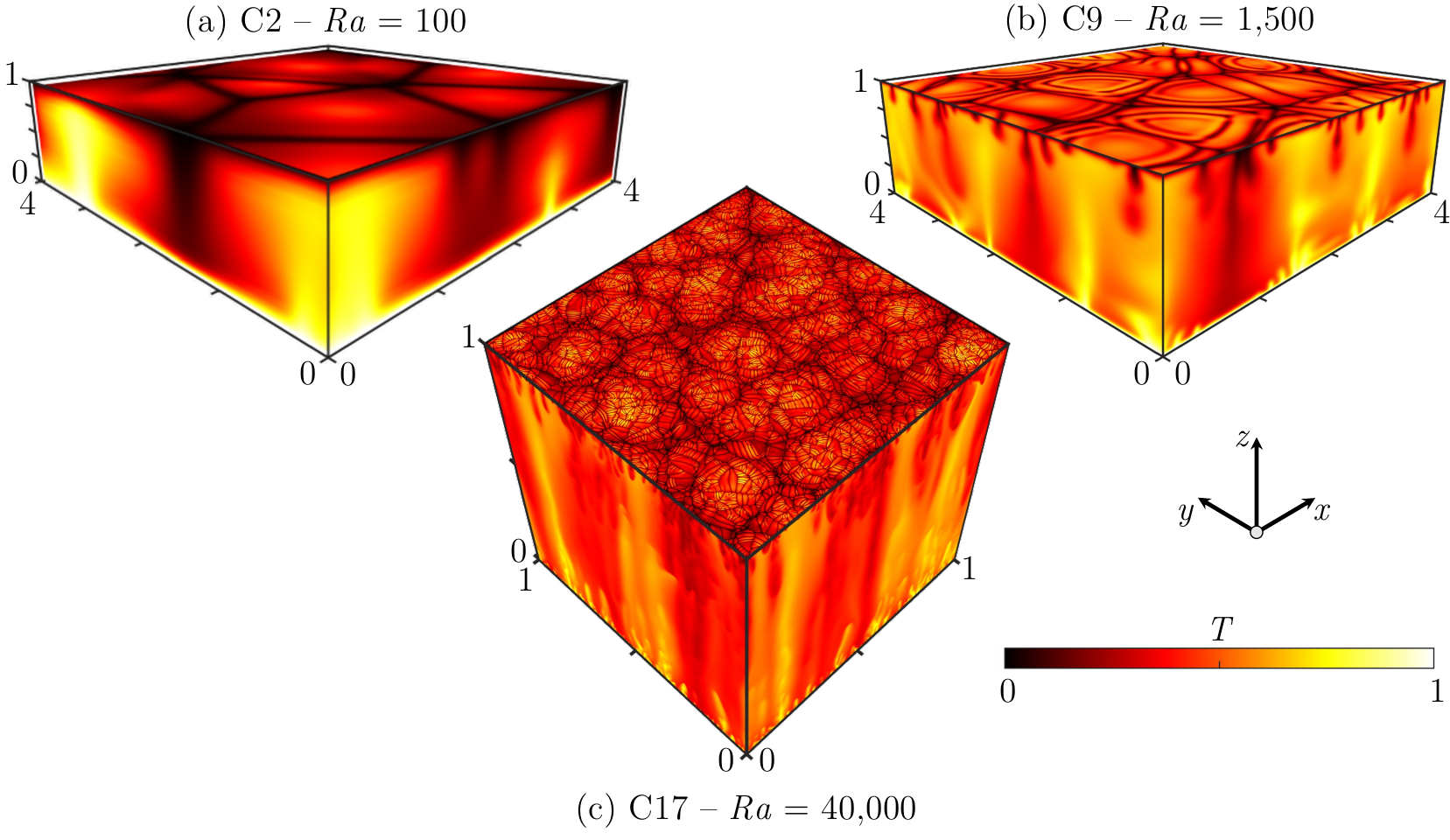}
    \caption{
    Temperature distribution over a portion of the domains considered, namely $0\le x \le4$, $0 \le y \le 4$ and $0\le z \le z_t$ (with $0.95\le z_t \le 0.999$, depending on $\ra$). Distributions shown are relative to simulations C2~(a), C9~(b) and C17~(c) (see table~\ref{tab:rb} for additional details).
    }
    \label{fig:3Dcells}
\end{figure}

Figure~\ref{fig:3Dcells} shows a few examples of the temperature distributions for three values of $\ra$, illustrating the changes of the flow morphology. To highlight the complex near-wall flow pattern, the temperature field is shown on a horizontal slice located near the top boundary and located at $z=z_t$ ($0.95\le z_t \le 0.999$, depending on $\ra$), focusing on $\ra \ge 100$. For smaller values (specifically, for $\ra < 4\pi^2$) any instability is suppressed, and the flow is purely conductive. When $\ra$ is slightly increased, a 2D unstable mode appears ($4\pi^2 \le \ra \lesssim  4.5\pi^2$), and dominates the flow also at larger $\ra$, when multiple modes are present ($4.5\pi^2 \lesssim \ra \lesssim  97$). In these regimes, temperature structures spanning the entire system height, from one boundary layer to the other, populate the domain. The flow eventually develops a 3D steady structure ($97 \lesssim \ra \lesssim  300$, see figure~\ref{fig:3Dcells}a), and multiple possible states exist.  Finally, the flow is unsteady for $\ra \gtrsim 300$: small flow instabilities develop and grow from the thermal boundary layers, move parallel to the horizontal walls, and eventually merge into larger plumes, still covering the whole domain height, but now being unsteady (figure~\ref{fig:3Dcells}b). For sufficiently large driving, namely $\ra\ge1750$, the system enters the so-called high-$\ra$ regime and the flow morphology does not exhibit any regular background structure \citep{hewitt2014high}, see e.g., figure~\ref{fig:3Dcells}(c). In this regime, the flow can be separated into three different parts: (i)~a thin and time-dependent thermal boundary layer at the walls; (ii)~an intermediate region populated by very dynamical sheet-like plumes, originated from the time-dependent wall layer; and (iii)~the bulk, controlled by large columnar structures of hot (cold) rising (sinking) fluid, labelled megaplumes.

The dynamics at larger $\ra$ ($\ra\lesssim 10^5$) has been recently explored in detail by \citet{pirozzoli2021towards} and \citet{depaoli2022strong}. These authors observed that when $\ra\gtrsim10^4$, the temperature distribution in the intermediate region is characterized not only by the presence of the small plumes originated from the boundary layer, but also by the existence of large structures, entraining many plumes and representing the footprint of the megaplumes populating the bulk of the flow \citep{depaoli2022strong}. These structures, labelled as supercells, have been identified visually as the large loops gathering many smaller cells (see e.g. the dark loops on the horizontal cut in figure~\ref{fig:3Dcells}c), and have been observed to be: (i)~persistent in time, and (ii)~correlated in space with the megaplumes \citep{depaoli2022strong}. Time persistence has been identified by averaging the temperature distribution in the near-wall region and observing that a pattern matching that of the supercells emerges.  Spatial correlation with the megaplumes was demonstrated by filtering the near-wall temperature field to remove flow structures with wavelengths smaller than the dominant wavelength at the mid-plane. 

Despite these efforts, a precise definition of the supercells and an accurate determination of the conditions required for their appearance remain elusive. The following question remains unanswered: Can we provide a more detailed description of the supercells? Can we describe their presence/formation as a function of $\ra$? Do we have to rely on the bulk temperature field (i.e., the temperature field far from the top and bottom boundaries) to determine the existence and the morphology of the supercells? In this work, we aim at precisely answering these questions: we provide a robust criterion to identify supercells and describe their formation across a wide range of $\ra$.

\begin{figure}
    \centering
    \includegraphics[width=1.0\linewidth]{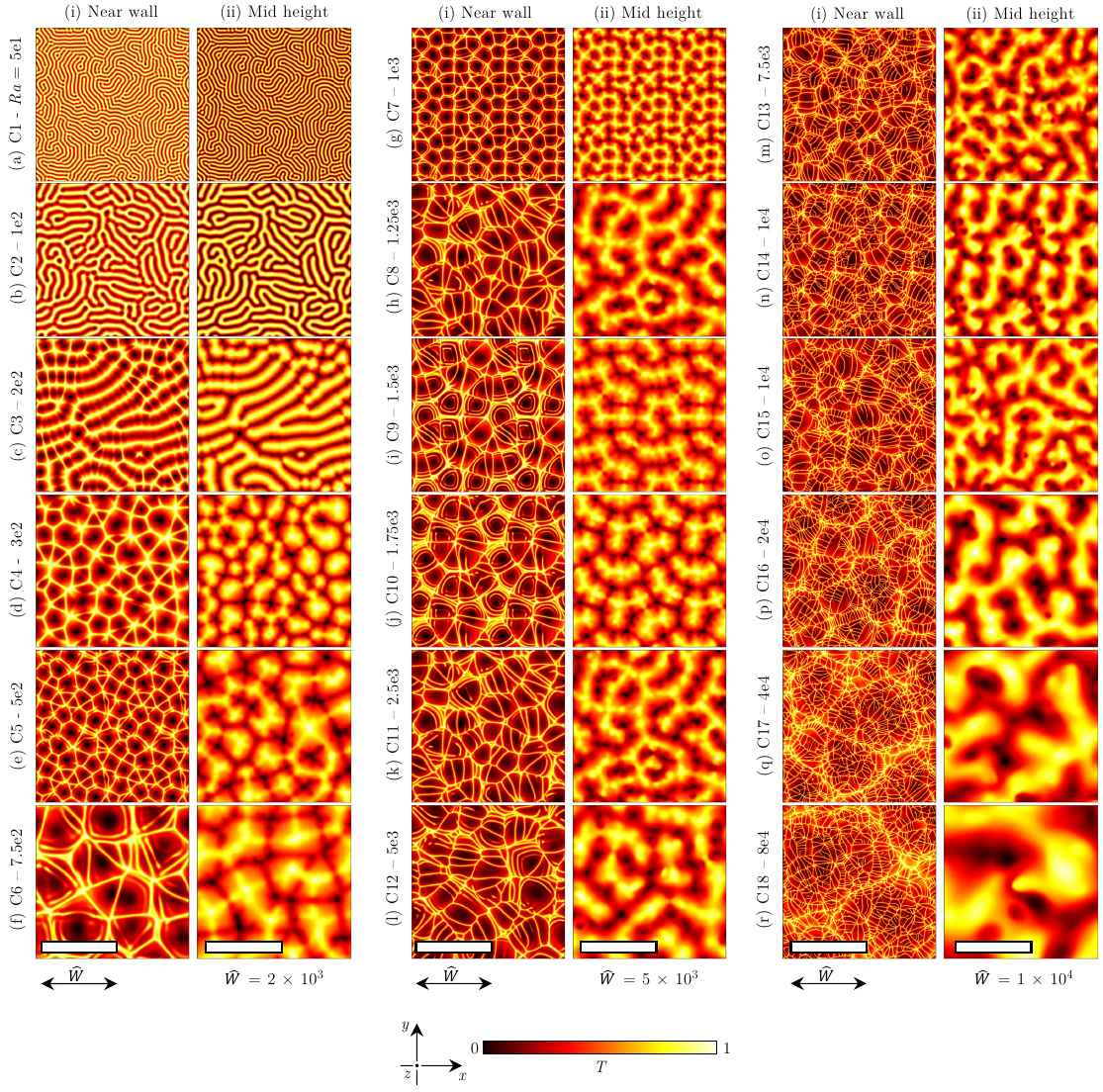}
    \caption{
    Visualization of the temperature fields on horizontal ($x,y$) planes near the bottom wall (panels~i) and at the mid height ($z=1/2$, panels~ii). The simulations label C\# and the Rayleigh numbers $\ra$ are indicated, for all other details we refer to table~\ref{tab:rb} (see~Appendix~\ref{sec:appA}).     The size of the domain shown is constant within each column, but varies from left to right for visualization purposes. A scale bar (width $\widehat{W}$ expressed in diffusive units) is reported for reference. In some cases (e.g., C18), only a small portion of the domain is shown, while in others (e.g., C7), the field is replicated to match the desired size. Note the difference between the cell structure in the present results and in those of \citet{hewitt2014high} (see the results labelled as IC1 in \citep{hewitt2014high}, corresponding to the same initial condition employed here), suggesting the strong effect of the domain size on the determination of the flow structure at low Rayleigh numbers $(\ra\le500)$. 
    The data analysed relative to simulations~C are available via \citet{databaseC}.
    }
    \label{fig:cells}
\end{figure}

\subsection{Qualitative analysis of the flow morphology}\label{sec:morphology}

We analyse the flow morphology for $5\times 10^1\le\ra\le8\times10^4$, by considering the horizontal temperature distributions in the near-wall region (panels~(i) in figure~\ref{fig:cells}) and at the midplane (panels~(ii) in figure~\ref{fig:cells}). For ease of comparison, the size of the portion of domains shown in figure~\ref{fig:cells} is constant (in diffusive units) within each column (scale bar is reported as a reference). The simulations considered are either those by \citet{pirozzoli2021towards} and \citet{depaoli2022strong}, or presented here for the first time, and the data are made available via \citet{databaseC}.

At $\ra\le1750$, the flow is also dependent on the initial condition, and it exhibits hysteresis effects \citep{otero2004high}. \citet{hewitt2014high} explored the dynamics following two initial conditions: IC1, corresponding to a perturbed linear temperature profile, and IC2, obtained starting from a steady-state temperature field obtained from a simulation at a lower $\ra$. Therefore, the initial condition adds here to the flow governing parameters ($\ra,\widehat{L}_1,\widehat{L}_2$). To reduce the parameters space, here we will consider only one initial condition \citep[corresponding to IC1 of][]{hewitt2014high} characterized by a perturbed form of the stable state, Eq.~\eqref{eq:bcrb2}, and we will consider very large domains ($\widehat{L}_1=\widehat{L}_2\ge4\times10^3$), such that the flow is independent of the values of $\widehat{L}_1$ and $\widehat{L}_2$. As a result, the only remaining governing parameter is $\ra$. Simulations performed in this work have been initialized with a linear temperature distribution, Eq.~\eqref{eq:bcrb2}, which in dimensionless terms reads $T=1-z$, and have been run for $t\ge1000$, to make sure the steady state (determined by keeping track of the time-averaged value of the Nusselt number) is achieved. Then, the temperature fields used to analyse the flow pattern are saved approximately every 10 or 20 convective time units.  All the relevant simulation details are indicated in table~\ref{tab:rb} (see~Appendix~\ref{sec:appA}).

Before quantifying the flow pattern using persistent homology techniques, we notice that a visual inspection reveals some interesting features previously unobserved. In particular, at low Rayleigh numbers the flow organization described in the literature \citep[see figure~2b of][]{hewitt2014high} consists of cells forming a very regular pattern that for $\ra\le 2\times 10^2$ is one-dimensional (rolls spanning across the entire domain width) and for $2\times10^2\le\ra\le 3\times10^2$ is two-dimensional and still very regular (equally squared cells). The domains considered in figure~\ref{fig:cells} are about 100 times larger than in \citet{hewitt2014high}, and the cells are organized in a more chaotic manner: For $\ra\le2\times10^2$, corresponding to figures~\ref{fig:cells}(a-c), we still observe the formation of rolls, which in contrast to \citet{hewitt2014high} do not span across the entire domain in the horizontal directions. Also for $\ra= 3\times10^2$ (figure~\ref{fig:cells}d) there are differences compared to previous works, where the cells were all equal: we do observe the transition from rolls to cells, but such cells have an irregular shape and are, in general, different from each other (polygons with 3 to 6 sides).
These differences suggest that the previously observed regular flow pattern emerging at small $\ra$ is a genuine domain size effect, which is induced by the periodic boundary conditions applied in the horizontal directions (additional details are provided in Appendix~\ref{sec:appBlow}).
These findings highlight the fact that at small $\ra$ the influence of the domain size on the flow morphology is considerable.
Finally, we observe that at very large Rayleigh numbers ($\ra\ge2\times10^4$, figures~\ref{fig:cells}p-r), where the supercells are clearly visible, their size increases with $\ra$. This is not surprising, since the temperature structures in the bulk \citep[megaplumes of size $\sim\ra^{1/2}$ in diffusive units, see][]{hewitt2014high,pirozzoli2021towards} are responsible for the formation of the supercells \citep{depaoli2022strong}. However, this highlights once again the importance of domain size in capturing the large-scale structures in the flow.

\begin{figure}
    \centering
    \vspace{0.5cm}
    \includegraphics[width=1\linewidth]{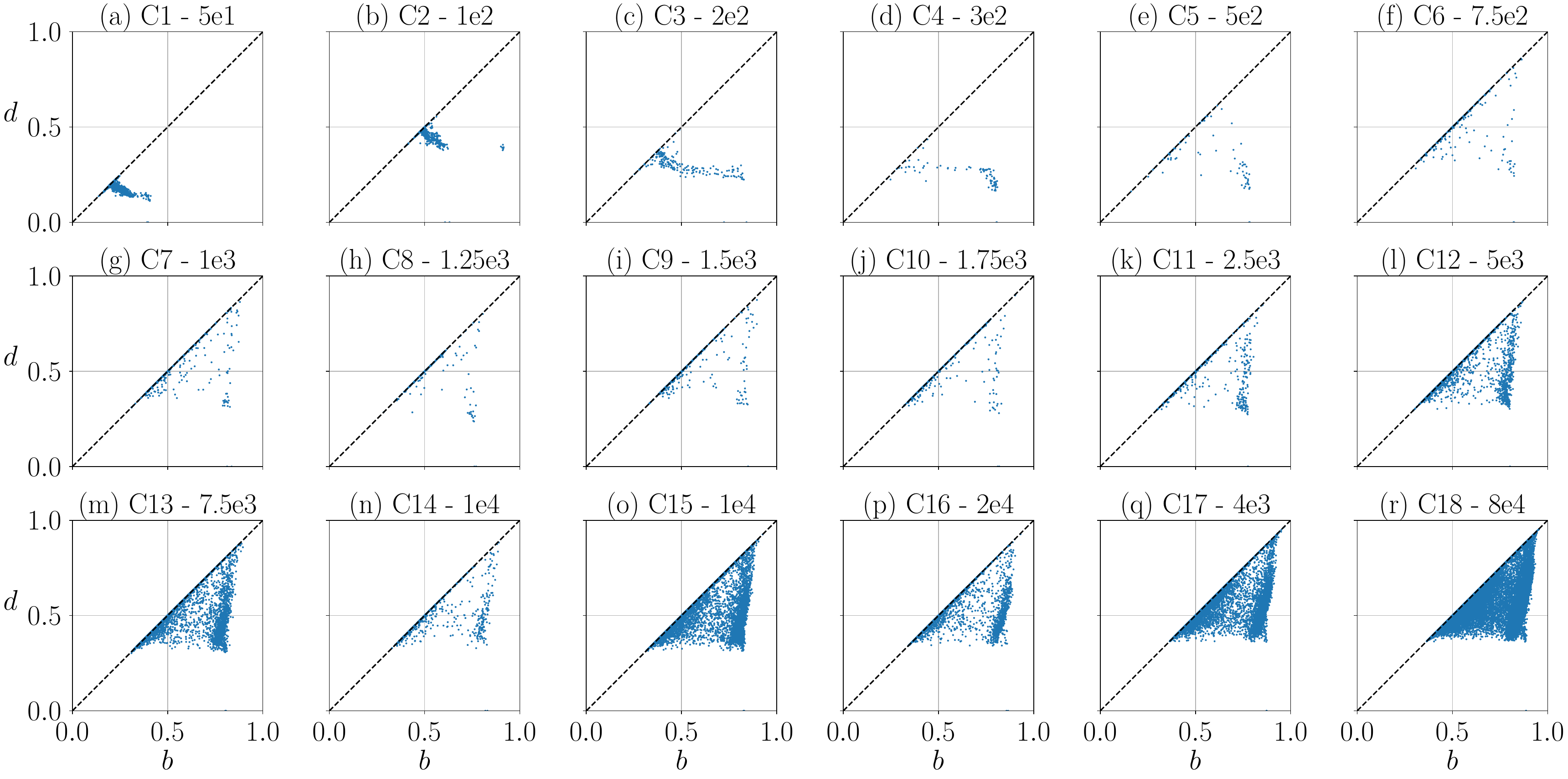}    
    \caption{ 
    Loop persistence diagrams, $\beta_1~\pd$s, in two-sided convection simulation near the wall (simulations~C in table~\ref{tab:rb}). As the Rayleigh number ($\ra$) increases, the morphology of the boundary develops significant changes, leading to variations in the corresponding $\beta_1~\pd$. Between $\ra=300$ and 500, we observe a morphological pattern change, with a clear transformation from a maze to a cellular network (see also figure~\ref{fig:cells}). Proceeding from $\ra=1,000$, a division emerges, forming two distinct clusters of points in the $\pd$s. This indicates an increasing number of loops within the boundary layer, appearing at high temperatures ($T\approx 0.75$). For $\ra \ge 2,500$, the cluster at high temperatures becomes denser, suggesting an increased number of loops in the patterns shown in figure~\ref{fig:cells}.}
    \label{fig:b1_multiplots}
\end{figure}

\subsection{Quantification of pattern formation} 

We focus the presentation that follows on the identification of cells and supercells and, more generally, on the influence of $\ra$ on the pattern-formation process. We discuss loop-related measures only, as they provide a clear description of the patterns of interest.

Figure~\ref{fig:b1_multiplots} shows the loop ($\beta_1$ $\pd$s) for the simulations listed in table~\ref{tab:rb}.  The emerging behaviour, as $\ra$ increases, is an increase in the number of generators, and, in particular, the gradual appearance of a band of generators with the birth coordinates of $T\gtrsim 0.75$ starting at $\ra \gtrsim 1,000$.  For larger $\ra$, the number of points in the $\pd$s increases substantially, and their structure becomes difficult to visualize. For this reason, we consider cumulative measures, which simplify the interpretation of the results, as discussed next.

\begin{figure}
    \centering 
    \includegraphics[width=1.00\linewidth]{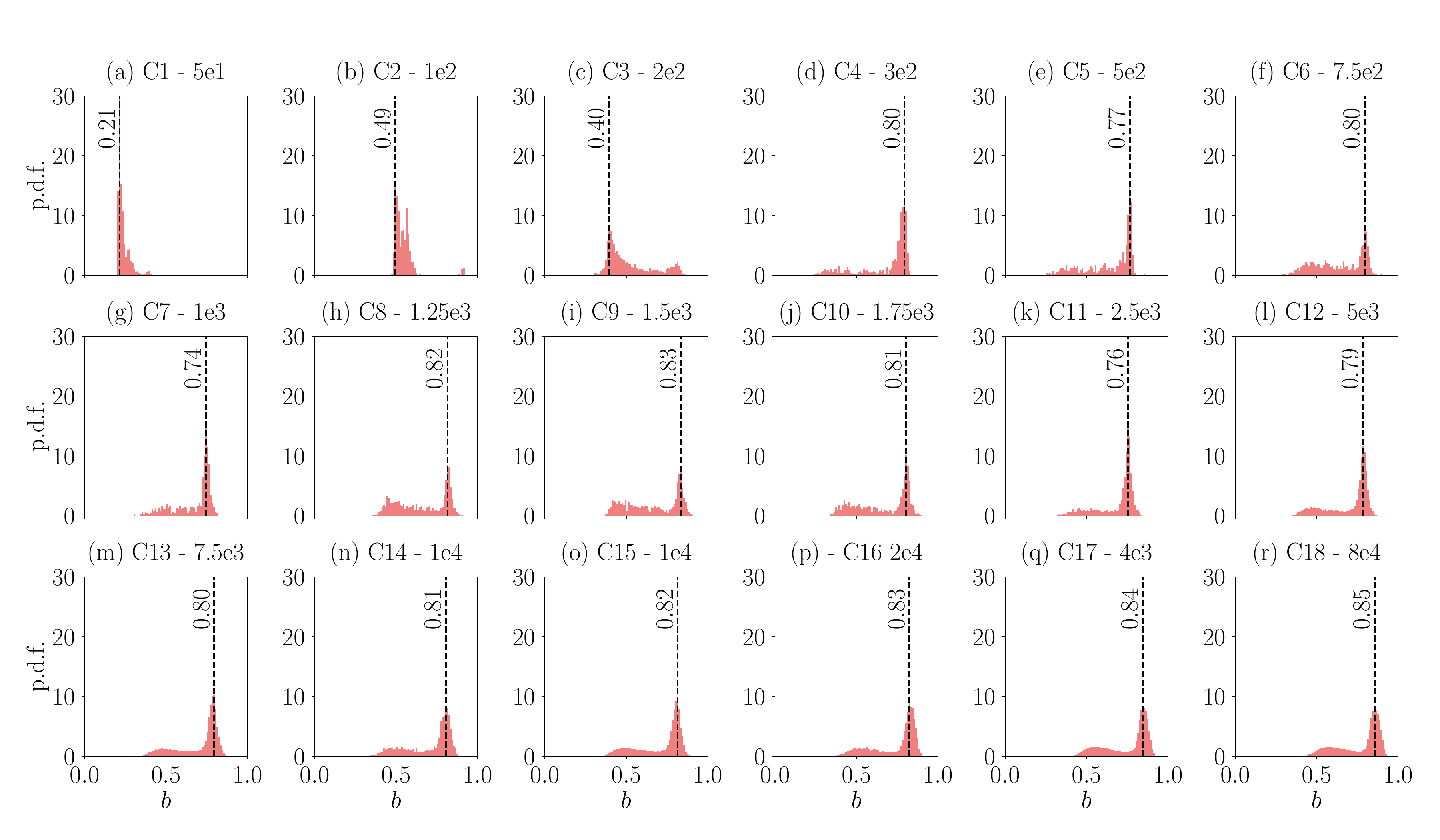} \caption{\label{fig:birth_distributions_removed_noise}
    Probability density function (p.d.f.) of the birth values of $\beta_1$ topological generators, accumulated over 10 time outputs for each simulation (noise band 0.01 is removed).
    The birth coordinate corresponding to the maximum value of the p.d.f. is indicated by the dashed line and is reported in each panel.
    A transition from rolls (a~-~c) to cells/supercells (d~-~r) is clearly observed.
    }
    \label{fig:birth_distributions}
    \end{figure}

Figure~\ref{fig:birth_distributions} shows the normalized distribution (p.d.f.) of the birth coordinate of the loops, using the data from $\pd$s shown in figure~\ref{fig:b1_multiplots}, but averaged over 10 time outputs (saved approximately every 10 or 20 convective time units) in order to improve statistics. We recall that the birth coordinate represents the temperature at which the loops close and, therefore, at which the cells (or supercells) form.
At low values of $\ra$, we note a dramatic change in the p.d.f. as $\ra$ increases from 200 (c) to 300 (d), illustrating the topological change between rolls (c) and cells (d).
This topological change can be visualized in figure~\ref{fig:cells}: the rolls initially present in the system (figure~\ref{fig:cells}a-i) start to connect, forming cells (figure~\ref{fig:cells}b-i,c-i).
These cells fully develop and are clearly visible at larger $\ra$ (starting with figure~\ref{fig:cells}d-i).
Further increase of $\ra$ leads to development of a clear and well-defined structure of the p.d.f.'s shown in figure~\ref{fig:birth_distributions} with a dominant peak at $T\approx 0.8$ corresponding to supercells (see, e.g., figure~\ref{fig:cells}o-i to r-i), and broad distribution of the birth values for $0.5 \lesssim T \lesssim 0.7$, corresponding to cells. We note that the portions of the flow in which $T \approx 0.75$  roughly correspond to the regions in which the horizontal temperature gradient is maximum~\citep{depaoli2022strong}. We also comment that the results for p.d.f.'s of connected components do not provide any useful information; therefore, considering loops is crucial for describing cells and supercells. 

\begin{figure}
    \centering
    \includegraphics[height=0.33\linewidth]{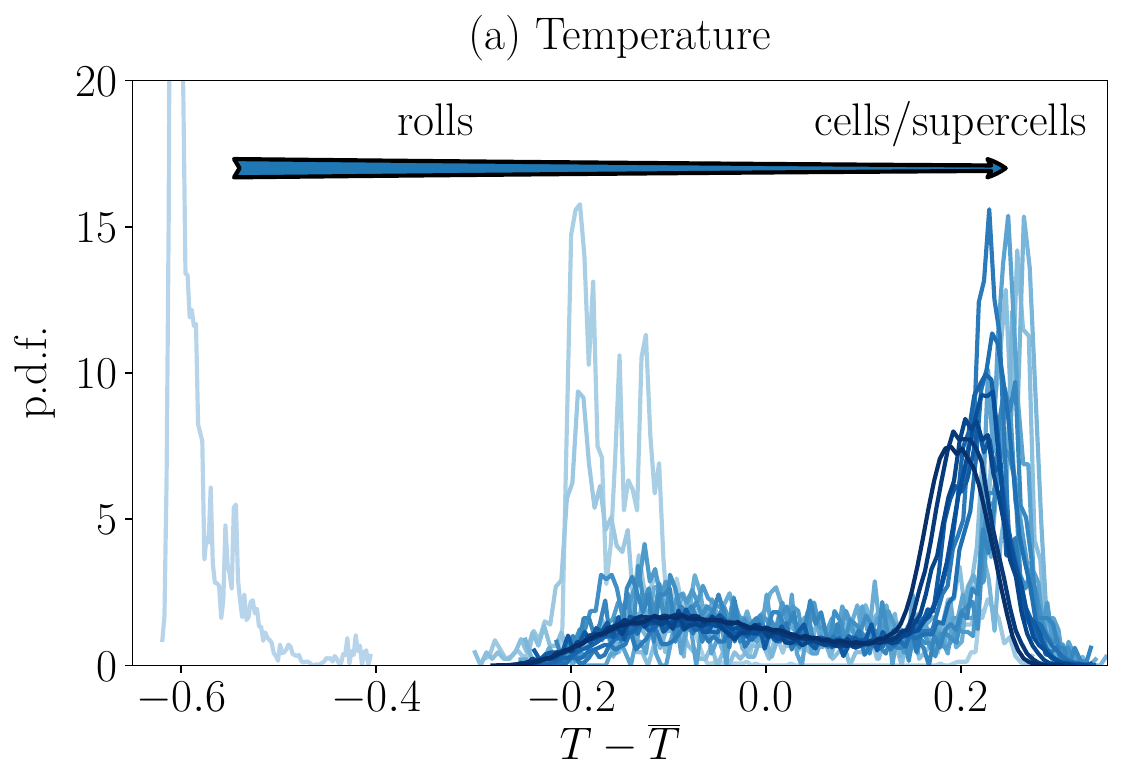}
    \includegraphics[height=0.33\linewidth]{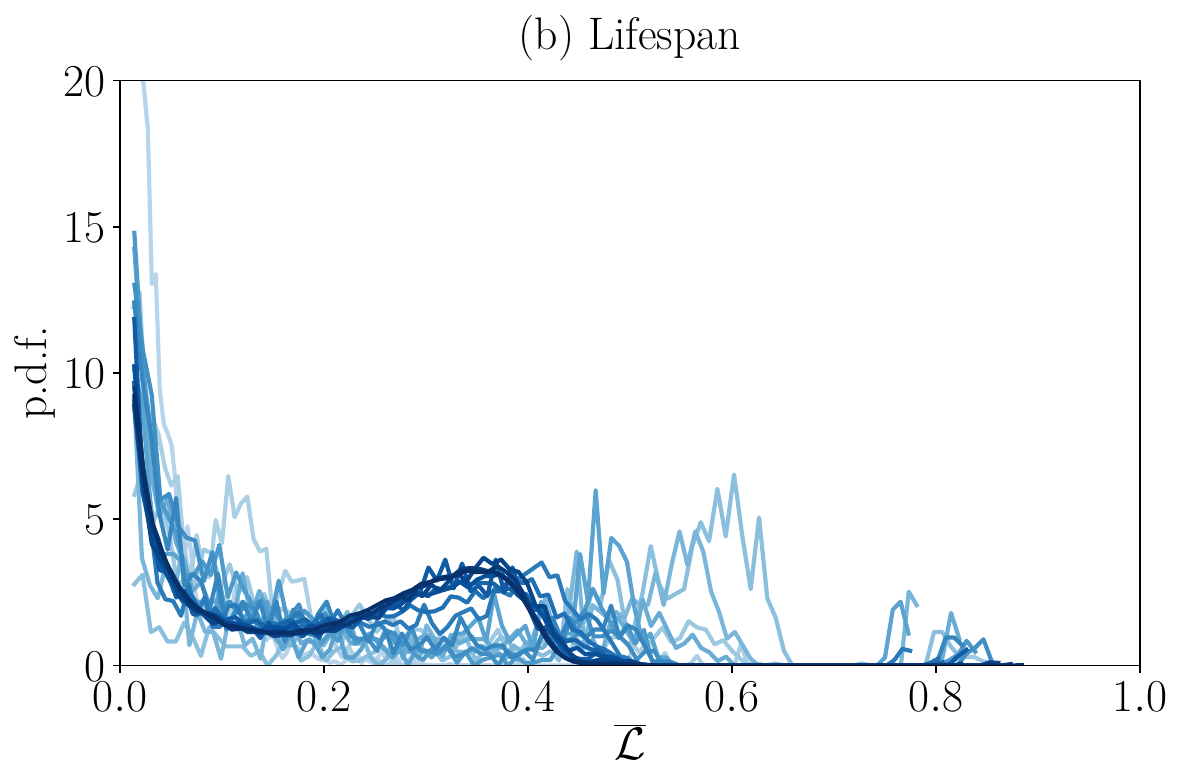}    
    \includegraphics[width=1\linewidth]{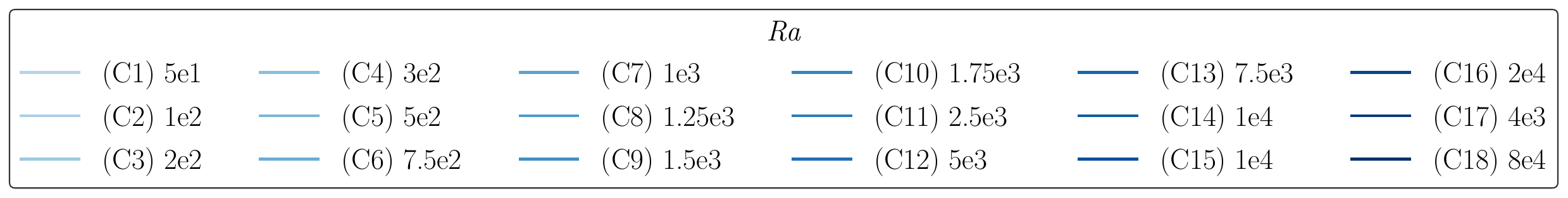}    
    \caption{\label{fig:rescaled}
 (a) Probability density function (p.d.f.) of $\beta_1$ topological generators as a function of the shifted temperature, with $\overline{T}$ corresponding to the space and time averaged temperature in the considered horizontal slice; (b) p.d.f of the lifespan, $\overline{\cal L}$ (noise band 0.01 is removed). Note that the lifespan is independent of the temperature shift. We observe approximate self-similarity of the results for large values of $\ra$.
 }
\end{figure}

Further insight can be obtained by considering the loops' birth coordinate as a function of $T-\overline{T}$, where $\overline{T}$ is the temperature averaged in space (over the horizontal plane considered) and in time (over 10 time outputs as also implemented for the plots shown in figure~\ref{fig:birth_distributions}).  Figure~\ref{fig:rescaled} plots combined results, showing (a)~the $\beta_1$ birth coordinate distribution and (b)~the $\beta_1$ lifespan.  Most importantly, the approximate collapse of the curves for $\ra\gtrsim 10^4$ for both birth values in \ref{fig:rescaled}(a) and lifespans~\ref{fig:rescaled}(b), illustrates the 
validity of the topological properties describing the temperature field across the wide range of parameters considered.
More precisely, the collapse in figure~\ref{fig:rescaled}(a) shows that the supercells appear at the similar values of the shifted temperature $T - \bar{T}$, while the collapse visible in figure~\ref{fig:rescaled}(b) shows that the cells also die (merge with other cells) at similar values of the shifted temperature (since lifespan measures difference between the birth and death coordinates). For smaller values of $\ra$, while the lifespans shown in figure~\ref{fig:rescaled} are too noisy to reach and useful insight, the birth coordinate distribution shows a clear shift in the peak of p.d.f.'s from $T - \bar{T} \sim -0.6$ for $\ra = 50 $ to $T - \bar{T} \sim -0.2$ for $100 \le \ra \le 200$, and then to $T - \bar{T} \sim 0.2$ for larger values of $\ra$.  
Additional details are provided in Appendix~\ref{sec:appBhigh}, where we prove the robustness of our findings, with respect to the number of samples and the different aspect ratios considered.


\section{Summary and conclusions}\label{sec:concl}
We examine the morphology of flow patterns arising in convection in porous media by combining large-scale numerical simulations with tools from topological data analysis, in particular, persistent homology (PH). 
The PH results we present are based on planar near-wall temperature fields and describe the topology of those slices, not the full 3D topology of the flow.
By analysing both one-sided and two-sided flow configurations over a broad range of Rayleigh–Darcy numbers and domain sizes, we demonstrate that PH provides an objective, quantitative framework for characterising flow structures. 
In particular, given the temperature distributions, we observe the emergence of complex temperature structures, which are described in the present work. Unlike classical approaches previously employed, based on threshold selection \citep{depaoli2022strong}, Fourier analysis \citep{pirozzoli2021towards}, or cell-size measurements \citep{fu2013pattern}, PH quantifies structures across all temperature levels simultaneously and captures their number, connectivity, and persistence. It is important to note that PH analysis shows that the emerging length scales depend on the threshold selection, with only a specific threshold range reproducing the existing results. This analysis allows us to relate flow morphology directly to the system's transport properties, such as the Nusselt number $\nus$, and to identify 
features that are common across different regimes.

For the one-sided configuration, PH measures reveal clear signatures associated with the canonical stages of convective dissolution. During the diffusive stage, the number and prominence of topological features remain small, reflecting the smooth gradients observed across the temperature field. As convection initiates and thermal plumes form, both the number of topological generators and their lifespans increase, capturing the onset of multi-scale structures. Subsequent plume merging and coarsening are reflected by a reduction in the number of generators but an increase in their lifespans, consistent with the emergence of larger coherent plumes. These trends correlate strongly with the temporal evolution of the Nusselt number, highlighting the tight coupling between topology and heat flux.
In particular, the rapid decay of PH lifespans correlates with the onset of shutdown and may provide a useful post-processing signature of this transition.

For the two-sided configuration, where the flow morphology is richer than in the one-sided case, PH successfully identifies the hierarchical organisation of structures near the walls. We show that the transition from near-wall maze-like structures to cellular supercells corresponds to systematic variations in the distributions of birth and death coordinates of loop generators.
At sufficiently high Rayleigh numbers, the probability density functions of both birth values and lifespans collapse onto each other, demonstrating a self-similar behaviour in the high-$\ra$ regime.  The topological measures, therefore, show generic, self-similar characteristics of the emerging patterns in this regime.  
This finding suggests that, despite the apparent visual complexity of the patterns, their near-wall topological organization becomes $\ra$-independent once convection is sufficiently vigorous, the domain is sufficiently large and relative to the initial condition considered. In addition, we also find that large aspect-ratio simulations at low $\ra$ (namely at $\ra\le10^2$) lead to a maze-like flow organization that differs from the more regular cell pattern previously observed in smaller domains \citep{hewitt2014high}.

Overall, our results show that PH is a powerful tool for analysing convective patterns in porous media, enabling us to uncover structural transitions, quantify multi-scale behaviour, and establish links between morphology and macroscopic transport. Unlike Fourier-based or cell-size analyses, persistent homology captures the connectivity and merging of plume structures, allowing identification of topological transitions (e.g. plume coalescence into supercells) that are not directly accessible from spectral measures.
The measures introduced here are fully objective, computationally efficient, and well-suited for large datasets. 
They represent a possible direction for developing predictive reduced-order models or data-driven modelling of heat and solute transport, and can be extended to pore-scale or heterogeneous media \citep{blunt2017multiphase}.
The availability of a large open database of simulations further enables the community to build upon this work. Future research directions include extending PH analyses to three-dimensional structures and integrating topological features into machine-learning frameworks for real-time prediction of heat-transport properties in convective flows.

 The topological measures, therefore, show generic, self-similar characteristics of the emerging patterns for large values of $\ra$.

\backsection[Acknowledgements]{
We acknowledge the contribution by Zhaoshu Cao, who (while at NJIT) carried out some of the topological data analyses performed during the initial stage of this work.
} 

\backsection[Supplementary movies]{
Supplementary movies \textcolor{red}{S1} and \textcolor{red}{S2} are available in the 
\textcolor{blue}{electronic supplementary material}.
}

\backsection[Supplementary material]{
An example of a script required to compute the PH statistics is available in the 
\textcolor{blue}{electronic supplementary material}.
}

\backsection[Funding]{
Funded by the European Union (ERC, MORPHOS, 101163625). Views and opinions expressed are however those of the author(s) only and do not necessarily reflect those of the European Union or the European Research Council. Neither the European Union nor the granting authority can be held responsible for them.
We acknowledge the EuroHPC Joint Undertaking for awarding the projects EHPC-EXT-2024E02-122 and EHPC-BEN-2024B08-060 to access the EuroHPC supercomputer MareNostrum5 hosted the Barcelona Supercomputing Center (Spain).  LK acknowledge partial funding by the NSF grants DMR-2410985 and DMS-2201627. 
M.D. and L.K acknowledge the Global Fellowship Program of TU Wien.
} 

\backsection[Declaration of interests]{The authors report no conflict of interest.}
\backsection[Author ORCID]{
\\Marco De Paoli, \href{https://orcid.org/0000-0002-4709-4185}{https://orcid.org/0000-0002-4709-4185};
\\Sergio Pirozzoli, \href{https://orcid.org/0000-0002-7160-3023}{https://orcid.org/0000-0002-7160-3023};
\\Catherin Neena Lalu, \href{https://orcid.org/0009-0005-5706-8414}{https://orcid.org/0009-0005-5706-8414};
\\Lou Kondic, \href{https://orcid.org/0000-0001-6966-9851}{https://orcid.org/0000-0001-6966-9851}.
}

 \appendix
 \section{Additional details on numerical simulations}\label{sec:appA}
 A summary of the simulations considered is presented in tables~\ref {tab:os} and~\ref {tab:rb}, corresponding to the one-sided and two-sided cases, respectively.
For each simulation, the Rayleigh number $\ra$, the domain extension in convective ($L_1,L_2,H$) and diffusive ($\widehat{L}_1,\widehat{L}_2,\widehat{H}$) units and the grid resolution $N_{x}\times N_{y}\times N_{z}$ are indicated. 
The data analysed relative to simulations~A and~C are available via \citet{databaseC}.
 \afterpage{\clearpage
 \begin{landscape}
   \begin{plaintable}
\footnotesize
  \begin{center}
\def~{\hphantom{0}}
\begin{tabular}{c c c c c c c c c c c c c c c c c} 
 \hline
 &&& \multicolumn{4}{c}{conv. units} &&  \multicolumn{3}{c}{diffusive units} &&  \multicolumn{1}{c}{grid resolution}&&\multicolumn{1}{c}{Ref.}\\
Case && $\ra$ && $L_1$ & $L_2$ & $H$ && $\widehat{L}_1$ & $\widehat{L}_2$ & $\widehat{H}$ && $N_x\times N_y \times N_z$ && \\\\ 
  A1 && $1.0 \times 10^4$ && 1  &   1 & 1   && $1.0\times 10^4$  & $1.0\times 10^4$  & $1.0\times 10^4$ && $768\times768\times256$ && this work\\
  A2 && $1.0 \times 10^4$ && 1  &   1/2 & 1   &&  $1.0\times 10^4$  & $5.0\times 10^3$  & $1.0\times 10^4$ && $768\times384\times256$ && this work\\
  A3 && $1.0 \times 10^4$ && 1  &   1/4 & 1   &&  $1.0\times 10^4$  & $2.5\times 10^3$  & $1.0\times 10^4$ && $768\times192\times256$ && this work\\  
  A4 && $1.0 \times 10^4$ && 1  &   1/8 & 1   &&  $1.0\times 10^4$  & $1.25\times 10^3$  & $1.0\times 10^4$ && $768\times96\times256$ && this work\\
  A5 && $1.0 \times 10^4$ && 1  &   1/16 & 1  &&  $1.0\times 10^4$  & $6.25\times 10^2$  & $1.0\times 10^4$ && $768\times48\times256$ && this work \\
  A6 && $1.0 \times 10^4$ && 2  &   2 & 1  &&  $2.0\times 10^4$  & $2.0\times 10^4$  & $1.0\times 10^4$ && $1536\times1536\times256$ && this work \\   
  A7 && $1.0 \times 10^4$ && 1/2  &   1/2 & 1  &&  $5.0\times 10^3$  & $5.0\times 10^3$  & $1.0\times 10^4$ && $384\times384\times256$ && this work \\
  A8 && $1.0 \times 10^4$ && 1/4  &   1/4 & 1  &&  $2.5\times 10^3$  & $2.5\times 10^3$  & $1.0\times 10^4$ && $192\times192\times256$ && this work \\
  A9 && $1.0 \times 10^4$ && 1/8  &   1/8 & 1  &&  $1.25\times 10^3$  & $1.25\times 10^3$  & $1.0\times 10^4$ && $96\times96\times256$ && this work \\ 
  A10 && $1.0 \times 10^4$ && 1/16  &   1/16 & 1 && $6.25\times 10^3$  & $6.25\times 10^3$ & $1.0\times 10^4$ && $48\times48\times256$ && this work \\   
  \\
        B1 &&  $1.0 \times 10^2$ && 5 & 5 & 1 && $5.0\times10^2$ & $5.0\times10^2$ & $1.0 \times 10^2$ && $64\times 64 \times 16$ && \citet{depaoli2025grl}\\
        B2 &&  $2.0 \times 10^2$ && 5 & 5 & 1 && $1.0\times10^3$ & $1.0\times10^3$ & $2.0 \times 10^2$ && $128\times 128 \times 24$ && \citet{depaoli2025grl}\\ 
        B3 &&  $5.0 \times 10^2$ && 5 & 5 & 1 && $2.5\times10^3$ & $2.5\times10^3$ & $5.0 \times 10^2$ && $256\times 256 \times 32$ && \citet{depaoli2025grl}\\ 
        B4 && $1.0 \times 10^3$ && 5 & 5 & 1 && $5.0\times10^3$ & $5.0\times10^3$ & $1.0 \times 10^3$ && $512\times 512 \times 48$ && \citet{depaoli2025grl}\\ 
        B5 && $2.0 \times 10^3$ && 5 & 5 & 1 && $1.0\times10^4$ & $1.0\times10^4$ & $2.0 \times 10^3$ && $1024\times 1024 \times 64$ && \citet{depaoli2025grl} \\ 
        B6 && $5.0 \times 10^3$ && 2 & 2 & 1 && $1.0\times10^4$ & $1.0\times10^4$ & $5.0 \times 10^3$ && $1024\times 1024 \times 128$ && \citet{depaoli2025grl}\\ 
        B7 && $1.0 \times 10^4$ && 1 & 1 & 1 && $1.0\times10^4$ & $1.0\times10^4$ & $1.0 \times 10^4$ && $1024\times1024 \times 256$ && \citet{depaoli2025grl}\\     
        B8 && $2.0 \times 10^4$ && 1/2 & 1/2 & 1 && $1.0\times10^4$ & $1.0\times10^4$ & $2.0 \times 10^4$ && $1024\times1024 \times 512$ && \citet{depaoli2025grl} \\ 
        B9 && $4.0 \times 10^4$ && 1/2 & 1/2 & 1 && $2.0\times10^4$ & $2.0\times10^4$ & $4.0 \times 10^4$ && $2048\times 2048 \times 1024$ && \citet{depaoli2025grl} \\
        B10 && $8.0 \times 10^4$ && 1/2 & 1/2 & 1 && $4.0\times10^4$ & $4.0\times10^4$ & $8.0 \times 10^4$ && $4096\times 4096 \times 2048$ && \citet{depaoli2025grl}\\
\hline        
 \end{tabular}
 \caption{\label{tab:os}
Summary of the simulations considered in the one-sided case.
For each simulation, the Rayleigh number $\ra$, the domain extension in convective ($L_1,L_2,H$) and diffusive ($\widehat{L}_1,\widehat{L}_2,\widehat{H}$) units and the grid resolution $N_{x}\times N_{y}\times N_{z}$ are indicated. 
Simulations~A are carried out in domains of variable sizes, assuming a constant Rayleigh number ($\ra=1\times 10^4$).
Simulations B \citep[presented by][]{depaoli2025grl} are carried out in square domains ($\widehat{L}_1/\widehat{L}_2=L_1/L_2=1$) while $\ra$ varies. 
The data analysed relative to simulations~A are available via \citet{databaseC}.
}
\end{center}
\end{plaintable}
\end{landscape}
\clearpage
}
\afterpage{\clearpage
 \begin{landscape}
   \begin{plaintable}
\footnotesize
  \begin{center}
\def~{\hphantom{0}}
\centering
\begin{tabular}{c c c c c c c c c c c c c c c c} 
 \hline
 &&& \multicolumn{4}{c}{conv. units} &&  \multicolumn{3}{c}{diffusive units} &&  \multicolumn{1}{c}{grid resolution}&&\multicolumn{1}{c}{Ref.}\\
Case && $\ra$ && $L_1$ & $L_2$ & $H$ && $\widehat{L}_1$ & $\widehat{L}_2$ & $\widehat{H}$ && $N_x\times N_y \times N_z$ && \\ \\
  C1 &  & $5.00 \times 10^1$ && 80 & 80 & 1  &&  $4.0\times 10^3$  & $4.0\times 10^3$  & $5.0\times 10^1$ && $384\times384\times16$ && this work \\
  C2 &  & $1.00 \times 10^2$ && 40 & 40 & 1  &&  $4.0\times 10^3$  & $4.0\times 10^3$  & $1.0\times 10^2$ && $384\times384\times16$ && this work \\
  C3 &  & $2.00 \times 10^2$ && 20 & 20 & 1  &&  $4.0\times 10^3$  & $4.0\times 10^3$  & $2.0\times 10^2$ && $384\times384\times16$ && this work \\
  C4 &  & $3.00 \times 10^2$ && 40/3 & 40/3 & 1  &&  $4.0\times 10^3$  & $4.0\times 10^3$  & $3.0\times 10^2$ && $384\times384\times16$ && this work \\
  C5 &  & $5.00 \times 10^2$ && 8 & 8 & 1  &&  $4.0\times 10^3$  & $4.0\times 10^3$  & $5.0\times 10^2$ && $384\times384\times32$ && this work \\
  C6 &  & $7.50 \times 10^2$ && 16/3 & 16/3 & 1  &&  $4.0\times 10^3$  & $4.0\times 10^3$  & $7.5\times 10^2$ && $384\times384\times32$ && this work \\\\  
  C7 &  & $1.00 \times 10^3$ && 4 & 4 & 1  &&  $4.0\times 10^3$  & $4.0\times 10^3$  & $1.0\times 10^3$ && $384\times384\times32$ && \citet{depaoli2022strong} \\
  C8 &  & $1.25 \times 10^3$ && 4 & 4 & 1  &&  $5.0\times 10^3$  & $5.0\times 10^3$  & $1.25\times 10^3$ && $512\times512\times64$ && this work \\  
  C9 &  & $1.50 \times 10^3$ && 10/3 & 10/3 & 1  &&  $5.0\times 10^3$  & $5.0\times 10^3$  & $1.50\times 10^3$ && $512\times512\times64$ && this work \\  
  C10 &  & $1.75 \times 10^3$ && 20/7 & 20/7 & 1  &&  $5.0\times 10^3$  & $5.0\times 10^3$  & $1.75\times 10^3$ && $512\times512\times64$ && this work \\    
  C11 &  & $2.50 \times 10^3$ && 4 & 4 & 1 &&  $1.0\times 10^4$  & $1.0\times 10^4$  & $2.5\times 10^3$ && $768\times 768\times 64$ && \citet{pirozzoli2021towards}  \\ 
  C12 &  & $5.00 \times 10^3$ && 4 & 4 & 1 &&  $2.0\times 10^4$  & $2.0\times 10^4$  & $5.0\times 10^3$ && $1536\times 1536\times 128$ && \citet{pirozzoli2021towards}  \\\\   
  C13 &  & $7.50 \times 10^3$ && 4 & 4 & 1 &&  $3.0\times 10^4$  & $3.0\times 10^4$  & $7.5\times 10^3$ && $2304\times 2304\times 192$ && \citet{depaoli2022strong}  \\     
  C14 &  & $1.00 \times 10^4$ && 1 & 1 & 1 && $1.0\times 10^4$  & $1.0\times 10^4$ & $1.0\times 10^4$ && $768\times768\times256$ && \citet{pirozzoli2021towards}  \\   
  C15 &  & $1.00 \times 10^4$ && 4 & 4 & 1 && $4.0\times 10^4$  & $4.0\times 10^4$  & $1.0\times 10^4$ && $3072\times3072\times256$ && \citet{depaoli2022strong}   \\
  C16 &  & $2.00 \times 10^4$ && 1 & 1 & 1 && $2.0\times 10^4$  & $2.0\times 10^4$  & $2.0\times 10^4$ && $1536\times1536\times512$ && \citet{pirozzoli2021towards}   \\ 
  C17 &  & $4.00 \times 10^4$ && 1 & 1 & 1 && $4.0\times 10^4$  & $4.0\times 10^4$ & $4.0\times 10^4$ && $3072\times3072\times1024$ && \citet{pirozzoli2021towards}  \\ 
 C18 &  & $8.00 \times 10^4$ && 1 & 1 & 1 && $8.0\times 10^4$  & $8.0\times 10^4$ & $8.0\times 10^4$ && $6144\times6144\times2048$ && \citet{pirozzoli2021towards} \\   
\hline        
 \end{tabular}
 \caption{\label{tab:rb} 
Summary of two-sided flow configuration at different Rayleigh numbers $\ra$ (data are taken from \citet{pirozzoli2021towards} and \citet{depaoli2022strong}). The domain dimensions ($H$, $L_1$ and $L_2$) are indicated. 
Quantities are expressed in dimensionless convective units ($H=1,L_1=L_1^*/H^*,L_2=L_2^*/H^*$, defined as in \S\ref{sec:dimeq34}) and dimensionless diffusive units ($\widehat{H}=\ra,\widehat{L}_1=L_1^*/\ell^*$, $\widehat{L}_2=L_2^*/\ell^*$, defined as in~\S\ref{sec:dimeq35}).
The data analysed relative to simulations~C are available via \citet{databaseC}.
}
\end{center}
\end{plaintable}
\end{landscape}\clearpage
}

\clearpage
 \section{Effect of the domain size }\label{sec:appB}
 
 \subsection{Low Rayleigh numbers}\label{sec:appBlow}
\begin{figure}
     \centering
     \includegraphics[width=0.99\linewidth]{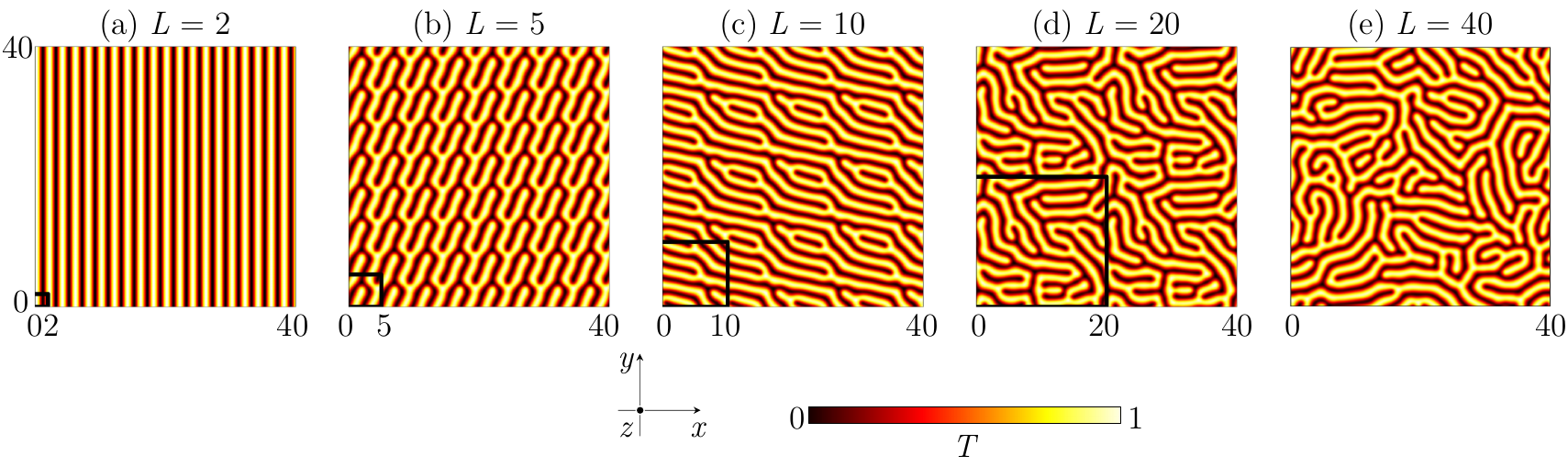}
     \caption{Simulations performed at $\ra=10^2$ and different value of domain width ($L_1=L_2=L$), indicated in each panel. The black, square frame indicates the domain simulated, which is repeated for visualization purposes to match the size of the largest simulation considered, C2, corresponding to $L=40$ (panel~e).}
     \label{fig:infldom}
 \end{figure}

 To illustrate the combined effect that, at low $\ra$, domain size and periodic boundary conditions in the horizontal directions have on the flow morphology, we consider the case at $\ra=10^2$, corresponding to simulation C2 in table~\ref{tab:rb}. In addition to C2 ($L=40$), we perform additional simulations at smaller values of $L_1=L_2=L$. To ensure that steady state is achieved, the simulations are run for $t>5,000$.  Figure~\ref{fig:infldom} then shows the temperature field in the centre of the domain at the final time instant. A gradual increase of the domain size $L$ allows us to observe the change from structures spanning the entire domain in the horizontal direction (figure~\ref{fig:infldom}a) to a more chaotic flow organization (figure~\ref{fig:infldom}d-e). Such transition, however, only mildly affects the Nusselt number, which varies between and 2.54 ($L=5$, figure~\ref{fig:infldom}b) and 2.61 ($L=2$, figure~\ref{fig:infldom}a). 

 \subsection{High Rayleigh numbers}\label{sec:appBhigh}
\begin{figure}
     \centering
     \includegraphics[width=0.49\linewidth]{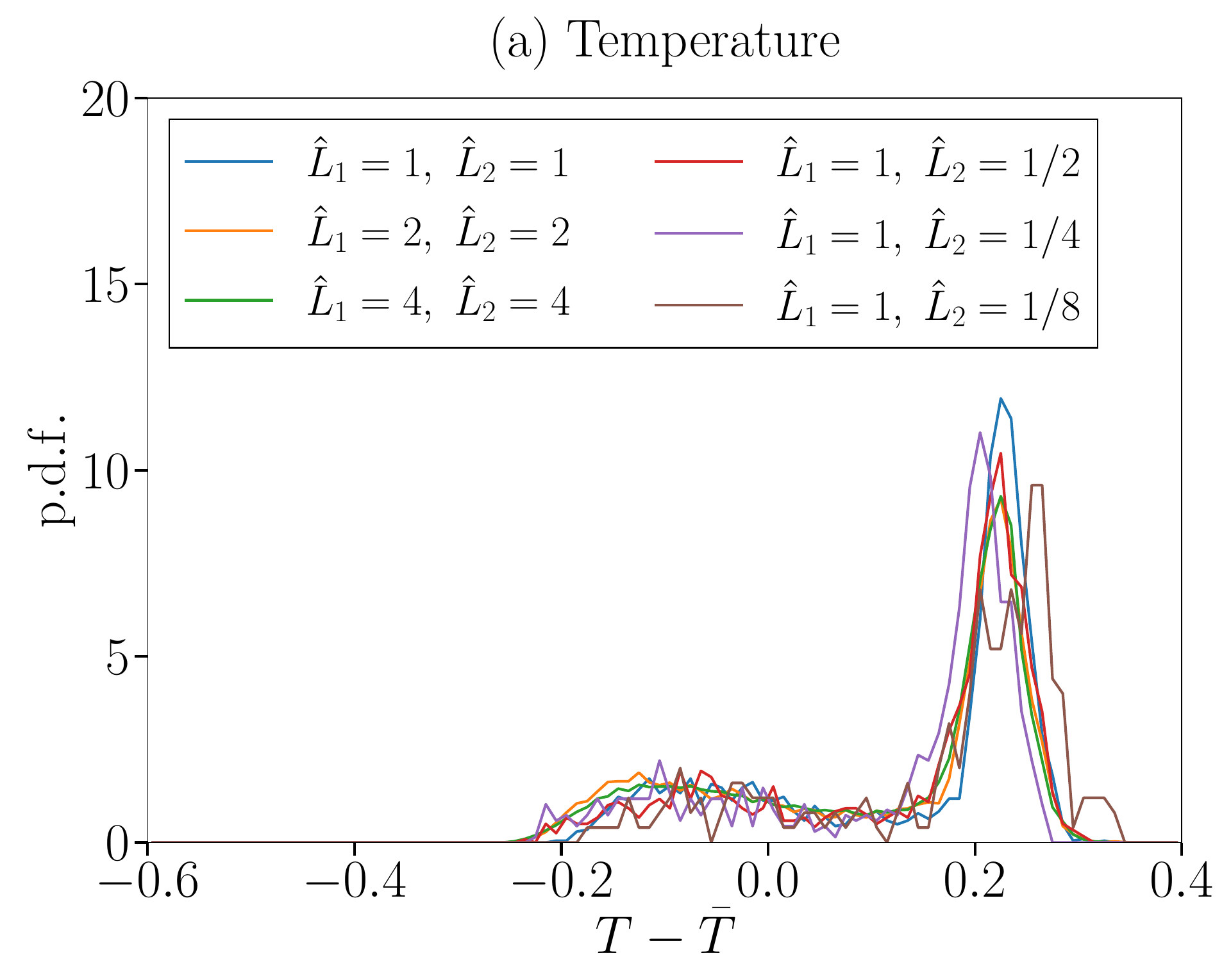}
     \includegraphics[width=0.49\linewidth]{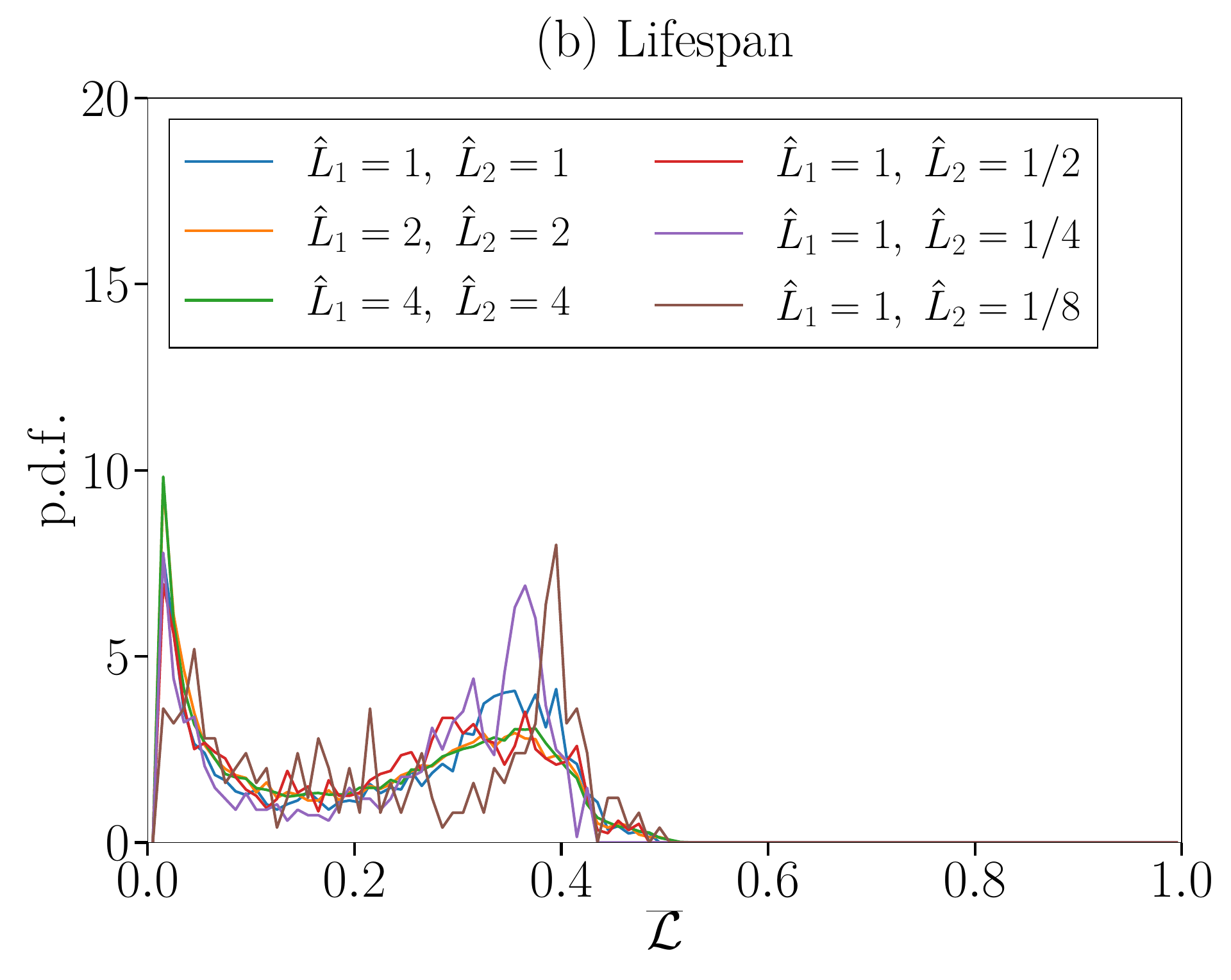}
    \caption{
       Influence of the domain size aspect ratio for $\ra = 10^4$ (compare figure~\ref{fig:rescaled}).  
    (a)
     Probability density function (p.d.f.) of $\beta_1$ topological generators as a function of the shifted temperature, with $\overline{T}$ corresponding to the space and time averaged temperature in the considered horizontal slice; (b) p.d.f. of the lifespan, $\overline{\cal L}$ (noise band 0.01 is removed). }     
     \label{fig:domain_size}
 \end{figure}

\begin{figure}
     \centering
     \includegraphics[width=0.49\linewidth]{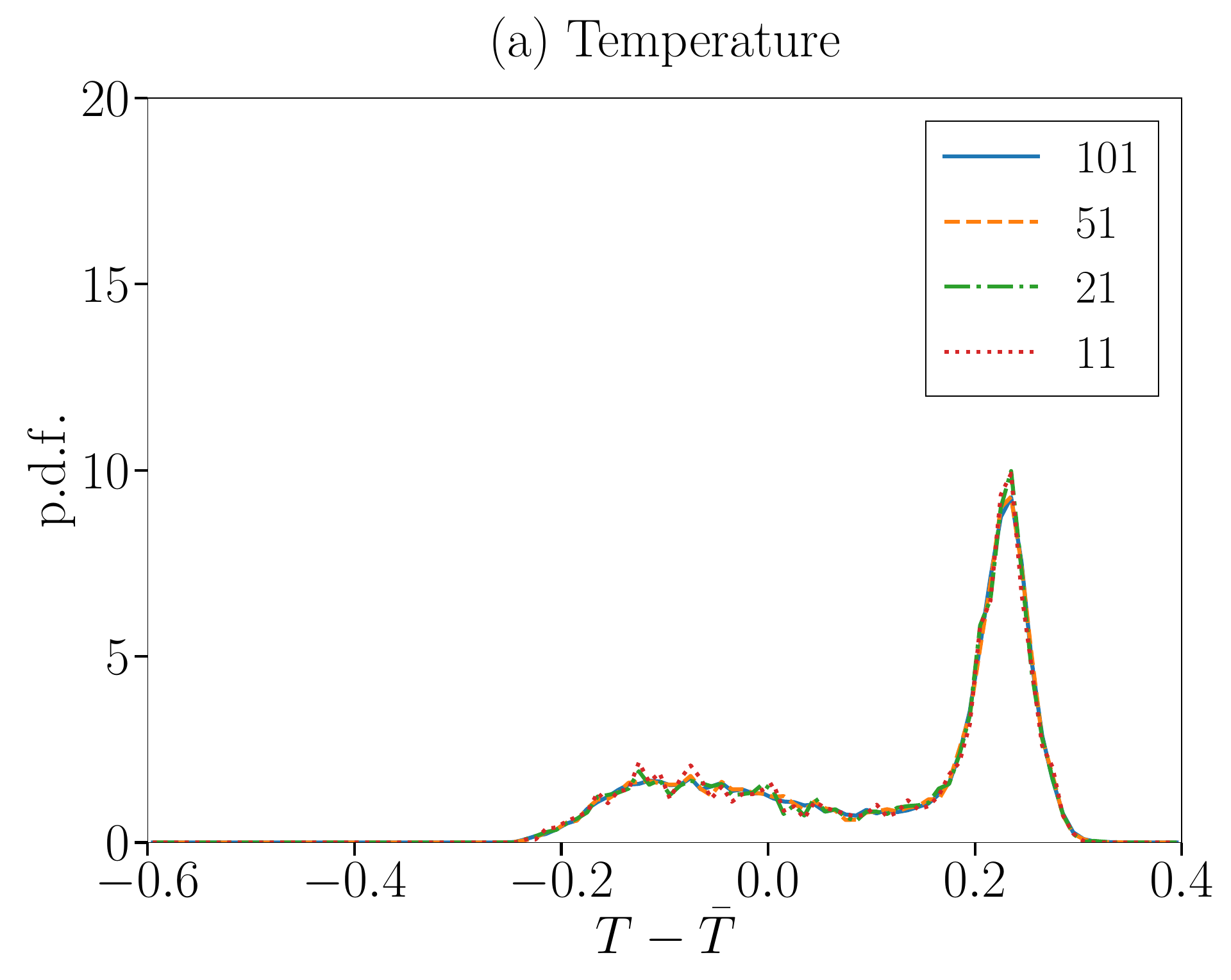}
     \includegraphics[width=0.49\linewidth]{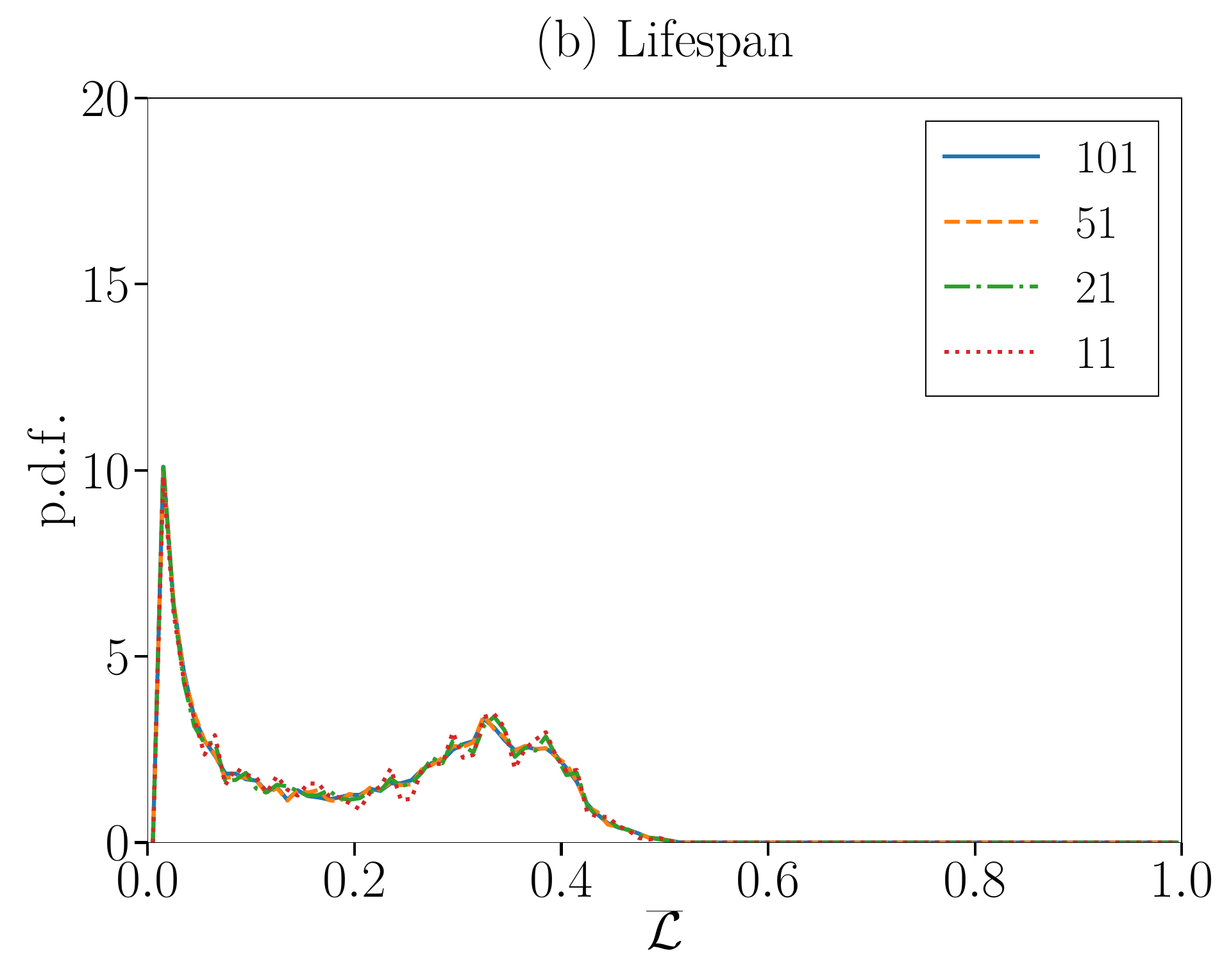}
    \caption{
       Influence of the number of considered samples for $\ra = 10^4$ (compare figure~\ref{fig:rescaled}).  
    (a)
     Probability density function (p.d.f.) of $\beta_1$ topological generators as a function of the shifted temperature, with $\overline{T}$ corresponding to the space and time averaged temperature in the considered horizontal slice; (b) p.d.f of the lifespan, $\overline{\cal L}$ (noise band 0.01 is removed). }     
     \label{fig:no_of_samples}
 \end{figure}

To illustrate the robustness of the results, here we include additional results for $\ra = 10^4$ as the aspect ratio is modified, see figure \ref{fig:domain_size}.  We have also confirmed that the results are robust with respect to the number of samples considered, see figure~\ref{fig:no_of_samples}.  These results confirm the robustness of our findings regarding the collapse of the PH measures for large values of $\ra$. 

 \section{Influence of the removed band width}\label{sec:appC}

\begin{figure}
    \centering
    \includegraphics[height=0.39\linewidth]{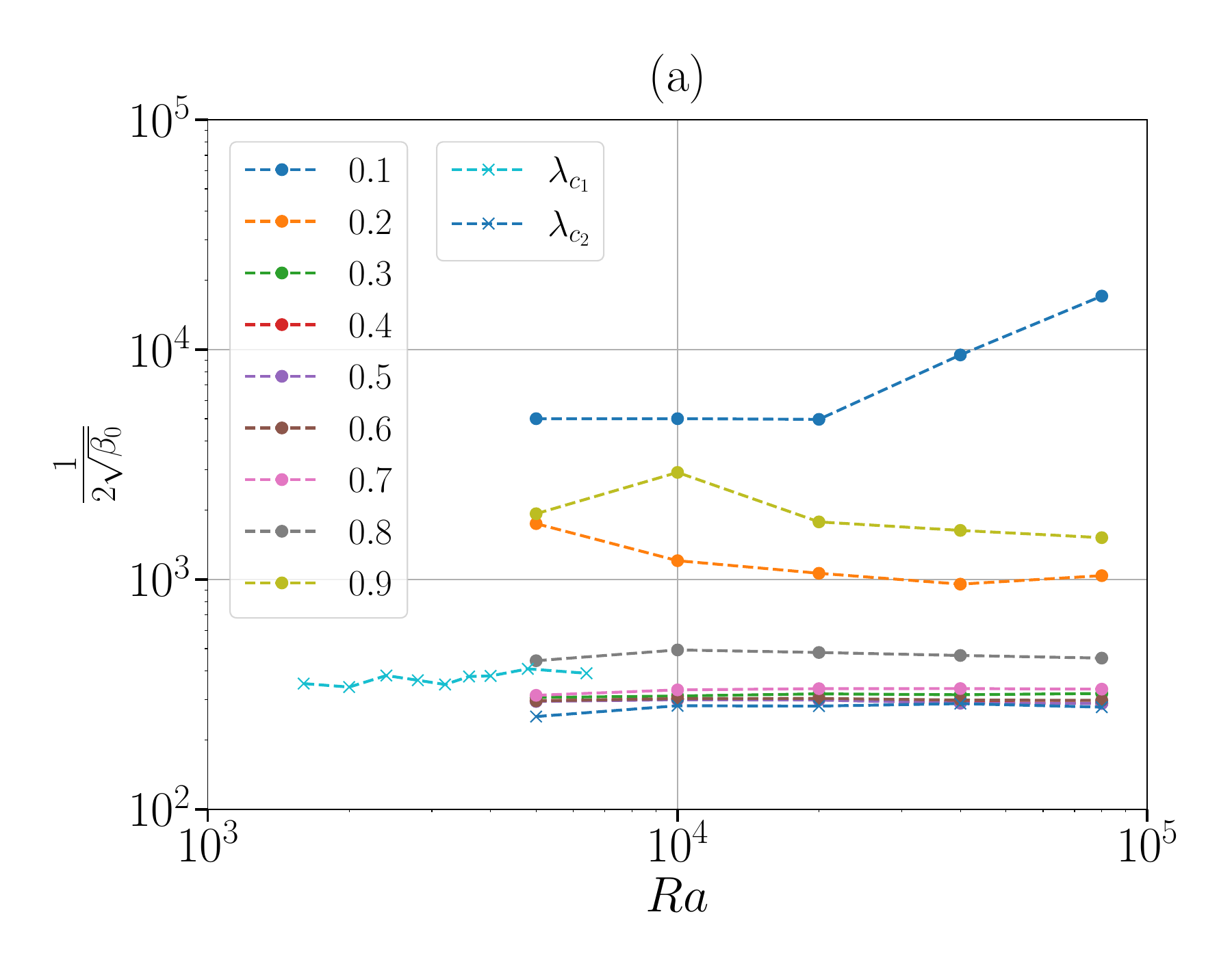}
    \includegraphics[height=0.39\linewidth]{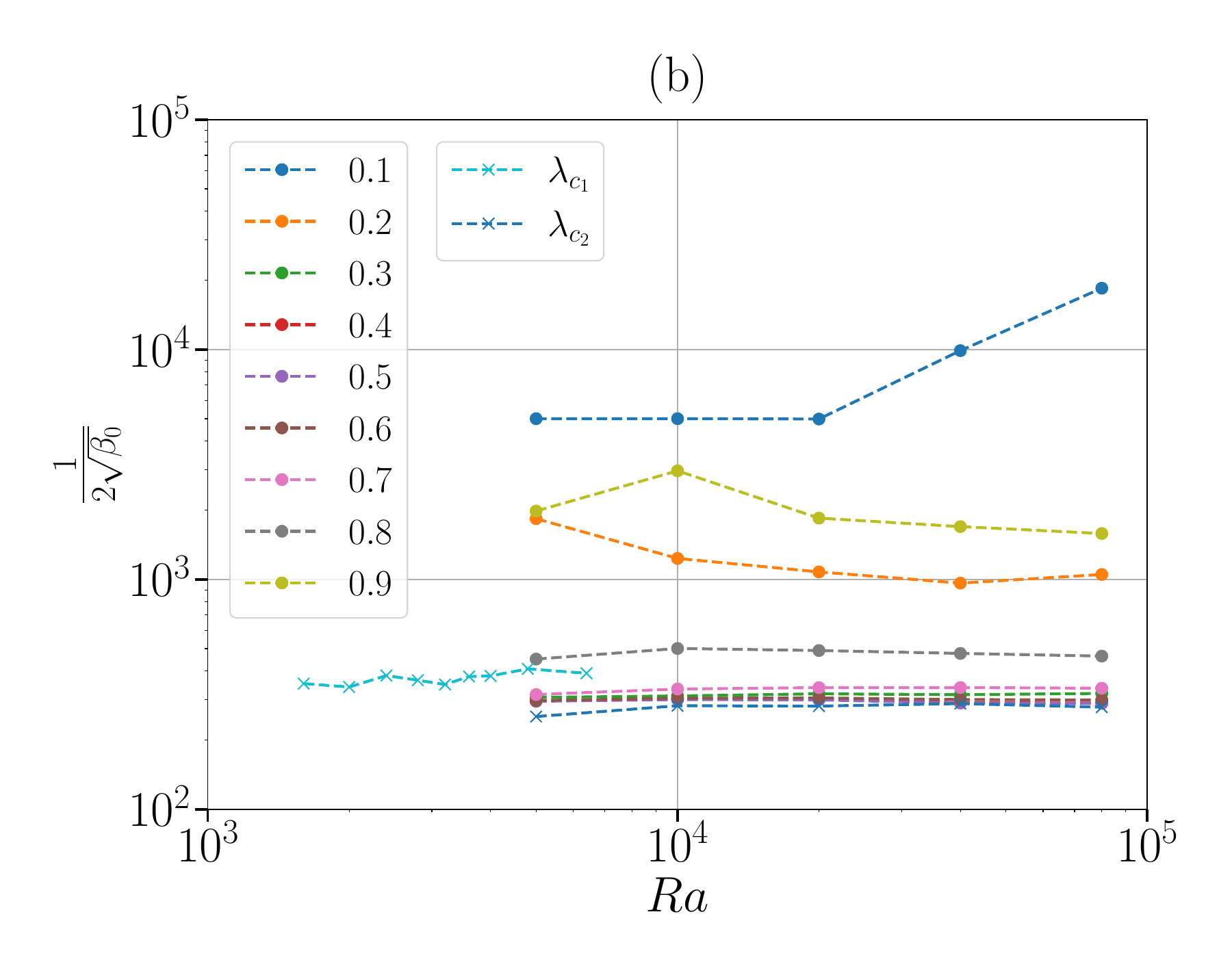}\\   
    \caption{Betti numbers (per area) for B6~-~B10 simulations for different values of the band width; compare with figure~\ref{fig:B6-B10lengthscale}(a). (a) removing the 0.005 band, and (b) removing the 0.02 band.}
    \label{fig:noise1}
\end{figure}

 Here we briefly comment regarding the influence of the width of the excluded band used when computing the quantities of interest, such as emerging lengthscales. As an illustrative example, figure~\ref{fig:noise1} reproduces figure~\ref{fig:B6-B10lengthscale}(a) from Sec.~\ref{sec:Ra} obtained by changing this width; direct comparison shows that the results are essentially independent of the band width.  
 
\bibliographystyle{jfm}
\bibliography{bibliography}

\end{document}